\documentclass[sn-basic]{sn-jnl}% Math and Physical Sciences Author Year Reference Style
\setcitestyle{aysep={,}}
\usepackage{graphicx}%
\usepackage{multirow}%
\usepackage{multicol}%
\usepackage{amsmath,amssymb,amsfonts}%
\usepackage{amsthm}%
\usepackage{mathrsfs}%
\usepackage[title]{appendix}%
\usepackage{xcolor}%
\usepackage{textcomp}%
\usepackage{manyfoot}%
\usepackage{booktabs}%
\usepackage{algorithm}%
\usepackage{algorithmicx}%
\usepackage{algpseudocode}%
\usepackage{listings}%
\usepackage{url}%
\usepackage{rotating}%
\usepackage{siunitx}%
\usepackage{caption}
\usepackage{placeins}

\newcommand*\fdg{\ensuremath{\overset{\circ}{.}}}

\newcommand{\reftab}[1]{{\footnotesize\textcolor{red}{\,$[#1]$}}}

\usepackage{float}
\usepackage[T1]{fontenc}

\begin{document}

\title[Accreting stellar-mass black holes]{Accreting stellar-mass black holes}


\author*[1]{\fnm{Greg} \sur{Marcel}}\email{gregoiremarcel26@gmail.com}

\author[2]{\fnm{Bailey} \sur{Tetarenko}} %\email{iiauthor@gmail.com}
\equalcont{These authors contributed equally to this work.}

\author[3,4]{\fnm{Adam} \sur{Ingram}} %\email{iiiauthor@gmail.com}
\equalcont{These authors contributed equally to this work.}

\author[5]{\fnm{Tom} \sur{Maccarone}} %\email{iiiauthor@gmail.com}
\equalcont{These authors contributed equally to this work.}

\author[1]{\fnm{Alexandra} \sur{Veledina}} %\email{iiiauthor@gmail.com}
\equalcont{These authors contributed equally to this work.}

\author[6]{\fnm{Phil} \sur{Charles}} %\email{iiiauthor@gmail.com}

\affil*[1]{\orgdiv{Department of Physics and Astronomy}, \orgname{University of Turku}, \orgaddress{\city{Turku}, \postcode{FI-20014}, \country{Finland}}}

\affil[2]{\orgdiv{Trottier Space Institute at McGill}, \orgname{McGill University}, \orgaddress{\street{3550 University Street}, \city{Montr\'eal}, \postcode{H3A 2A7}, \state{Quebec}, \country{Canada}}}

\affil[3]{\orgdiv{School of Mathematics, Statistics, and Physics}, \orgname{Newcastle University}, \orgaddress{\city{ Newcastle upon Tyne}, \postcode{NE1 7RU}, \country{United Kingdom}}}

\affil[4]{\orgdiv{Centre for Extragalactic Astronomy, Department of Physics}, \orgname{Durham University}, \orgaddress{\city{ South Road, Durham}, \postcode{DH1 3LE}, \country{United Kingdom}}}

\affil[5]{\orgdiv{Department of Physics \& Astronomy}, \orgname{Texas Tech University}, \orgaddress{\street{Box 41051}, \city{Lubbock}, \postcode{79409-1051}, \state{Texas}, \country{United States of America}}}

\affil[6]{\orgdiv{Department of Physics \& Astronomy}, \orgname{University of Southampton}, \orgaddress{\city{Southampton}, \postcode{SO17 1BJ}, \country{United Kingdom}}}

%%================================%%
%% Sample for structured abstract %%
%%================================%%

\abstract{

Accreting stellar-mass black holes exhibit dramatic variability across the electromagnetic spectrum, including spectral state transitions, outbursts, and jet production, making them unique laboratories for understanding accretion processes in strong gravitational fields. This review synthesizes recent progress in understanding these systems, focusing on their continuum emission, timing properties, emission lines, and X-ray polarization.
A complex interplay between the accretion disk, the so-called corona, and jet underlies the observed spectral and timing behavior, with quasi-periodic oscillations and broadband noise providing windows into the dynamics of the innermost accretion flow. Emission lines across all wavelengths serve as critical diagnostics of the disk structure, outflows, and reprocessing, while iron K lines in the X-ray band probe the properties of the inner disk through relativistic reflection. Polarization studies suggest that the corona is likely extended perpendicular to the jet axis in the hard state, while the soft state remains poorly understood, with observations that do not yet conform to simple theoretical expectations; a puzzle that continues to challenge our interpretation of accretion geometry.
Despite significant advances, fundamental questions remain about the physical origins of state transitions, the role of magnetic fields in driving outflows and shaping the accretion flow, and the connection between disk instabilities and jet launching. This review underscores the need for future multi-wavelength, timing, and polarimetric studies to deepen our understanding of accretion physics in strong-gravity environments.

}




\maketitle

%\FloatBarrier
\section{Introduction} \label{sec:intro}

\subsection{A brief history of black hole X-ray binaries}

X-ray binaries are binary systems consisting of a compact object, either a black hole or a neutron star, and a companion star. Their observational appearance is entirely driven by accretion: gravitational energy released as matter flows from the companion onto the compact object powers the bright, variable emission that makes these systems detectable across the electromagnetic spectrum. The first detected extra-solar X-ray source \citep[Sco~X-1,][]{1962PhRvL...9..439G} was later identified as hosting a neutron star, and the first confirmed black hole X-ray binary (hereafter BHXB), Cyg~X-1, was discovered shortly afterwards in the Cygnus constellation \citep{1965Sci...147..394B}. These discoveries spurred a rapid growth of X-ray observations \citep[e.g.,][]{1964AJ.....69Q.135B, 1964ApJ...140..460M, 1967Natur.215...38H, 1967Natur.216..773F, 1968ApJ...152L..45C, 2003IJMPA..18.3127G}, as well as the identification of their radio \citep[e.g.,][]{1968Natur.218..855A, 1971ApJ...164L...1H} and optical \citep[e.g.,][]{1966ApJ...146..316S, 1968SvA....11..749S} counterparts, laying the groundwork for multi-wavelength astrophysical studies.

Over the past five decades, it has become clear that black hole X-ray binaries cycle through distinct spectral states, each characterized by a markedly different emission pattern. A pivotal early example was reported for Cyg~X-1 by \citet{1972ApJ...177L...5T}:

\begin{quote}
\it We report on a remarkable transition which occurred during 1971 March and April. The average X-ray intensity in the 2-6\,keV energy range decreased by about a factor of 4, the average X-ray intensity in the 10-20\,keV band increased by a factor of 2, and a weak radio source suddenly appeared.
\end{quote}

This observation, now over 50 years old, remains at the heart of the mystery surrounding X-ray binaries. It not only demonstrated the dynamic behavior of these systems but also posed a critical question: what physical processes could explain such dramatic, correlated variations across the electromagnetic spectrum?
Theoretical considerations soon followed, and viscosity emerged as the central concept. It was first formalized in the seminal $\alpha$-disk model of \citet{SS73}, providing the framework to explain how angular momentum transport drives accretion and hence the observed luminosity and variability \citep{1966Natur.212..885M, 1967IAUS...31..463B, 1967ApJ...148L...1S}. These ideas became the building blocks of accretion physics \citep[see also][]{PR72, SLE76}, the process that remains central to our understanding of all accreting objects: from proto-stars to active galactic nuclei.

\begin{figure}[h!]
\centering
\includegraphics[width=0.49\textwidth]{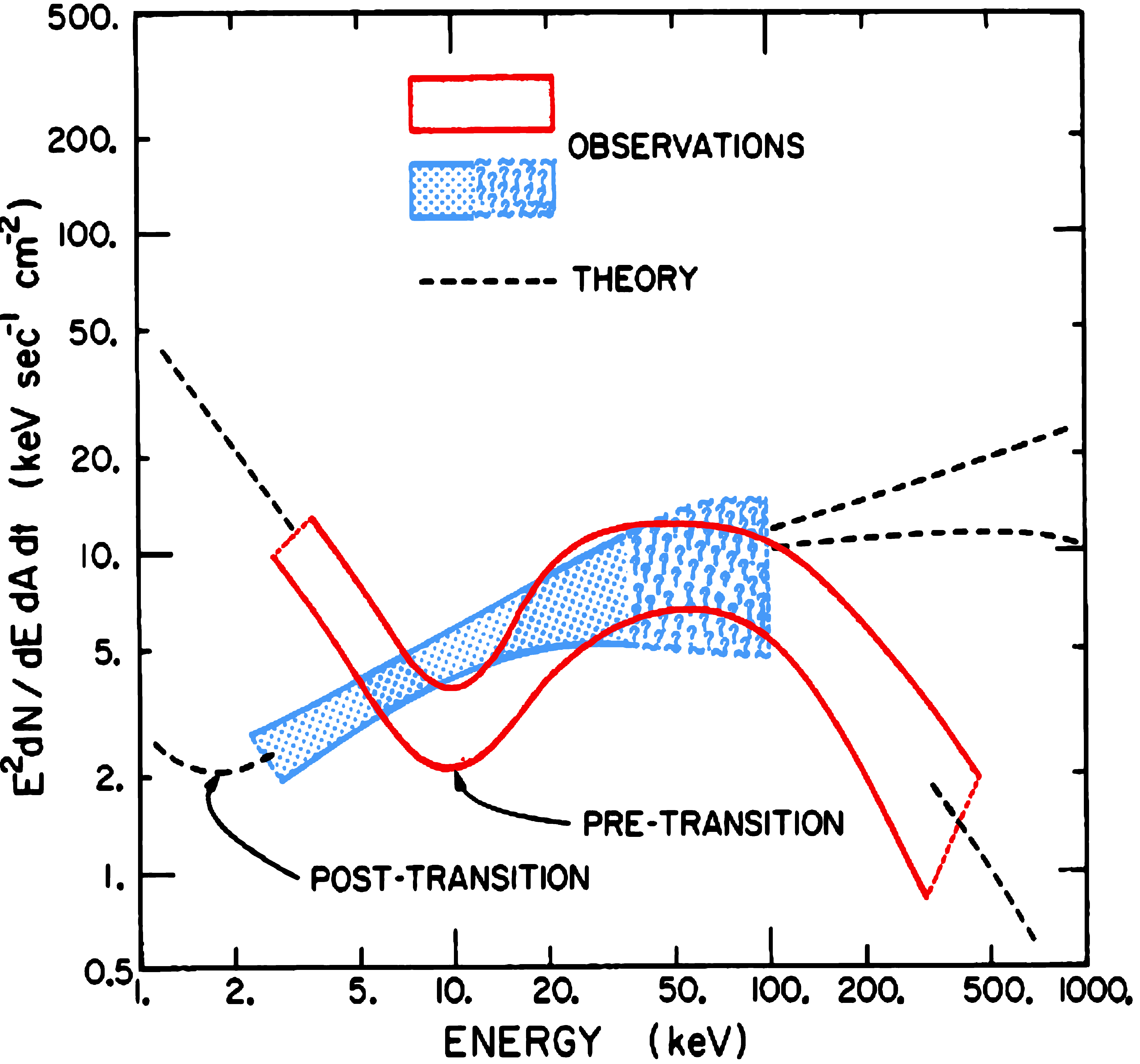} \includegraphics[width=0.49\textwidth]{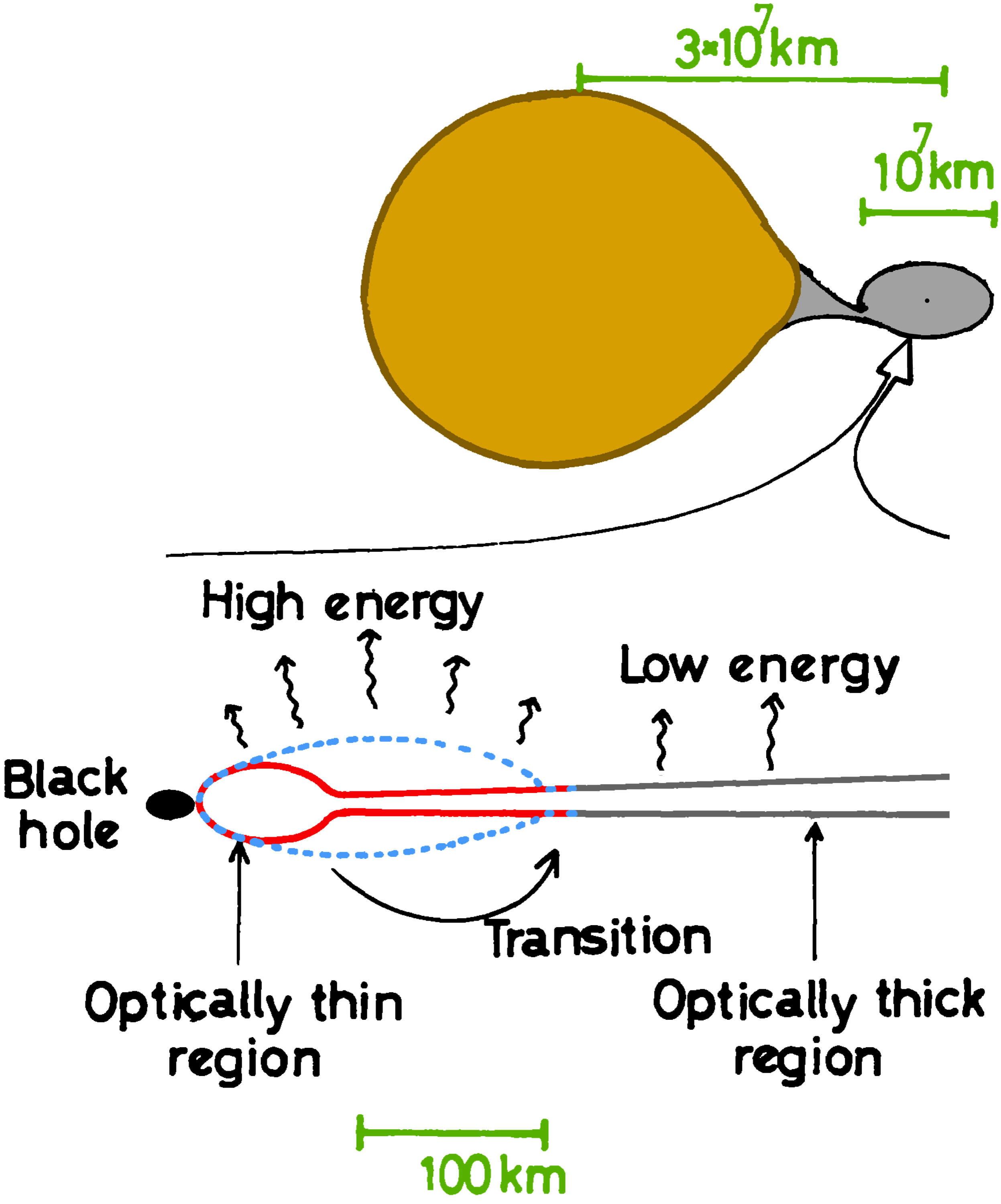}
\caption{Left: Spectral transition from Cyg~X-1 as observed by \citet{1972ApJ...177L...5T}, where the data is represented in colored-regions, adapted from \citet{1975ApJ...195L.101T}. Right: Tentative two-state explanation, where a change in the transition radius would change the relative contributions of the inner optically thin and outer optically thick regions, adapted from \citet{Oda77}.}
\label{fig:ThorneOda}
\end{figure}

Early attempts to explain the spectral transitions of BHXBs included models by \citet{1975ApJ...195L.101T}, \citet{Oda77}, and \citet{1977ApJ...214..840I}, illustrated in Fig.~\ref{fig:ThorneOda}. The left panel shows the two spectral states of Cyg~X-1 during its 1971 transition between a high state (red) and a low state (blue), named after their soft X-ray flux. The black portions of these spectra are used for illustrative purposes and are speculative (though realistic considering today's knowledge). The right panel illustrates the proposed two-zone accretion disk model, where the disk is divided into two distinct regions separated by a transition radius, $r_{\rm t}$. The optically thick outer disk generates a softer spectrum (the lower spectral bump around and below 1\,keV), while the optically thin inner region is responsible for the emission above 1--10\,keV (depending on the state). Variations in $r_{\rm t}$ shift the spectral ``notch'', explaining the observed transitions.
In this proposition, the mechanism responsible for variations in $r_{\rm t}$, and hence the transition between these two regimes, remained unclear. The authors suggested that:

\begin{quote}
\it In the context of the standard model, the 1971 transition might have been caused by a change in any of several variables—e.g., the time-averaged accretion rate, the average magnetic field strength, the turbulent viscosity, etc.
\end{quote}

These ideas were remarkably prescient. As we review in this paper, variations in the accretion rate \citep{1997ApJ...489..865E, 1997MNRAS.292L..21P}, viscosity \citep{1996A&A...314..813L, 2014ApJ...782L..18B}, and magnetic field strength \citep{Contopoulos98, 2006A&A...447..813F} are now considered among the leading ingredients driving state transitions in BHXBs.

\subsection{Confirmed black hole X-ray binaries and their parameters}

We now know hundreds of X-ray binaries, with dedicated catalogs, for example Binary rEvolution \citep{fortin2023, fortin2024}, XRBcats \citep{2023A&A...675A.199A} or BlackCAT for transients with confirmed/candidate black holes \citep{corralsantana2016}. These systems harbor a compact object of mass $M_{\rm X}$ that can either be a black hole or a neutron star, and are usually classified depending on the mass $M_2$ of their companion: high-mass (HMXB, $M_2 \gtrsim 8-10\,M_{\odot}$) or low-mass (LMXB, $M_2 \lesssim 1\,M_{\odot}$). An important difference is that most HMXBs are persistent while most LMXBs are transient, likely owing to continuous mass transfer from stellar wind in the former \citep{2012MNRAS.424.1991C}. As a result, HMXBs are often treated separately in the literature, but we do not make this distinction here and include all XBs. For a review of neutron-star X-ray binaries, we refer the reader to Baglio et al. (submitted to SSRv).

\begin{table}[h!]
\caption{\label{tab:BHXBs} {\sc Dynamically confirmed black hole X-ray binaries and their parameters, ordered by mass of companion star}}
\begin{tabular}{llllll}%{lccccc}
\hline \hline
Object & $P_{\rm orb}$\reftab{i} & $d$ & $M_{\rm X}$ & $M_2$ & $i_{\rm orbit}$ \\
 & days & kpc & $M_\odot$ & $M_\odot$ & $(^\circ)$ \\
\hline
M33~X-7\reftab{6}  & 3.45\reftab{8}    & 750--1017\reftab{13}      & 15.6$\pm$1.5\reftab{10}            & 70.0$\pm$6.9\reftab{10}    & 74.6$\pm$1.0\reftab{10} \\
Cyg~X-1\reftab{1}  & 5.60\reftab{7}    & 2.2$\pm$0.2\reftab{30}    & 21.2$\pm$2.2\reftab{ii,30}        & 41$\pm$7\reftab{ii,30}       & 27.5$\pm$0.8\reftab{ii,14,30} \\
LMC~X-1\reftab{2}  & 3.90\reftab{12}    & 55\reftab{11}              & 10.9$\pm$1.4\reftab{12}            & 31.8$\pm$3.5\reftab{12}    & 36.3$\pm$1.9\reftab{12} \\
SS~433\reftab{4}   & 13.08\reftab{9}   & $\sim$4.5\reftab{27}      & 4.2$\pm$0.4\reftab{27}            & 11.3$\pm$0.6\reftab{27}   & $\sim$80\reftab{5} \\
\hline
LMC~X-3\reftab{3}  & 1.70\reftab{17}   & 55\reftab{11}              & 7.0$\pm$0.6\reftab{17}            & 3.6$\pm$0.6\reftab{17}    & 69.2$\pm$0.7\reftab{17} \\
SAX~J1819.3--2525$^a$ & 2.82            & 4.7\reftab{iii}           & 6.4$\pm$0.6                       & 2.9$\pm$0.4               & 72$\pm$4 \\
4U~1543--475        & 1.12              & 6.1--12.4                 & 9.4$\pm$2.0\reftab{24}            & 2.7$\pm$1.0               & 20.7$\pm$1.5 \\
GRO~J1655--40       & 2.62              & 3.2$\pm$0.2               & 6.0$\pm$0.4                       & 2.5$\pm$0.2               & 69$\pm$2 \\
GRO~J0422+32        & 0.21              & 2.5$\pm$0.3               & 2--15                             & 0.2--2.4                  & 10--50 \\
GRS~1124--684$^b$   & 0.43              & 5.0$\pm$0.7\reftab{18}    & 9.6--13.1\reftab{18}              & 0.78--1.07\reftab{18}     & 40.5--45.3\reftab{18} \\
GX~339--4           & 1.76              & $\gtrsim$5\reftab{19,iv}     & 2.3--9.5\reftab{19,iv}               & 0.4--1.7\reftab{19,iv}       & 37--78\reftab{19,iv} \\
GS~1354--64$^c$     & 2.55              & $\sim$25                  & $\leq$7.0$\pm$0.7                 & $\leq$0.9$\pm$0.1         & $\leq$79 \\
Swift~J1727.8--1613\reftab{33} & 0.45\reftab{35} & 3.4$\pm$0.3\reftab{35,36} & $>$3.1$\pm$0.1\reftab{35}   & $\lesssim 0.7$ \reftab{v}               & $<$74\reftab{35} \\
XTE~J1650--500      & 0.32              & 2.6$\pm$0.7               & 4.0--7.3                          & $\lesssim 0.7$ \reftab{v}               & $>$47 \\
GRS~1915+105        & 33.83             & 9.4$\pm$1.4\reftab{34}    & 12.4$\pm$2.0\reftab{34}           & 0.28--1.00\reftab{21}                 & 64$\pm$4\reftab{34} \\
GS~2023+338$^d$     & 6.47              & 2.39$\pm$0.14             & 8.4--9.2                          & 0.56--0.62                & 66--70 \\
XTE~J1859+226       & 0.27              & 12.5$\pm$1.5              & 7.8$\pm$1.9\reftab{32}            & 0.5$\pm$0.2\reftab{32}    & 67$\pm$4\reftab{32} \\
MAXI~J1820+070\reftab{22,23} & 0.69\reftab{24,26}   & 3.0$\pm$0.3\reftab{25}    & 5.73--10.50\reftab{26,28}          & 0.5$\pm$0.1\reftab{31}     & 66--81\reftab{28} \\
A~0620--00          & 0.32              & 1.1$\pm$0.1               & 5.9$\pm$0.3\reftab{20}            & 0.49$\pm$0.02             & 54.1$\pm$1.1\reftab{20} \\
MAXI~J1305--704\reftab{15,16}& 0.40\reftab{29}   & 6.1--9.3\reftab{29}   & 7.9--10.5\reftab{29}              & 0.4$\pm$0.2\reftab{29}    & 65--77\reftab{29} \\
H~1705--250$^e$     & 0.52              & 8.6$\pm$2.1               & 4.9--7.9                          & $\leq$0.4                 & 48--80 \\
XTE~J1550--564      & 1.54              & 4.5$\pm$0.5               & 7.8--15.6                         & 0.23--0.47                & 74.7$\pm$3.8 \\
GS~2000+25          & 0.34              & 2.7$\pm$0.7               & 5.5--8.8                          & 0.22--0.35                & 54--60 \\
GRS~1009--45$^f$    & 0.28              & 3.8$\pm$0.3               & $\geq$4.4                         & $\geq$0.24                & 37--80 \\
XTE~J1118+480       & 0.17              & 1.7$\pm$0.1               & 6.9--8.2                          & 0.17--0.20                & 68--79 \\
\hline \hline
\end{tabular}
\footnotesize
The horizontal line separates what are usually called high-mass (above) and low-mass (below) X-ray binaries. \reftab{i}\,No error-bars are shown but all are far below $0.01$ days. \reftab{ii}\,More recent work found $M_{\rm X} = 12.7-17.8\,M_\odot$, and $M_2 \approx 29\,M_\odot$, and $i_{\mathrm{orbit}} \approx 33.7$ \citep{2025A&A...698A..37R}, consistent with previous estimates \citep{2008MNRAS.390.1762D}. \reftab{iii}\,From Gaia DR3. \reftab{iv}\,See also \citet{2019MNRAS.488.1026Z}. \reftab{v}\,No constraints on the mass, but the spectral type (K4V, \citealt{2004ApJ...616..376O} for XTE~J1650--500, \citealt{2025A&A...693A.129M} for Swift~J1727.8--1613) suggests a mass in this range. Alternative names: $^a$V4641~Sgr, $^b$N~Mus~1991 or GU~Mus, $^c$BW~Cir, $^d$V404~Cyg, $^e$N~Oph~1977, $^f$N~Vel~1993 or MM~Vel. References when different from \citet{corralsantana2016}:
[1]~\citet{1965Sci...147..394B}
[2]~\citet{1969ApJ...155L.143M},
[3]~\citet{1971ApJ...170L..67L},
[4]~\citet{1978IAUC.3314....2M},
[5]~\citet{1980ApJ...241..306M},
[6]~\citet{1989ApJ...336..140P},
[7]~\citet{1999A&A...343..861B},
[8]~\citet{1999MNRAS.302..731D},
[9]~\citet{2007A&A...474..903B},
[10]~\citet{Orosz07},
[11]~\citet{2008MNRAS.390.1762D},
[12]~\citet{2009ApJ...697..573O},
[13]~\citet{2010AIPC.1314..285V},
[14]~\citet{2011ApJ...742...84O},
[15]~\citet{2012ATel.4024....1S},
[16]~\citet{2013ApJ...779...26S},
[17]~\citet{2014ApJ...794..154O},
[18]~\citet{2016ApJ...825...46W},
[19]~\citet{heida2017},
[20]~\citet{2017MNRAS.472.1907V},
[21]~\citet{2017MNRAS.469.3315Z},
[22]~\citet{2018ATel11399....1K},
[23]~\citet{2018ApJ...867L...9T},
[24]~\citet{2019MNRAS.485.2642G},
[25]~\citet{2019ApJ...882L..21T},
[26]~\citet{2020MNRAS.493L..81A},
[27]~\citet{2020A&A...640A..96P},
[28]~\citet{TorresMAXI1820},
[29]~\citet{2021MNRAS.506..581M},
[30]~\citet{MillerJones2021},
[31]~\citet{2022ApJ...930....9M},
[32]~\citet{2022MNRAS.517.1476Y},
[33]~\citet{2023ATel16205....1N},
[34]~\citet{2023ApJ...959...85R},
[35]~\citet{2025A&A...693A.129M},
[36]~\citet{2025ApJ...994..243B}.
\end{table}

We show the current list of dynamically confirmed black-hole X-ray binaries in Table~\ref{tab:BHXBs}. When no reference is given, we refer the reader to \citet{corralsantana2016}. It is important to note that some estimates are mutually dependent, and are thus subject to caveats. One example concerns SAX~J1819.3--2525, whose mass was estimated assuming $d=6.2\pm0.7$\,kpc \citep{2014ApJ...784....2M}, significantly different from the recently obtained Gaia distance $d=4.7$\,kpc. Another example is MAXI~J1820+070, where the constraints on jets $63\pm3^\circ$ \citep{2020MNRAS.493L..81A} and orbital inclination $66^\circ$--$81^\circ$, marginally consistent, provide different constraints on the masses \citep{2022ApJ...930....9M}. Moreover, inclinations can be obtained via different methods \citep[see Table~1 in][]{2024A&A...681A..49P}, and while some disk-jet systems appear well aligned \citep[e.g., Cyg~X-1,][]{Kravtsov2025}, others are misaligned, such as MAXI~J1820+070 \citep{2022Sci...375..874P}. Inclination estimates should be interpreted with this caveat in mind.

The values given in Table~\ref{tab:BHXBs} are constrained via the mass function. It has also been demonstrated that the FWHM of the H$\alpha$ emission line correlates tightly with the radial velocity semi-amplitude of the donor star, $K_2 \propto \mathrm{FWHM}$ \citep{casares2015}, providing an independent route to constraining system parameters directly from spectroscopy; see Sect.~\ref{sec:OIRlines}. This method has been applied to, e.g., MAXI~J1659--152 \citep[$M_{\rm X} = 3.3-7.5\,M_{\odot}$,][]{2013A&A...552A..32K, 2018MNRAS.475.1036C, torres2021} and TrA~X-1 \citep[also called KY~TrA or 3A~1524--617, $M_{\rm X} = 5.8^{+3.0}_{-2.4}\,M_{\odot}$,][]{yanesrizo2024}. Though purely empirical, this relation yields mass constraints consistent with those derived from the mass function, and we consider it reliable.

Other mass estimation methods exist, but their reliability remains controversial; either because they lack a clear physical motivation or because they rest on model assumptions that have not been independently confirmed. One such approach uses the state-transition luminosity, applied for example to MAXI~J0637--430 \citep[$M_{\rm X}=5$--$12\,M_{\odot}$,][]{2021MNRAS.504.4793J}, MAXI~J1727--203 \citep[$M_{\rm X}\geq 11.5\,M_{\odot}$,][]{2022MNRAS.514.5320W}, or 1E~1740.7--2942 \citep[$M_{\rm X}\approx 5\,M_{\odot}$,][]{2020MNRAS.493.2694S}. Another is direct spectral modeling, as employed for Swift~J1728.9--3613 \citep[$M_{\rm X} \sim 4.6\,M_{\odot}$,][]{2023MNRAS.519..519S}, or MAXI~J1910--057 \citep[$M_{\rm X}=6.31-13.65\,M_{\odot}$,][]{2023AdSpR..71.1045N}. Finally, some sources exhibit behavior strongly indicative of a black hole, either through their spectral shape \citep{2007A&ARv..15....1D}, timing properties \citep{2000A&A...358..617S}, or multi-wavelength behavior. Examples include 4U~1630--47 \citep{1998NewAR..42..613K}, MAXI~J1848--015 \citep{2022ApJ...927..190P}, MAXI~J1803--298 \citep{2022ApJ...927..151S}, XTE~J1728--295 \citep[also called IGR~J17285--2922,][]{2021MNRAS.507..330S}, Swift~J174510.8--262411 \citep{2013MNRAS.432.1133M}, MAXI~J1836--194 \citep{2014MNRAS.439.1381R}, Cyg~X-3 \citep{2000ApJ...541..308H, 2009A&A...501..679V, 2022ApJ...926..123A}, MAXI~J1810--222 \citep{2022MNRAS.513.6196R}, or 4U~1957+115 \citep{2011ApJ...730...43B, 2015ApJ...809....9G, 2021RAA....21..214S, Marra2024}.
While these methods have contributed significantly to our understanding of BHXBs, they carry limitations we do not discuss further here. For the purposes of this review, we treat these systems as unconfirmed black holes, while acknowledging the need for further investigation.

\subsection{A field of knowns and unknowns}

Six decades of multi-wavelength observations have established a broad picture in which the emission from BHXBs is shaped by the interplay of several distinct physical components: the accretion disk, the hot inner accretion flow (also called corona), disk winds, and relativistic jets. Energy and momentum are exchanged between these components in ways that leave clear, measurable imprints across the electromagnetic spectrum. Outbursts, state transitions, and the launching and quenching of winds and jets all reflect this coupling, even if its detailed physics remains debated.

Yet, fundamental mysteries persist. The physical origin of spectral state transitions is not understood from first principles. The nature of the so-called corona\footnote{A name borrowed from solar physics, not a claim of physical kinship.}—its geometry, heating mechanism, and relationship to winds and jets—remains an open question despite decades of study. The conditions that determine whether a given system produces a powerful jet and/or a disk wind, and why outbursts can differ so dramatically from source to source, are still not settled. These unknowns are not peripheral: they sit at the core of accretion physics in the strong-gravity regime.

Progress has come, and continues to come, from probing these systems across the full range of available observational tools. The \textit{Rossi X-ray Timing Explorer} \citep[\textit{RXTE},][]{1993A&AS...97..355B} revolutionized our understanding of rapid X-ray variability over its 16-year lifetime, providing the timing and spectral foundation on which much of the current paradigm rests. More recently, \textit{XMM-Newton} \citep{2001A&A...365L...1J} and \textit{Chandra} \citep{2000SPIE.4012....2W} have enabled high-resolution spectroscopy of emission and absorption features, while \textit{NuSTAR} \citep{2013ApJ...770..103H} and \textit{INTEGRAL} \citep{2003A&A...411L...1W} have extended broadband coverage into the hard X-ray and soft gamma-ray bands. Both \textit{NICER} \citep{2016SPIE.9905E..1HG} and \textit{Insight-HXMT} \citep{InsightHXMT} have brought unprecedented timing precision in the X-rays, opening new windows on quasi-periodic oscillations and reverberation mapping of the inner accretion flow. In the radio, facilities such as the VLA, VLBI networks, and MeerKAT \citep{2016mks..confE...1J} have mapped jet morphology and proper motions with unprecedented resolution. Together, broadband continuum spectroscopy constrains the geometry and thermodynamics of the disk and corona; timing analysis reveals variability on timescales approaching the innermost stable circular orbit; emission lines in the infrared (IR), optical (O), and ultraviolet (UV) trace disk structure, outflows, and reprocessing; and X-ray polarization, now accessible with the \textit{Imaging X-ray Polarimetry Explorer} \citep[\textit{IXPE},][]{Weisskopf2022}, provides a new window on the geometry of the emitting regions.

In this review, we synthesize recent advances in the study of accretion processes in BHXBs. We first focus on the continuum emission in both the radio and X-ray bands, with particular attention to the mechanisms driving outbursts and state transitions (Sect.~\ref{sec:continuum}). We then discuss the insights gained from variability and timing at all wavelengths (Sect.~\ref{sec:timing}). We also review recent results on the emission lines observed in IR, Opt, UV, and X-rays (Sect.~\ref{sec:lines}), and discuss the new constraints from X-ray polarization enabled by the \textit{IXPE} (Sect.~\ref{sec:polar}). Finally, we address the broader implications of these findings for our understanding of accretion-ejection physics. By integrating observations, theoretical models, and numerical simulations, we aim to provide a comprehensive overview of the current state of the field and to identify the key open questions driving future research.

\section{Continuum emission} \label{sec:continuum}

Black-hole X-ray binaries are studied at many different wavelengths, from radio to $\gamma$-rays, which allow us to probe different parts of the binary system.
The challenges in undertaking pure spectral modeling are most severe in the UV, Opt, and IR bands, where numerous components can contribute to the emission. These include the donor star, the reprocessed emission from the outer accretion disk \citep{vanParadijsMcClintock}, synchrotron emission from the relativistic jet \citep{Jain2001,Markoff2001} and from the hot material near the black hole \citep{Kanbach2001,Veledina2013}. There is also potential emission from the hot spot where the stream from the donor star impacts the outer accretion disk, as was seen in CVs \citep{Smak1971}, in particular at low-luminosity (i.e., in quiescence), although it is probably too faint to be relevant during outburst \citep{froning2011}. 
We discuss in this section the continuum emission in three different spectral bands: X-ray, O/IR, and radio.

\subsection{X-ray emission}

\begin{figure}[h!]
\centering
\includegraphics[width=0.99\textwidth]{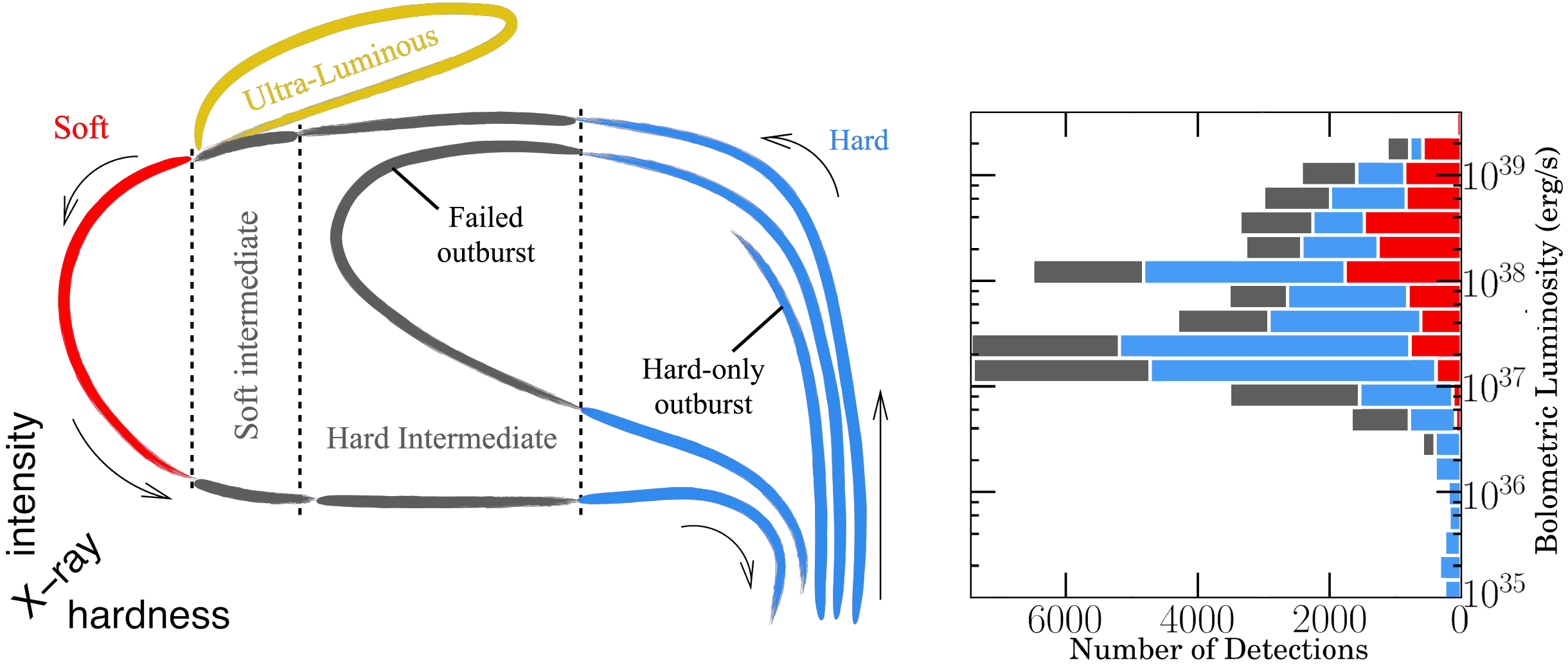}
\begin{tabular}{@{}p{0.45\textwidth}p{0.45\textwidth}@{}}
\raisebox{-\height}{\includegraphics[width=0.45\textwidth]{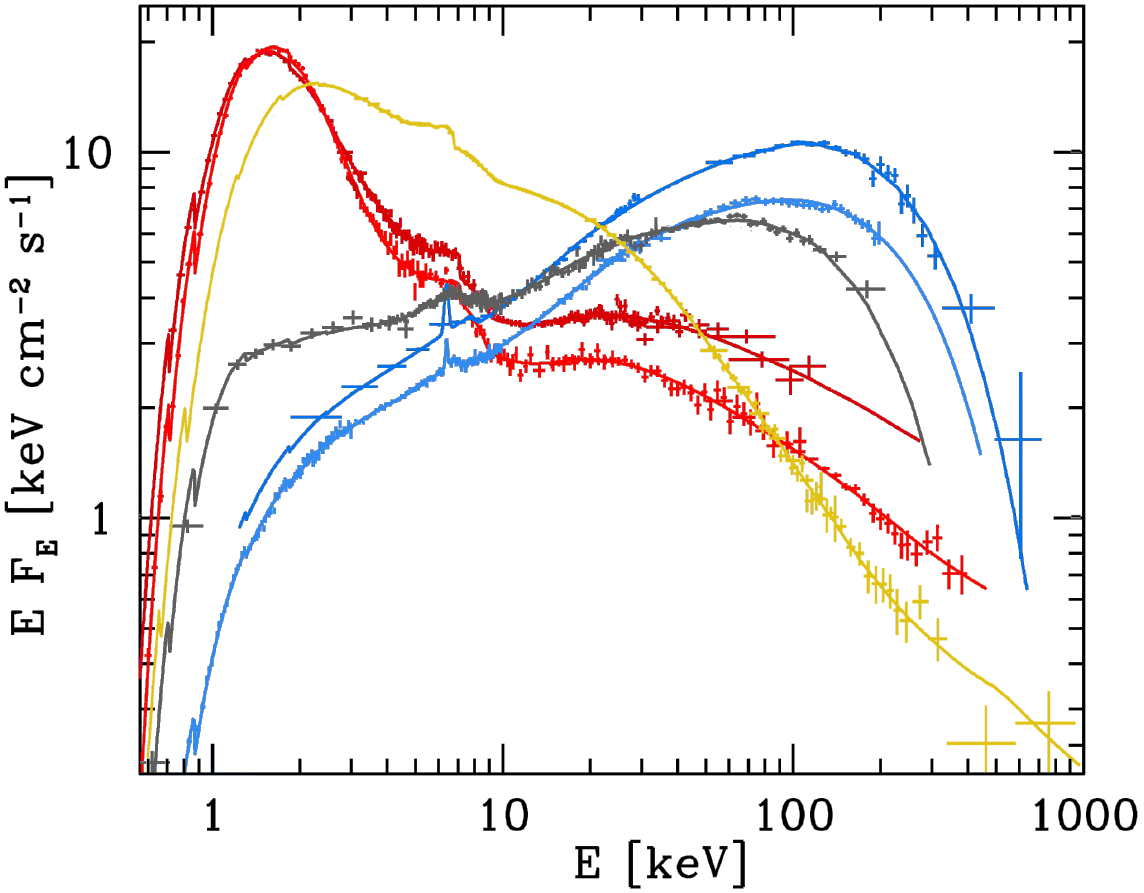}}
&
\parbox[t]{0.48\textwidth}{%
\caption{
Top-left: Typical hardness-intensity diagrams, adapted from \citet{2021NewAR..9301618M}. Top-right: Distribution of spectral states in a selected sample, adapted from \citet{tetarenko2016}. Bottom: sample of spectral states of GX~339--4 (very high state) and Cyg X-1 (all others), adapted from \citet{2002ApJ...578..357Z} and \citet{2004PThPS.155...99Z}.}
\label{fig:Zdz02-04}
}
\end{tabular}
\end{figure}

Most known low-mass X-ray binaries spend more than $90\%$ \citep{2015ApJ...805...87Y} of their life in a quiescent state and, from time to time, undergo outbursts where their bolometric luminosity rises by several orders of magnitude. During such outbursts, they display two canonical X-ray spectral states: a soft (or high) state dominated by optically thick emission around $\sim 1$\,keV, and a hard (or low) state dominated by a power-law extending to $\sim 10$--$100$\,keV, akin to what was initially reported by \citet{1972ApJ...177L...5T}. They can also go through intermediate states, usually labeled soft- and hard-intermediate, as well as more rare states such as the very-high or the ultra-luminous state; such as GRO J1655--40 during its 2005 outburst \citep{Motta2012, 2015MNRAS.451..475U}. In Fig.~\ref{fig:Zdz02-04}, bottom panel, we show a sample of the X-ray spectral shapes displayed by Cyg~X-1 and GX~339--4.

The evolution through these states during a successful outburst traces a well-known hysteresis pattern in the hardness-intensity diagram (HID), illustrated in the top-left panel of Fig.~\ref{fig:Zdz02-04}: sources rise in luminosity along the right branch in the hard state (blue), transition to the soft state (red) at high luminosity through the intermediate states (gray), and decay along the left branch before transitioning back to the hard state at a significantly lower luminosity than the original hard-to-soft transition, tracing a counter-clockwise loop. Occasionally, sources in outburst can also reach the very high or ultra-luminous state (yellow) at the top of the diagram before completing this cycle. This luminosity asymmetry between the two transitions is the defining feature of the HID hysteresis, and remains one of the key open questions in accretion physics. Not all outbursts complete this cycle, however: some never soften and instead decay directly within the hard state (`hard-only' outbursts in Fig.\,\ref{fig:Zdz02-04}), while others soften only partially before reverting to the hard state without ever reaching the soft state (`failed' outbursts). This diagram should nonetheless be read with caution, as it can give the misleading impression that the fate of an outburst can be determined by the luminosity reached. As discussed in Sect.~\ref{sec:outbursts}, this is not the case, and the transition luminosity varies considerably both between and within sources. We discuss the physical interpretation of this behavior, and the diversity of outburst morphologies, in Sect.~\ref{sec:outbursts}.

\subsubsection{The soft state} \label{sec:softXrays}

The soft state is characterized by a strong thermal component peaking around $0.1-1$\,keV, though the apparent absence of emission below $\approx 0.1$\,keV can largely be attributed to strong Galactic absorption. The spectra in this state are well described by a multi-color blackbody \citep{1984PASJ...36..741M, 1986ApJ...308..635M, GD04,2005ApJ...618..832Z}, a component that is generally interpreted as arising from an optically thick, geometrically thin, radiatively efficient accretion disk, see Table~1 in \citet{2026NewAR.10201746Z} for a list of the typically used relativistic disk models.
The soft state is only observed above a certain luminosity threshold, $\gtrsim 10^{-3}\,L_{\rm Edd}$, where $L_{\rm Edd} \approx 1.26 \times 10^{38} (M/M_{\odot})$~erg.s$^{-1}$ is the Eddington luminosity for a compact object of mass $M$, defined as the luminosity where radiation pressure balances gravity for spherical accretion. All BHXBs below this Eddington fraction appear to reside in the hard state (Fig.~\ref{fig:Zdz02-04}, top-right panel), despite thousands of observations across dozens of sources; see however cases such as XTE~J1752--223 \citep{tetarenko2016} and V4641~Sgr \citep{2026A&A...710A..28P}.

A key observational result is that the disk bolometric luminosity scales as $L_{\rm bol} \propto T^4$ \citep[e.g.,][]{2001ApJ...560L.147K, 2004MNRAS.347..885G, 2004ApJ...601..428K, 2004MNRAS.353..980K}, implying a fixed inner radius and a constant radiative efficiency, as expected for a disk extending to the innermost stable circular orbit (ISCO). Color-correction factors $f_{\rm col}$ are sometimes required to account for atmospheric effects (\citealt{1995ApJ...445..780S}, \citealt{2005ApJ...621..372D}; see reviews by \citealt{2007A&ARv..15....1D}, \citealt{2019ApJ...874...23D}).

Because the \citeauthor{SS73} $\alpha$-disk solution naturally produces an optically thick, geometrically thin disk that radiates all available accretion power, it is standardly used to model the soft-state disk \citep{2002apa..book.....F}. This assumption has proven remarkably successful in explaining soft-state spectra observed around $L \simeq (0.01$--$1)\,L_{\rm Edd}$ \citep[][Fig.~16]{2007A&ARv..15....1D}. However, the \citeauthor{SS73} solution becomes unstable once radiation pressure dominates over gas pressure \citep[][see also \citealt{1976MNRAS.175..613S}]{LightmanEardley}, expected at luminosities around and above a few percent of the Eddington luminosity \citep[e.g.,][]{arXiv.2502.08718}. This is problematic because thermal disks are predominantly observed precisely in this luminosity range and are rarely (if ever) observed outside of it. This is also somewhat inconsistent with the variability observed in the disk-dominated states, see Sect.~\ref{sec:timing}. One proposed resolution involves the inclusion of a globally organized magnetic field, either through wind-driven torques \citep{BP82, 1995A&A...295..807F, 1997A&A...319..340F} or through magnetic pressure support \citep{BP07}, which can stabilize the disk against the radiation-pressure instability; see \citet{2014ApJ...786....6L} and \citet{2024A&A...692A..99Z} for recent discussions. All in all, this remains an active area of research without consensus \citep[see, e.g.,][]{Sadowski16a, Sadowski16b, 2016MNRAS.460.3488S, ZhuStone18, 2021A&A...647A.192J, Mishra22, 2022ApJ...935L...1L, 2024MNRAS.527.1424S, 2024ApJ...965..175M}.

One important application of the soft-state disk emission is the \textit{continuum fitting method} for estimating black hole spin \citep[e.g.,][]{1997ApJ...482L.155Z, 2005ApJS..157..335L, 2014SSRv..183..295M, Reynolds21}. Under the assumption that the disk extends to the ISCO \citep{GD04, 2007A&ARv..15....1D, 2011MNRAS.416..941S} and follows the $\alpha$-prescription, the inner disk temperature and normalization constrain the spin parameter. The Novikov-Thorne solution \citep{NT73} is central here, as it accounts for the impact of spin on accretion efficiency spin and predicts the radial disk structure in the relativistic regime. This method has been applied to numerous sources: GX~339--4 \citep[e.g.,][]{2016ApJ...821L...6P, 2025ApJ...981L..15Z}, 4U~1543--47 \citep[e.g.,][]{2006ApJ...636L.113S, 2023A&A...677A..79Y}, Cyg~X-1 \citep[e.g.,][]{2021ApJ...908..117Z, 2024ApJ...967L...9Z}, and many others \citep{Reynolds21}. We refer the reader to \citet{2026NewAR.10201746Z} for a recent critical assessment of this method.

\subsubsection{The hard state} \label{sec:hardXrays}

The hard state is observed across an extraordinary luminosity range, from deep quiescence ($L \approx 10^{-9}\,L_{\rm Edd}$) to the brightest hard states ($L \gtrsim 10^{-1}\,L_{\rm Edd}$). Across this range, the X-ray spectrum is generally approximately described by a power law with photon index $\Gamma$ and an exponential cut-off at energy $E_{\rm cut}$ \citep{2014MNRAS.443.1733N, 2020MNRAS.492.5234Z}, $N(E) \propto E^{-\Gamma} \exp(-E/E_{\rm cut})$, where the cut-off is not always required or detected. Above this cut-off, which is typically attributed to Comptonization in a thermal plasma, non-thermal tails extending to $\sim$10\,MeV have been detected in several sources \citep[][see Fig.\,\ref{fig:Zdz02-04}]{Ling1997, McConnell2002, 2021ApJ...914L...5Z}. Two caveats apply to the power-law description. First, it may be an oversimplification, because in brighter sources multiple spectral components are required \citep[e.g.,][]{2021MNRAS.506.2020D, 2021ApJ...914L...5Z}. Second, the cut-off is often observed to be too abrupt for standard Comptonization models \citep[see, e.g.,][]{2003MNRAS.342..355Z}, and reported values of $E_{\rm cut}$ and $\Gamma$ should be interpreted with this caveat in mind.

\begin{figure}[h!]
\centering
\includegraphics[width=0.45\textwidth]{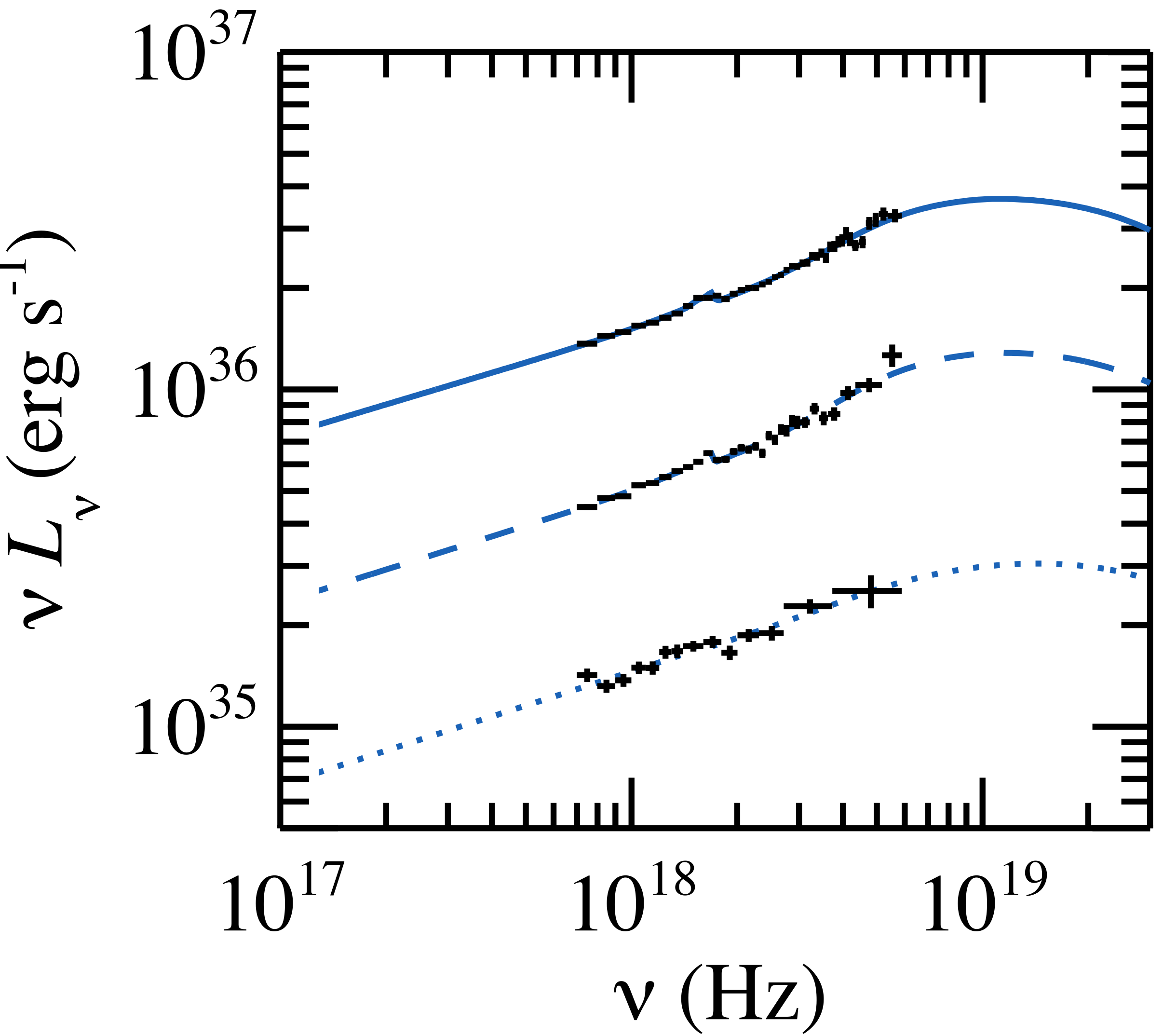}
\includegraphics[width=0.45\textwidth]{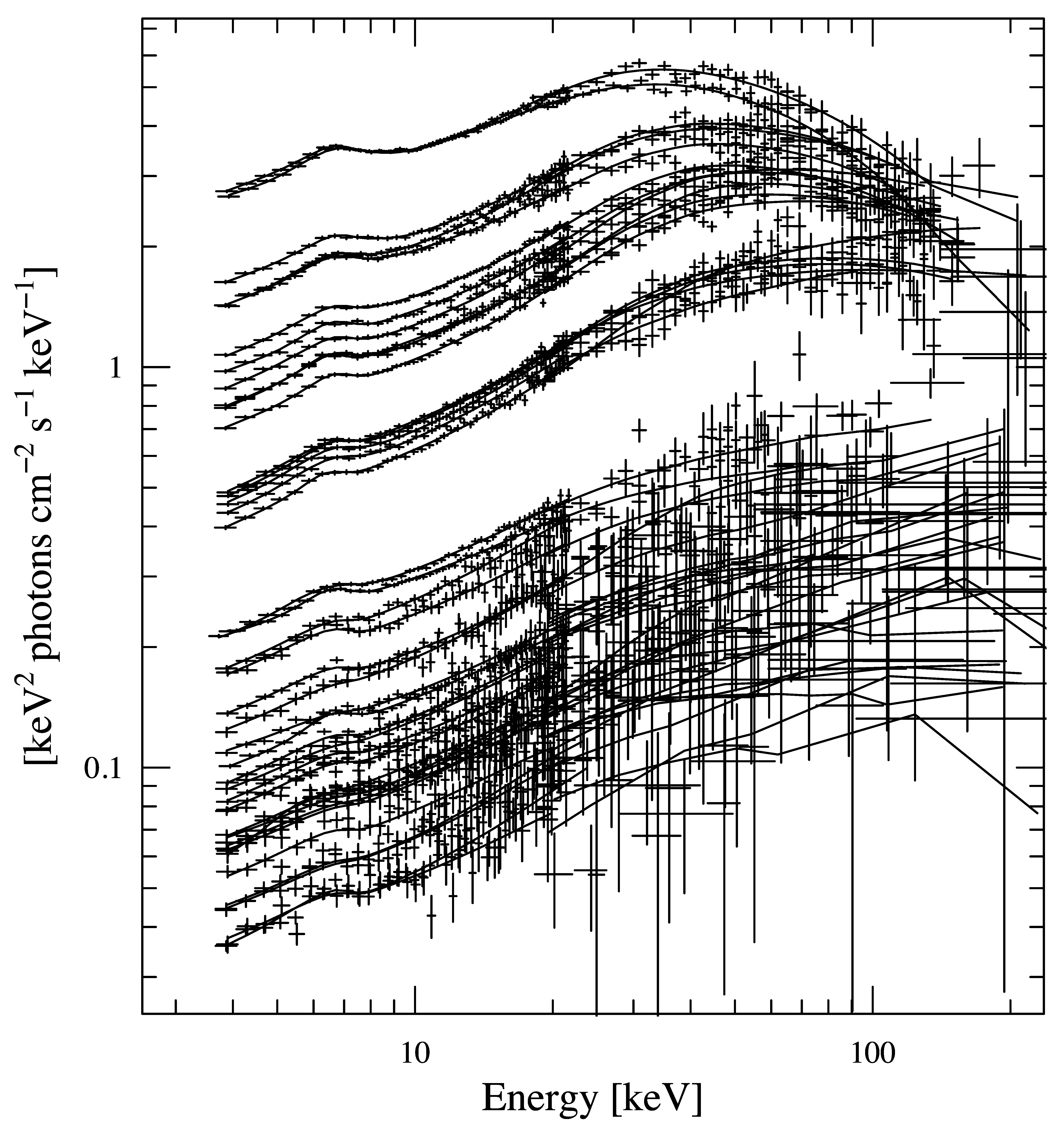}
\caption{Evolution of the X-ray spectra of XTE~J1550--564 (left) and GX~339--4 (right) during selected hard states, as observed by \textit{RXTE}, fitted with a cut-off power law. Left is adapted from \citet{2014MNRAS.445.3987P}, right is adapted from \citet{2019ApJ...871...26K}.}
\label{fig:Koljonen2019}
\end{figure}

We show in Fig.~\ref{fig:Koljonen2019} the spectral trends across multiple outbursts of XTE~J1550--564 and GX~339--4. The spectral index evolves with luminosity in a non-trivial way, and the overall behavior is broadly divided into three luminosity regimes:
\begin{itemize}
    \item In the \textit{quiescent state}, $L \lesssim 10^{-5}\,L_{\rm Edd}$, most sources display a relatively uniform photon index $\Gamma \simeq 2.1$ \citep{plotkin2013, 2015MNRAS.447.1692Y} with no detectable cut-off ($E_{\rm cut} \gtrsim 100-200$\,keV). Though constraints are often weak and exceptions exist \citep[e.g.,][]{2003A&A...399..631H, 2016ApJ...832..115F}, this consistency is supported by many individual source studies \citep[e.g.,][]{2002ApJ...570..277K, 2003ApJ...593..435M, 2006ApJ...636..971C, 2014MNRAS.444..902A, 2014MNRAS.441.3656R, 2015MNRAS.447.1692Y},
    \item In the \textit{low-luminosity hard state}, $L \sim 10^{-5}-10^{-3}\,L_{\rm Edd}$, the photon index lies in the range $\Gamma \approx 1.5-2$ and marginally decreases with increasing luminosity \citep{2011MNRAS.417..280S, 2014MNRAS.443.1733N, 2015MNRAS.447.1692Y, 2022MNRAS.517.3588J}, though the absence of a trend has also been reported \citep{2006ApJ...639..340K, 2014MNRAS.445.3987P}. A similar behavior is observed in AGN \citep{2011MNRAS.414.3330V, 2015MNRAS.447.1692Y, 2020ApJ...895..114H, 2023MNRAS.524.4670J},
    \item In the \textit{bright hard state}, $L \gtrsim 10^{-3}\,L_{\rm Edd}$, the behavior reverses: $\Gamma$ now increases with luminosity \citep{2011MNRAS.414.3330V, 2011MNRAS.417..280S, 2013ApJ...764....2Q}. This change in slope around $L \sim 0.1-1\%\,L_{\rm Edd}$ remains poorly understood \citep{2013MNRAS.435.3395R, 2015MNRAS.447.1692Y, 2019ApJ...871...26K, 2022A&A...659A.194M}. In this regime, the cut-off energy also becomes observationally accessible and decreases with luminosity, from $E_{\rm cut} \simeq 200$\,keV down to $\sim$50\,keV \citep[e.g.,][]{2009MNRAS.392..992D, 2014MNRAS.442.1767P, 2016A&A...591A..66K, 2017A&A...602A..40S, 2019ApJ...871...26K, 2021MNRAS.508..287S, 2023MNRAS.521.2692D}, and in some cases as low as 10--20\,keV \citep{2009ApJ...698.1398M, 2024ApJ...975..165S}; though such low values have also been attributed to Compton reflection features rather than a genuine thermal cut-off \citep[e.g.,][]{2003MNRAS.342..355Z}.
\end{itemize}

\begin{figure}[h!]
\centering
\includegraphics[width=.45\textwidth]{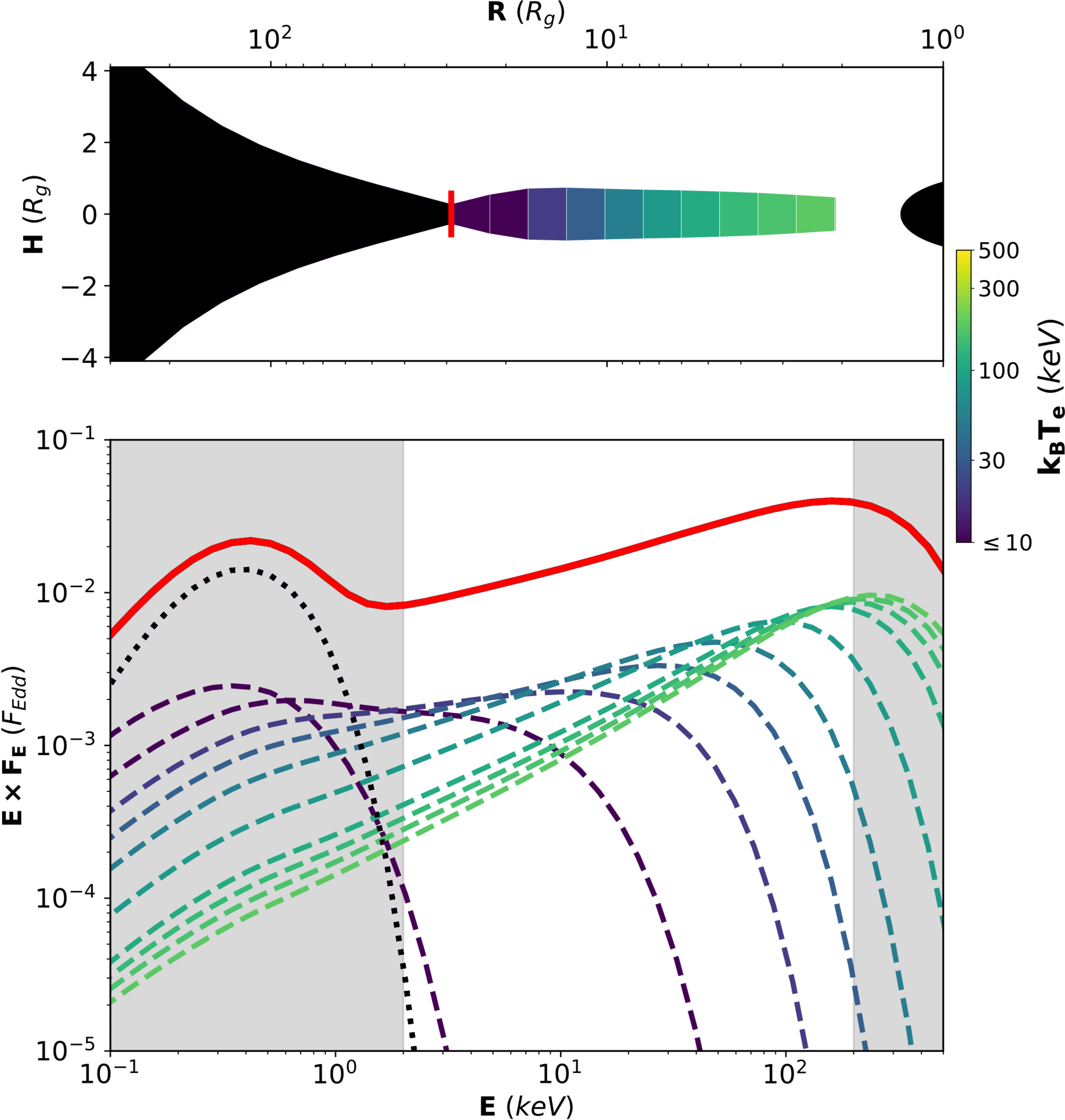}
\includegraphics[width=.50\textwidth]{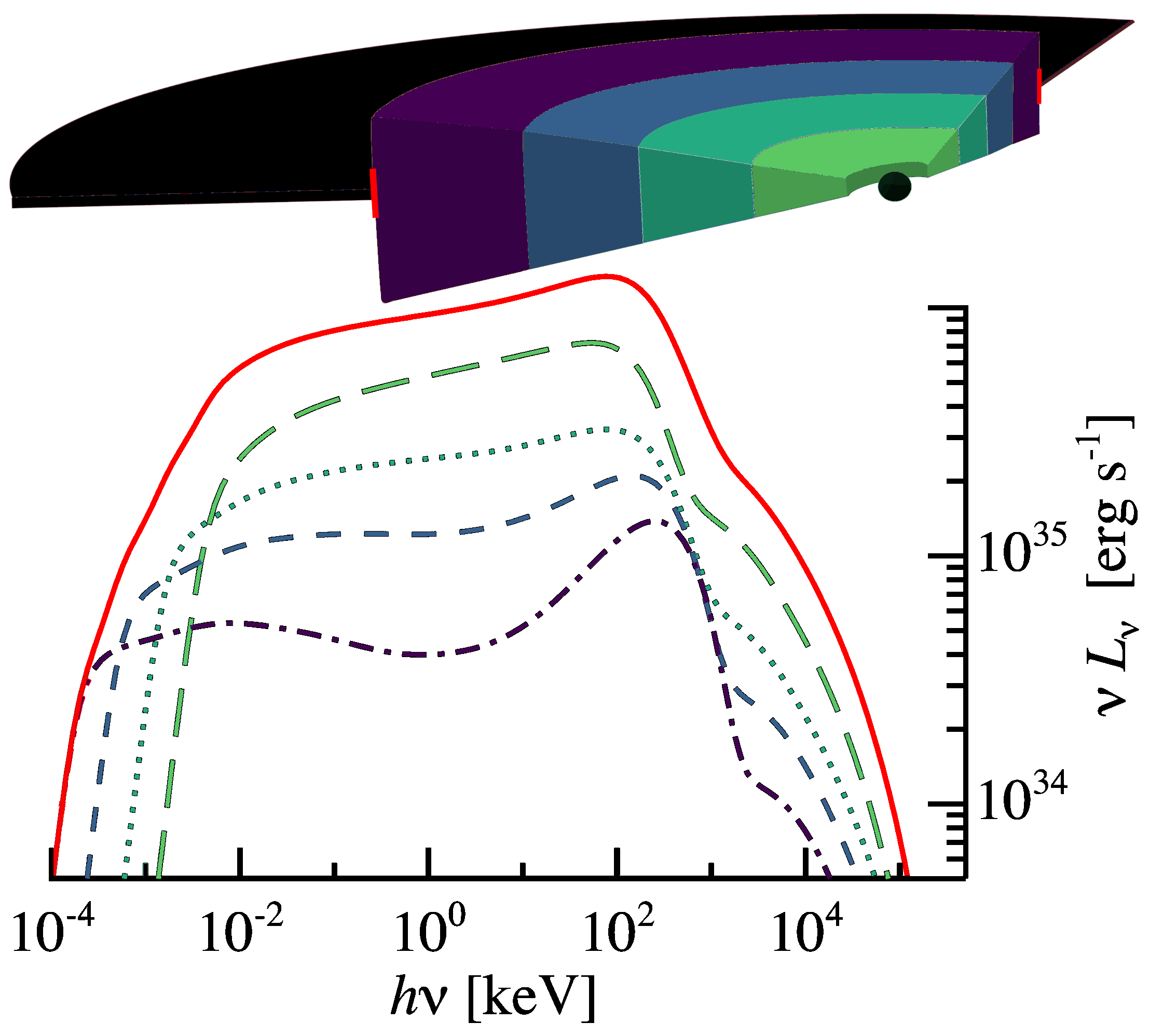}
\caption{Geometry (top) and spectra (bottom) for two hard state models (\citealt{2018A&A...615A..57M, 2018A&A...617A..46M,2022A&A...659A.194M}, left; \citealt{Veledina2013}, right). The vertical lines symbolizes the radius where the disk turns into a hot flow. Each color and line style corresponds to a given region (or annulus), the red represents the total spectrum. 
}
\label{fig:Veledina13}
\end{figure}

The physical origin of the hard X-ray emission has been debated for decades. There is now growing evidence that it arises from a hot, optically thin inner accretion flow located close to the black hole, as already envisioned in the 1970s and 1980s \citep[e.g.,][]{1975ApJ...195L.101T, Oda77, 1984SSRv...38..353L} and later formalized with the advection-dominated accretion flow \citep[ADAF;][]{1977ApJ...214..840I, 1995ApJ...452..710N, Narayan96, 1996ApJ...465..312E}. Within this framework, seed photons are Compton up-scattered by the hot electrons of the flow, producing the observed power law \citep{1984SSRv...38..353L, 1997MNRAS.288..958G, 1998MNRAS.301..435Z, 2020MNRAS.492.5234Z}. We show in Fig.\,\ref{fig:Veledina13} two recent hot flow models \citep{Veledina2011, 2018A&A...615A..57M, 2018A&A...617A..46M}. An alternative scenario proposes that the hard X-rays originate at the base of compact jets \citep{1991ApJ...374..741M,Markoff2001, 2003A&A...397..645M, 2005ApJ...635.1203M}, which would naturally link the X-ray and radio emissions (see Sect.~\ref{sec:radio}); however, this geometry appears inconsistent with polarization results in the hard state (see Sect.~\ref{sec:polarhard}), and the physical mechanism for heating the electrons to the required $\gtrsim 0.1$\,MeV energies remains unclear.

A key open question concerns the origin of the aforementioned seed photons. While disk photons have long been assumed to dominate, producing a sufficiently hard spectrum from disk seed photons alone is challenging \citep{2007A&ARv..15....1D, 2010PASJ...62..621K, 2018A&A...614A..79P}, and there is growing evidence for a significant contribution from local synchrotron emission \citep[e.g.,][]{Veledina2011, Veledina2013, 2016A&A...591A..66K}. The aforementioned change in $\Gamma$ behavior during the hard-state has so far only been explained through variations in the seed photon population \citep[e.g.,][]{2011MNRAS.417..280S, 2019ApJ...871...26K, 2022A&A...659A.194M}. Furthermore, timing studies and direct spectral fits suggest that the hard X-ray emission is not a single uniform component but likely comprises multiple contributions \citep[e.g.,][]{2001ApJ...546.1027F, 2005MNRAS.362.1435I, 2018MNRAS.480..751A, 2018MNRAS.480.4040M, 2021MNRAS.506.2020D, 2021ApJ...914L...5Z}, possibly arising from a radially stratified hot flow whose integrated spectrum mimics a single power law (Fig.\,\ref{fig:Veledina13}, left panel).

Two major theoretical challenges strongly bound hot-flow models. First, any solution must remain physically viable across at least eight orders of magnitude in luminosity \citep{2012MNRAS.427.1580X}. The canonical ADAF falls short of the brightest hard states by two to three orders of magnitude \citep{YN14, 2018A&A...615A..57M} and predicts temperatures exceeding observed values \citep{2004MNRAS.354..953Y}. The luminous hot accretion flow \citep[LHAF,][]{2001MNRAS.324..119Y, 2003ApJ...594L..99Y} partially alleviates this problem: as the accretion rate increases, the advected energy becomes insufficient to balance radiative cooling locally, yet the flow remains hot because of compression work that sustains the high temperature. This extends the maximum luminosity of the hot solution beyond the ADAF limit, but still falls short of the brightest observed hard states by roughly an order of magnitude \citep{2012MNRAS.427.1580X,YN14}. Including a globally organized magnetic torque that extracts angular momentum vertically can significantly extend the maximum luminosity of a hot solution further \citep{1997A&A...319..340F, 2006A&A...447..813F, 2016ApJ...817...71C, 2018A&A...617A..46M, 2019A&A...626A.115M, 2024A&A...692A.153S}. Second, the hot flow must form naturally in almost all systems across this range. Proposed formation mechanisms include disk evaporation \citep{1999ApJ...527L..17L, 2000A&A...361..175M}, magnetic field advection \citep{2006A&A...447..813F}, local dynamo action \citep{Contopoulos98, 2014ApJ...782L..18B, 2015A&A...574A.133K, 2024MNRAS.532.1522J}, and the strong-ADAF principle \citep{1997ApJ...489..865E}, none of which has yet achieved broad observational or numerical confirmation. See Sect.~\ref{sec:outbursts} for a discussion.

Another challenge is the heating mechanism. Gravitational energy is typically transformed into particle energization via magnetic reconnection or shocks. Early ab initio plasma simulations, relevant to accreting black hole environments, supported these mechanisms, producing power-law-like electron distributions as a generic outcome of plasma energization \citep{ZenitaniHoshino2001, SironiSpitkovsky2009, SironiSpitkovsky2014, Riquelme2012}.

These particles may explain the observed non-thermal high-energy tails. However, the success of phenomenological thermal Comptonization models, achieving $5\%$ accuracy in temperature measurements \citep[e.g.,][]{1997MNRAS.288..958G, 1998MNRAS.301..435Z}, has driven the search for a thermalization mechanism. Such a mechanism could produce the (mostly) Maxwellian electron distributions after initial acceleration.
One possible solution is the synchrotron boiler mechanism \citep{Ghisellini1988}. In this process, the Maxwellian part of the distribution rapidly forms due to repeated synchrotron emission and self-absorption. When applied to black hole environments, it produces spectra dominated by thermal Comptonization with an additional non-thermal tail, closely resembling observations \citep{PoutanenVurm2009, MalzacBelmont2009, Poutanen2014}.

Recent kinetic simulations suggest an alternative mechanism for replicating the observed near-thermal electron distributions: plasma processes. Large-scale reconnection and turbulent environments create an inhomogeneous plasma, with most matter concentrated in plasmoids surrounded by a rarefied, magnetized medium \citep{SironiSpitkovsky2014, SironiBeloborodov2020, Sridhar2021, Sridhar2023}. The plasmoid motion distribution resembles a Maxwellian, implying that the observed X-ray continuum could arise from bulk motion Comptonization of disk or synchrotron photons \citep{2004ApJ...601..405S, Beloborodov2017, Groselj2024, Nattila2024}. However, the electrons are cold in those plasmoids.

\subsubsection{Intermediate states}

Between the canonical hard and soft states, black-hole X-ray binaries exhibit intermediate states, commonly divided into the hard-intermediate state (HIMS) and soft-intermediate state (SIMS). These states are characterized by the increasing importance of a thermal disk component as the source softens, together with a still substantial Comptonized component. HIMS and SIMS are distinguished by their difference in variability, see Sect.~\ref{sec:timing}.

Intermediate states are often short-lived, although the duration and luminosity evolution within the intermediate state can vary substantially between outbursts. For example, the source may continue to brighten considerably after entering the intermediate state before reaching the soft state (see next section). This association with state transitions is not universal, however: Cyg~X-1 can spend extended periods in an intermediate state without undergoing a full hard--soft transition. Thus, while the intermediate states provide a useful phenomenological description of the transition region, they do not correspond to a unique luminosity or a single evolutionary stage.

\subsubsection{Non-canonical states}

The canonical hard-intermediate-soft sequence does not capture the full diversity of black-hole X-ray binary behavior. Historical labels such as the `very high state' overlap substantially with luminous intermediate-state phenomenology, while hypersoft or ultrasoft episodes, such as that seen in GRO J1655-40 \citep[e.g.,][]{2015MNRAS.451..475U}, appear spectrally and temporally distinct and are often associated with strong winds or obscuration.

GRS~1915+105 stands as the most extreme example of this departure from the canonical picture. Since its discovery \citep{1992IAUC.5590....2C} and the subsequent detection of superluminal radio ejections \citep{1994Natur.371...46M}, the source has remained persistently active at luminosities close to, and at times exceeding, the Eddington limit for several decades, rather than following the outburst-quiescence cycle typical of most BHXBs. Its X-ray light curves display an extraordinary range of variability patterns, famously classified into more than a dozen distinct classes based on their count rate and color-color evolution \citep{2000A&A...355..271B, 2002MNRAS.331..745K, 2005A&A...435..995H}. Several of these classes exhibit quasi-regular limit-cycle oscillations on timescales of seconds to minutes, the most iconic being the `heartbeat' (class~$\rho$) oscillations, generally interpreted as a direct manifestation of the radiation-pressure instability of the inner accretion disk \citep[][see Sect.~\ref{sec:softXrays}]{2011ApJ...737...69N, 2023Natur.615...45V}. This behavior has made GRS~1915+105 both a uniquely rich laboratory for disk instability physics and a long-standing challenge to any description of state evolution. More recently, the source has transitioned into an unusual, heavily obscured, and comparatively dim state, marking an apparent end to its three-decade-long active phase \citep{2020ApJ...902..152N, 2021MNRAS.503..152M}.

Systems such as GRS~1915+105, V404~Cyg, and V4641~Sgr further show that reprocessing, outflows, and changing accretion geometry can dominate the observed phenomenology \citep[e.g.,][see also Sect.~\ref{sec:polar}]{Koljonen2020}. More generally, these examples emphasize that the canonical state sequence should be regarded as a phenomenological framework rather than a unique evolutionary track. In particular, both the luminosity of state transitions and the path followed through spectral and timing parameter space can vary substantially between outbursts, while additional accretion-flow and outflow phenomena can become important at high accretion rates.

\subsubsection{Outbursts and spectral transitions} \label{sec:outbursts}

While the spectral states described above are now well characterized observationally, the mechanisms that drive transitions between them remain poorly understood. The disk instability model \citep[DIM;][]{2001NewAR..45..449L, 2001A&A...373..251D}, first developed for dwarf novae \citep{1971AcA....21...15S, 1974PASJ...26..429O, 1979PThPh..61.1307H} and later applied to X-ray binaries \citep{vanparadijs1996, 2012MNRAS.424.1991C, 2021ApJ...912..110B}, is broadly accepted as the trigger for outbursts, and has been thoroughly reviewed elsewhere \citep{2020AdSpR..66.1004H, Blaes+25}. In brief, once hydrogen in the outer disk becomes partially ionized, a local increase in viscosity propagates inward, driving the outburst. The DIM successfully predicts which systems are transient (though not the outburst timescale or recurrence times), see \citet{2018A&A...617A..26D} for the case of cataclysmic variables. However, the picture is certainly incomplete: some X-ray binary outbursts last far longer than any plausible viscous timescale of the disk, implying that enhanced mass transfer from the companion---perhaps irradiation-driven and triggered by the outburst itself---must play a significant role \citep{2014SSRv..183..101M}. While the DIM provides the essential framework, additional physics beyond the basic thermal-viscous cycle is needed to fully account for the observed outburst phenomenology.

\begin{figure}[h!]
\centering
\includegraphics[width=0.95\textwidth]{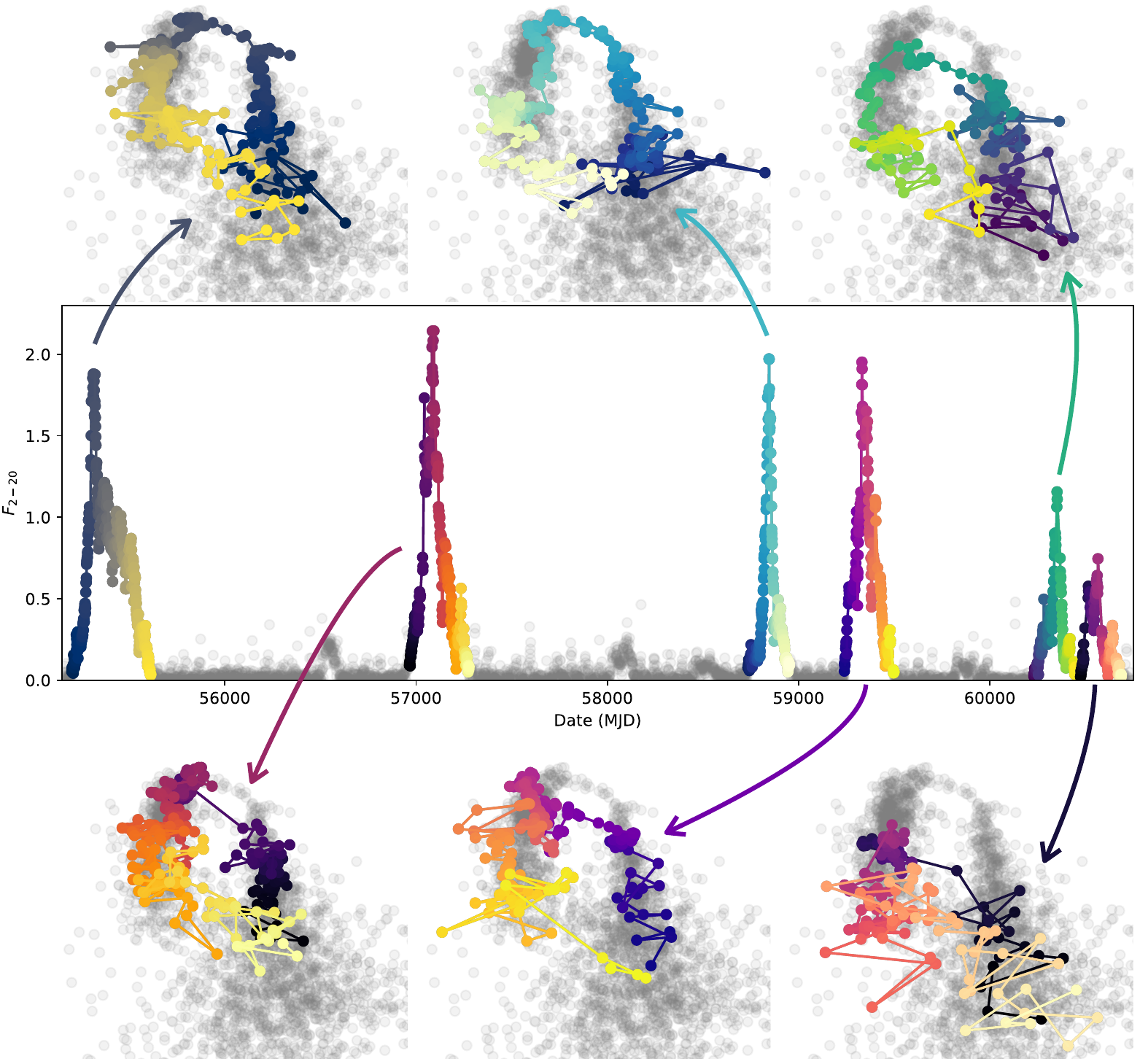}
\caption{(Middle panel) 16-year light-curve of GX~339--4 ($2-20$\,keV, in Photon\,cm$^{-2}$\,s$^{-1}$) as seen by MAXI. The top and bottom panels show the hardness-intensity diagrams (HID) of the six X-ray outbursts observed, with count rate (y-axis) plotted against hardness ($4-10$\,keV / $2-4$\,keV).}
\label{fig:GX339-4}
\end{figure}

Observationally, roughly $60-70\%$ of outbursts follow a ``successful'' or full pattern \citep{tetarenko2016, Alabarta21}: the source rises in the hard state, transitions through the intermediate states to the soft state at high luminosity, decays in the soft state, and then transitions back to the hard state at a significantly lower luminosity before returning to quiescence (see Fig.\,\ref{fig:Zdz02-04}). Perhaps the best example of these successful outbursts is GX~339--4 \citep[at least since 1998][]{2004MNRAS.351..791Z}, undergoing an outburst every two years on average, see Fig.~\ref{fig:GX339-4}. In the remaining $30-40\%$ of cases, the soft state is never reached; these are termed ``hard-only'' \citep{2004NewA....9..249B, 2021NewAR..9301618M} or ``failed-transition'' outbursts, observed in a significant fraction of BHXBs ($36\%$, \citealt{Alabarta21}). Some sources show comparable numbers of full and failed outbursts (e.g., H1743--322), while others seem exclusively to produce successful outbursts (e.g., 4U~1630--47, \citealt{tetarenko2016}) or have never been observed in a pure soft state, such as GRS~1716--249 \citep{2019MNRAS.482.1587B} or Cyg~X-1 \citep[see, however,][]{2016ApJ...829L..22S}. The reasons behind this diversity remain elusive \cite[see, e.g.,][]{2014SSRv..183..101M}.

It is regularly suggested that sources transition to the soft state once the accretion rate reaches a certain threshold. This is, however, hard to reconcile with individual cases such as MAXI~J1820+070, where the transition happened during a decrease in luminosity; the hard state had reached a higher luminosity earlier in the outburst without transitioning \citep[see, e.g.,][]{2021NatCo..12.1025Y}. Moreover, this picture struggles to explain the diversity of outbursts observed in a given source.
This diversity is illustrated in Fig.~\ref{fig:GX339-4}, which shows the six most recent complete outbursts of GX~339--4 as seen by MAXI. Each outburst differs in duration, peak luminosity, and preceding quiescence time, and we refer the reader to dedicated studies of these outbursts \citep[e.g.,][]{2011A&A...534A.119C, 2012A&A...542A..56N, 2015A&A...573A.120P, 2015ApJ...808..122F, Garcia15, 2016AN....337..435C}, including the failed outbursts around MJD\,56500 \citep{2015ApJ...808..122F} and MJD\,58000 \citep{Garcia19}.

A defining feature of the outburst cycle is its hysteresis in the HID: the hard-to-soft transition occurs at a significantly higher luminosity than the soft-to-hard transition. Both transitions unfold over days to weeks, which is infinitely longer than the dynamical timescale in the inner regions of the accretion flow. This implies that the system finds temporary equilibrium at each stage rather than jumping instantaneously between states. A notable exception is the rare ``flip-flop'' behavior, where the spectral state alternates on timescales of seconds to minutes \citep[e.g.,][]{1991ApJ...383..784M, 2020A&A...641A.101B, arXiv.2502.08718}.

\begin{figure}[h!]
\centering
\includegraphics[width=.95\textwidth]{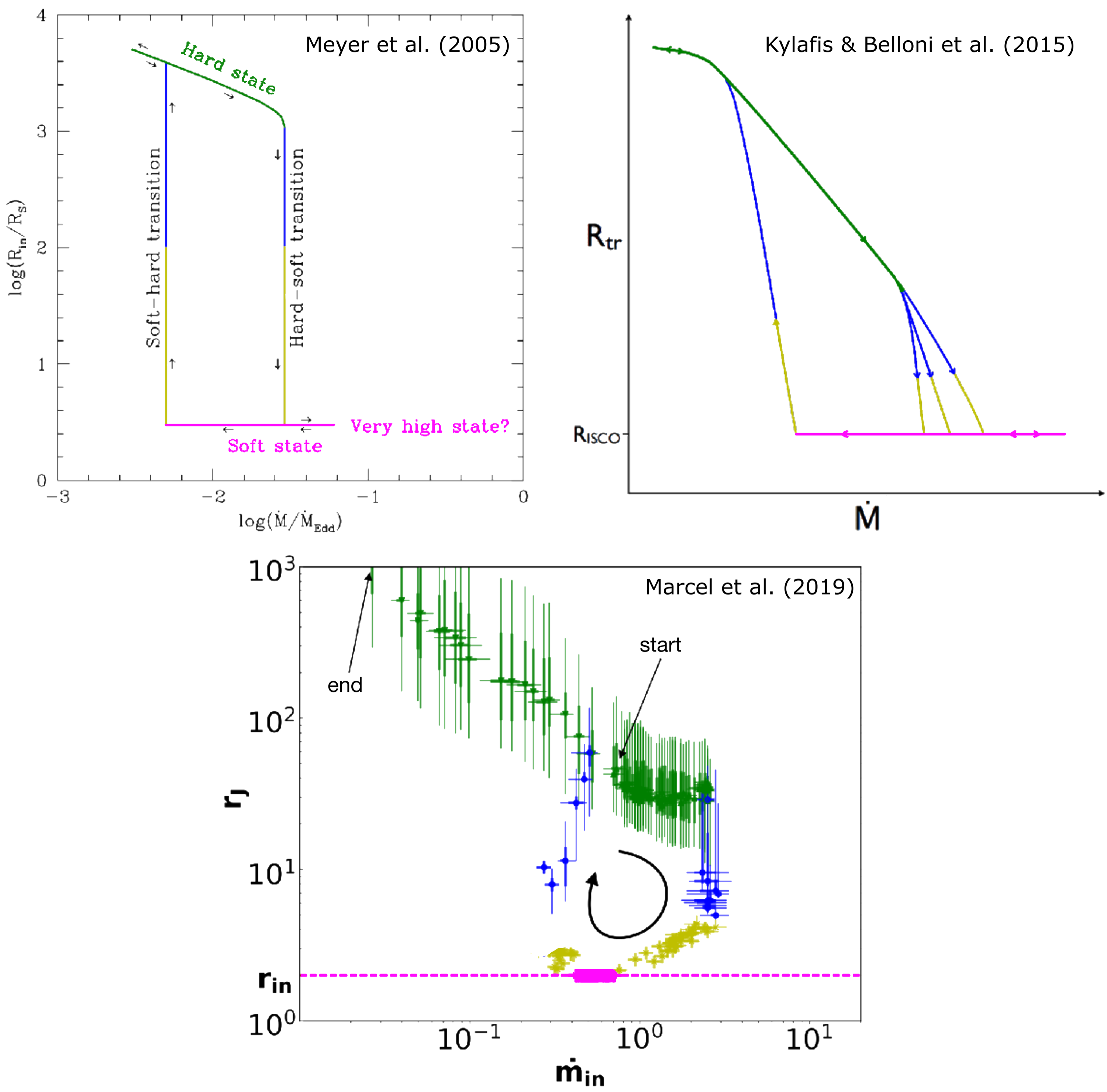}
\caption{Parametric evolution of the transition radius $R_{\rm t}$ as a function of accretion rate $\dot{M}$ in different works; adapted from \citet{2005A&A...432..181M, 2015A&A...574A.133K, 2019A&A...626A.115M}. Color scheme: hard state (green), hard-intermediate (blue), soft-intermediate (yellow), soft state (pink).}
\label{fig:OB}
\end{figure}

Two key asymmetries between the transitions stand out. First, the hard-to-soft transition can occur at a wide range of luminosities, while the soft-to-hard transition occurs at a relatively fixed value $L \simeq 2\%\,L_{\rm Edd}$ \citep{2003A&A...409..697M, 2019MNRAS.485.2744V},\footnote{This has been questioned by \citet{Dunn2010}, who present a comprehensive analysis of \textit{RXTE} outbursts and argue that the spread in hard-to-soft transition luminosities is no larger than that in soft-to-hard transitions. However, their conclusions are significantly affected by the assumption of $d = 5$\,kpc and $M_{\rm X} = 10\,M_\odot$ for all sources with unconstrained parameters. Sources with well-constrained system parameters are systematically closer than the Galactic Center, reflecting a selection bias whereby only relatively nearby systems are bright enough in quiescence to allow reliable mass and distance measurements. As a change in assumed distance from 5 to 10\,kpc modifies the inferred luminosity by a factor of 4, this can strongly bias the inferred distribution of transition luminosities.} see however cases mentioned in Sect.~\ref{sec:softXrays} as well as J1727.8--1613 \citep{2025MNRAS.542.1803H}. Second, the hard-to-soft transition is not inevitable, and many sources remain in the hard state throughout an outburst, while the soft-to-hard transition appears to be a necessary step once a source enters the soft state. These asymmetries suggest that the two transitions are not governed by the same physical process.
Moreover, the roughly constant luminosity during each transition \citep{2005MNRAS.360L..68M} has an important implication: the radiative efficiency of the soft X-ray emitting component (the disk) cannot be dramatically different from that of the hard X-ray emitting component (the corona or hot flow). This argues against a highly radiatively inefficient corona, and is consistent with a hot flow that is significantly more efficient than envisioned in classical ADAF models \citep{2012MNRAS.427.1580X, 2022A&A...659A.194M}.

The physical picture that has emerged from these observations is one in which the system is composed of an outer accretion disk and an inner hot flow, separated at a transition radius $R_{\rm t}$ (or $r_{\rm t}$) that evolves during the outburst cycle \citep[e.g.,][see Fig.\,\ref{fig:OB}]{1998ApJ...505..854E, 2005A&A...432..181M, 2008MNRAS.385L..88P, 2011MNRAS.414.3330V, Veledina2013, 2014ApJ...782L..18B, 2015A&A...574A.133K, 2019A&A...626A.115M}.
The exact value of this transition radius, sometimes referred to as 'truncation radius', remains a matter of hot debate. Some studies suggest this radius could be almost down to the ISCO throughout the hard state \citep[e.g.,][see Fig. 6 in \citet{2026Ap&SS.371...27L} for a compilation]{Garcia15, 2026ApJ..1003...73K}, while others suggest much further truncation \citep[e.g.,][]{2014MNRAS.442.1767P, 2014A&A...564A..37P}; sometimes on the same data-set \citep[e.g., MAXI~J1820+070,][]{Kara2019, 2019MNRAS.490.1350B, 2021MNRAS.507.2744A, 2021A&A...656A..63M, 2022ApJ...928...11Z}. Regardless of the exact value, there is broad agreement that the hard-to-soft transition corresponds to the inward contraction of the hot flow (decreasing $R_{\rm t}$), and the soft-to-hard transition to its outward expansion. Several hypotheses have been proposed to drive this evolution: disk evaporation \citep{1999ApJ...527L..17L, 2000A&A...361..175M}, the advection and diffusion of magnetic fields \citep{2006A&A...447..813F, 2008MNRAS.385L..88P, 2014ApJ...782L..18B, 2019A&A...626A.115M}, stability curves of ADAF/LHAF \citep{1997ApJ...489..865E, YN14} or magnetically arrested disk solutions \citep{2024A&A...692A.153S}, and even spin-orbit misalignment \citep{2014MNRAS.437.3994N, 2026A&A...710A.387M}. Magnetic-field-based scenarios have a natural advantage in that the advection-diffusion timescale of the field can substantially exceed local dynamical timescales, providing a plausible explanation for the gradual, secular nature of the transitions. Ongoing numerical efforts to model these processes \citep[e.g.,][]{2020A&A...641A.133S, 2022ApJ...935L...1L, 2024MNRAS.532.1522J} have not yet reached consensus, and no comprehensive model has emerged that simultaneously explains the hysteresis, the asymmetry between the two transitions, and the diversity of outburst behaviors.

\subsection{Optical and Infrared emission}

Early studies of BHXBs generally assumed that the UV, Opt, and IR continuum emission originates predominantly from a viscously heated and/or X-ray–irradiated accretion disk, especially for systems with long orbital periods, as the flux from reprocessed emission increases with size of the accretion disk \citep{vanParadijsMcClintock}. 
This interpretation was further supported by the presence of double-peaked emission lines, which naturally arise in the differentially-rotating accretion disk (see more details in Sect.~\ref{sec:lines}).
Within this framework, the evolution of the O-IR flux was expected to broadly track the X-ray outburst evolution, analogous to the multiwavelength behavior observed in white dwarf binaries \citep{2026SSRv..222...32S}.
In addition, the irradiated disk scenario makes clear predictions for short-timescale variability: if at all correlated with the X-ray emission, O-IR variations should appear delayed and temporally smeared owing to reprocessing in the disk \citep{obrien2002}.

However, observational studies have largely failed to confirm either of these expectations: multiple lines of evidence now challenge the view that the accretion disk dominates the O-IR emission throughout the outburst. 
Sometimes, the continuum appears remarkably featureless \citep{2014MNRAS.445.2424N,froning2014}, raising questions about the expected ionization structure of the disk material.
The broadband evolution of the O-IR flux neither follows the canonical fast-rise exponential-decay profile nor closely resembles the X-ray light-curve evolution \citep{Jain2001,Jain2001b,Kosenkov2020GX339,Baglio2025}.
Finally, the advent of high-time-resolution observations has largely ruled out a thermal disk-dominated origin of the O-IR emission in the hard and intermediate states \citep{Casella2010, 2019ApJ...887L..19V}, although such a contribution may still be important in the soft state or close to that (see Sect.~\ref{sec:mwtiming}).

\begin{figure}[h!]
\centering
\includegraphics[width=1.0\textwidth]{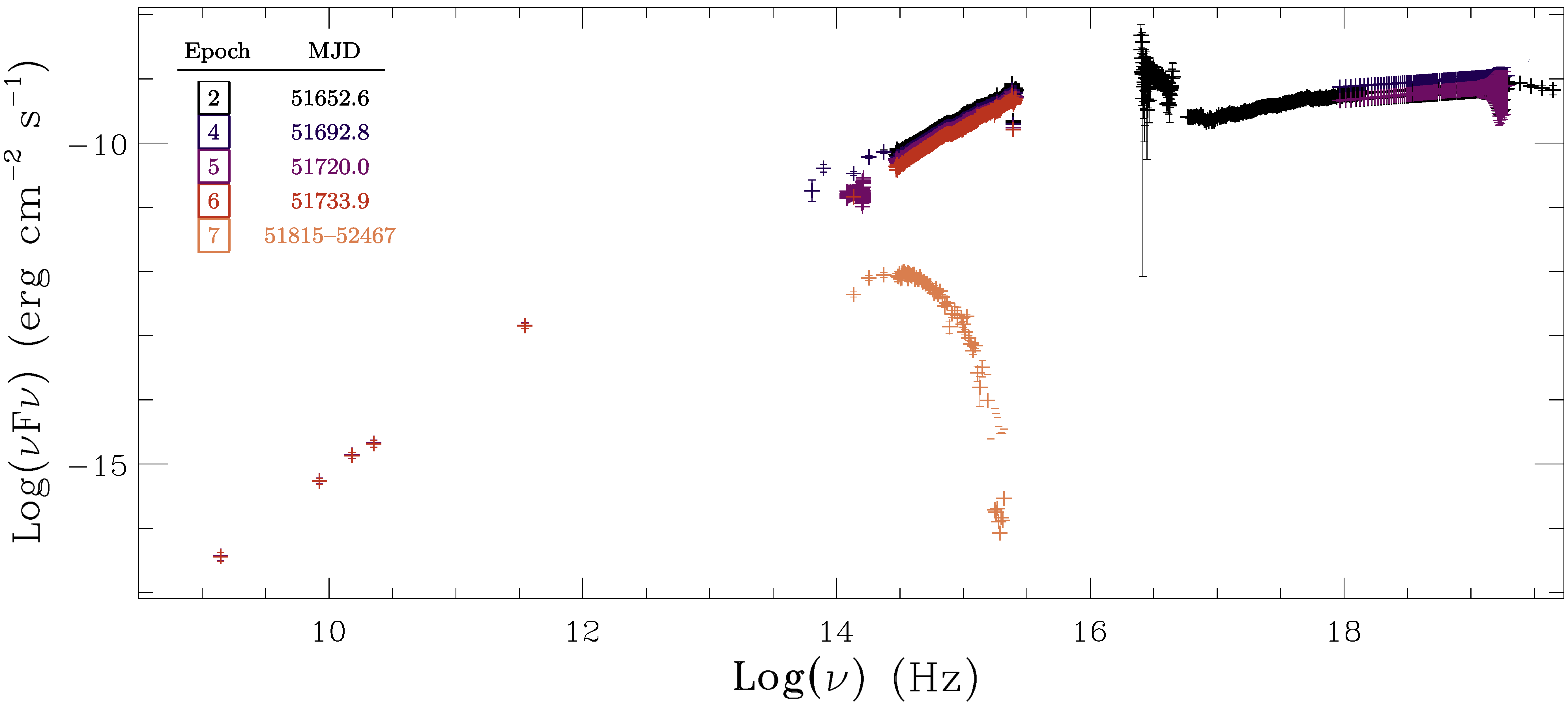}
\caption{Multi-wavelength observations from XTE~J1118+480 in 2000--2001. Observation were carried out in radio (VLA, Ryle Telescope, and JCMT, log$(\nu [{\rm Hz}])=9.0$--$11.6$), near IR (UKIRT, $13.78$--$14.48$), Opt and UV (HST, $14.47$--$15.41$; EUVE, $16.38$--$16.61$), X-ray (BeppoSAX, $16.61$--$19.68$; Chandra, $16.76$--$18.23$; XTE, $17.78$--$19.68$). Epochs 2 through 6 are during the hard state, while Epoch 7 was during quiescence. This figure is adapted from \citet{2003MNRAS.346..689C}.}
\label{fig:Chaty03}
\end{figure}

On the other hand, the UV-to-optical spectra of BHXBs in outburst are generally consistent with a thermal origin, although they are often steeper than predicted by the standard \citet{SS73} disk model \citep[][see Fig.\,\ref{fig:Chaty03}]{2003MNRAS.346..689C,hynes2005b,Zurita2006,froning2014}.
Toward IR wavelengths, the spectra frequently flatten, indicating an excess above the extrapolated thermal emission. 
This excess is commonly interpreted as the contribution of an additional component with red spectrum, likely of non-thermal origin.
Two main scenarios have been proposed: synchrotron emission from the jet \citep{Markoff2001,Malzac2013,Peer2014} or from the hot accretion flow \citep{Veledina2013,Poutanen2014}.
Additional contributions could arise from the accretion disk winds \citep{2026SSRv..222...39M} and, particularly in the mid-IR band, from circumbinary material \citep{Muno2006}.
Disentangling these components spectrally is challenging, but can be achieved by exploiting their distinct predictions for short-term variability, long-term evolution, and polarimetric properties.

\subsection{Radio emission} \label{sec:radio}

Radio emission was detected and associated with X-ray binaries already a few years after their discovery (see for example \citealt{1968Natur.218..855A}, \citealt{1969ApJ...155L..27A}, or \citealt{1971ApJ...164L...1H}). When detected, the radio is believed to be produced (or at least dominated) by synchrotron emission emanating from two jets \citep{BK79}. These jets are often called \textit{radio jets}, although their emission can extend all the way up to the optical ranges \citep{2003MNRAS.346..689C, 2015ApJ...814..139K, 2024ApJ...962..116E}, where much more energy is released.

During the life of an X-ray binary, the radio emission can change drastically, leading to three main different cases: compact jets, transient jets, absence of emission.

\subsubsection{Observations}

Compact jets are observed throughout the hard state \citep{2020MNRAS.493L.132T, 2026MNRAS.551g1022C}, when the radio spectrum is flat (spectral index $\alpha \approx 0$, where $F_\nu \propto \nu^\alpha$) or inverted to slightly inverted ($\alpha \approx 0.5$). Although there are some changes from quiescence to the hard state (see, e.g. \citealt{2015MNRAS.446.4098P,2019ApJ...874...13P,2021ApJ...907...34S}), only a handful of sources have been studied at low luminosity \citep{2020MNRAS.493L.132T} and the general picture depicted below remains relatively consistent.

\begin{figure}[h!]
\centering
\includegraphics[width=1.0\textwidth]{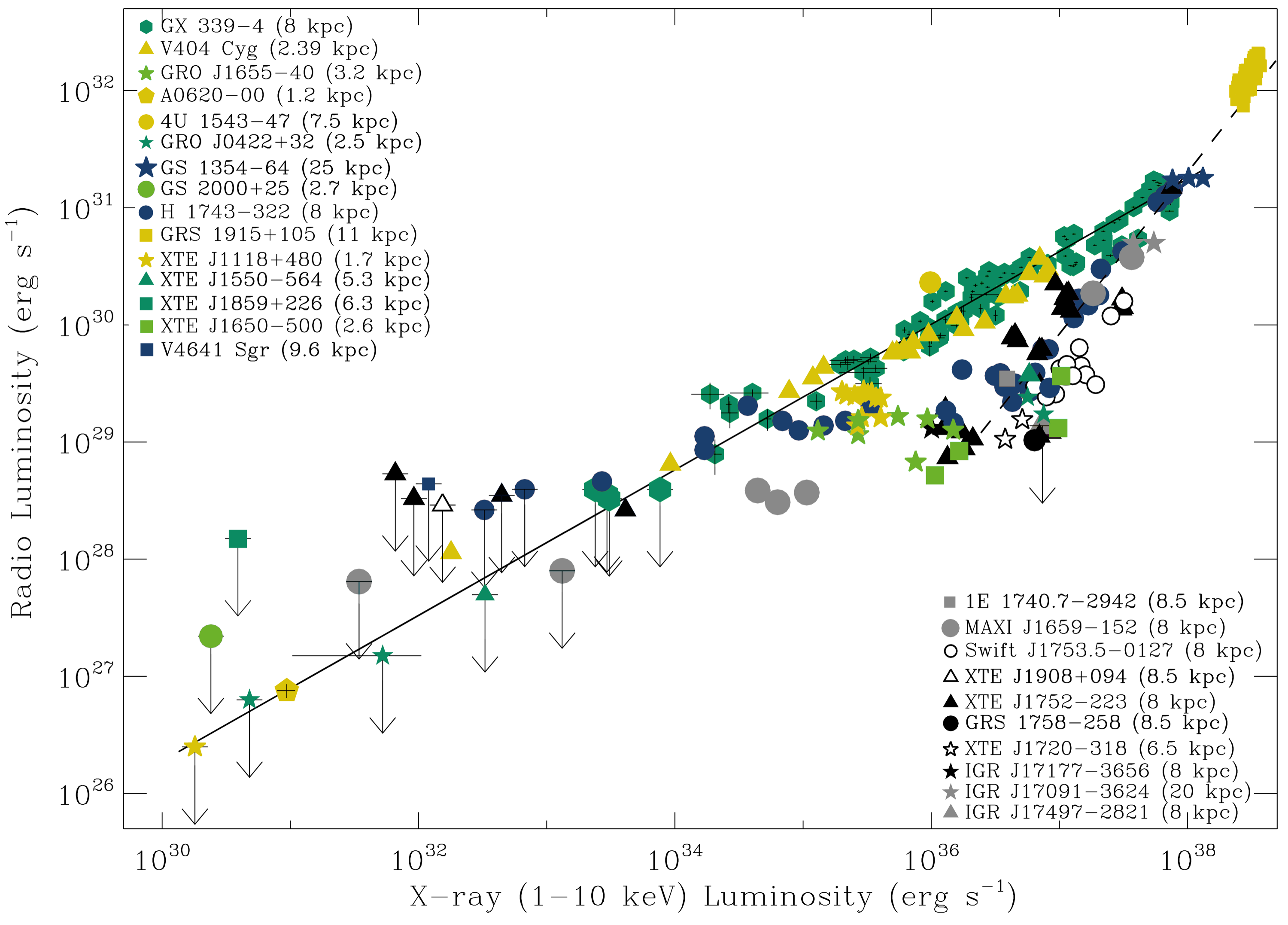}
\caption{Radio--X-ray correlation for a number of sources. Colors are for sources with a dynamically confirmed black hole, gray for non-confirmed ones. The two slopes illustrate the `standard' track $L_{\rm R} \propto L_{\rm X}^{\approx .6}$ (solid line) and the `outlier' track $L_{\rm R} \propto L_{\rm X}^{\approx 1.4}$ (dashed line). Adapted from \citet{2013MNRAS.428.2500C}.}
\label{fig:Corbel13}
\end{figure}

These jets are vertically stratified, with emitted photon energy decreasing with distance from the disk's plane and/or the black hole; the radio component thus being further away than the IR and optical components. Their spectral energy distribution is the sum of all contributions at the different regions, usually attributed to partially self-absorbed synchrotron emission due to local internal shocks \citep{BK79, HS03, 2014MNRAS.443..299M}.
When the jets are compact, the radio flux is tightly linked to the soft to medium X-ray ($1-10$ or $3-9$\,keV) luminosity. This is evidenced by the strong radio--X-ray correlation observed in all XRBs \citep[e.g.,][]{1998A&A...337..460H, 2004ApJ...617.1272C, FBG2004, FHB2009, 2010MNRAS.406.1471S, 2011ApJ...739L..18M, 2013MNRAS.428.2500C, 2014MNRAS.445..290G}. \citet{arash_bahramian_2022_7059313} collects published data on a repository, but note that they mix different X-ray and radio ranges. We show in Fig.\,\ref{fig:Corbel13} the correlation evidenced by \citet{2013MNRAS.428.2500C}, indicating that sources can harbor different correlations, i.e. different slopes. Some XRBs share the same track $L_{\rm R} \propto L_{\rm X}^{\approx 0.6}$, where $L_{\rm R}$ is the observed radio luminosity (usually at $5$ or $10$\,GHz) and $L_{\rm X}$ is the soft X-ray luminosity (usually in the range $3-9$\,keV). This is called the standard track, and it describes well the behavior of numerous sources, e.g., GX~339--4 \citep{2000A&A...359..251C} and V404~Cyg \citep{2008MNRAS.389.1697C}.

However, more and more sources seem to diverge from the track, showing instead $L_{\rm R} \propto L_{\rm X}^{\approx 1.4}$, and are called the `outliers'. For these systems, a change in X-ray luminosity, and by extension in accretion power, leads to a much broader change in radio flux. More recent work suggest that those outliers actually change track as the luminosity evolves, similarly to H1743--322 (in red on Fig.\,\ref{fig:Corbel13}), see e.g. MAXI~J1348--630 \citep{2021MNRAS.505L..58C, 2022MNRAS.517L..21C}. These track changes have been suggested to be due to changes in radiative efficiency of the accretion flow \citep{2011MNRAS.414..677C, 2019ApJ...871...26K, 2022A&A...659A.194M}, but no consensus has yet emerged.
Moreover, throughout this entire phase, the jets are relativistic when their velocity can be estimated, with Lorentz factors that can reach $\Gamma \geq 2$ \citep[][]{2006csxs.book..381F, Carotenuto2024, 2026NatCo..17.6477Z, 2026MNRAS.545f2102L}. Compact jets have been resolved in the radio in a handful of sources \citep{2000ApJ...543..373D, 2001MNRAS.327.1273S, 2014ApJ...796....2R, 2015MNRAS.450.1745R, 2021MNRAS.504.3862T}. In all those cases, they propagate up to $\lesssim 10^{15-16}$\,cm from the central black hole \citep{2024ApJ...967L...7Z, Wood2024}. These compact jets, are usually what is referred to when jets from X-ray binaries are mentioned \citep{2000MNRAS.317....1F}.

\begin{figure}[h!]
\centering
\includegraphics[width=1.0\textwidth]{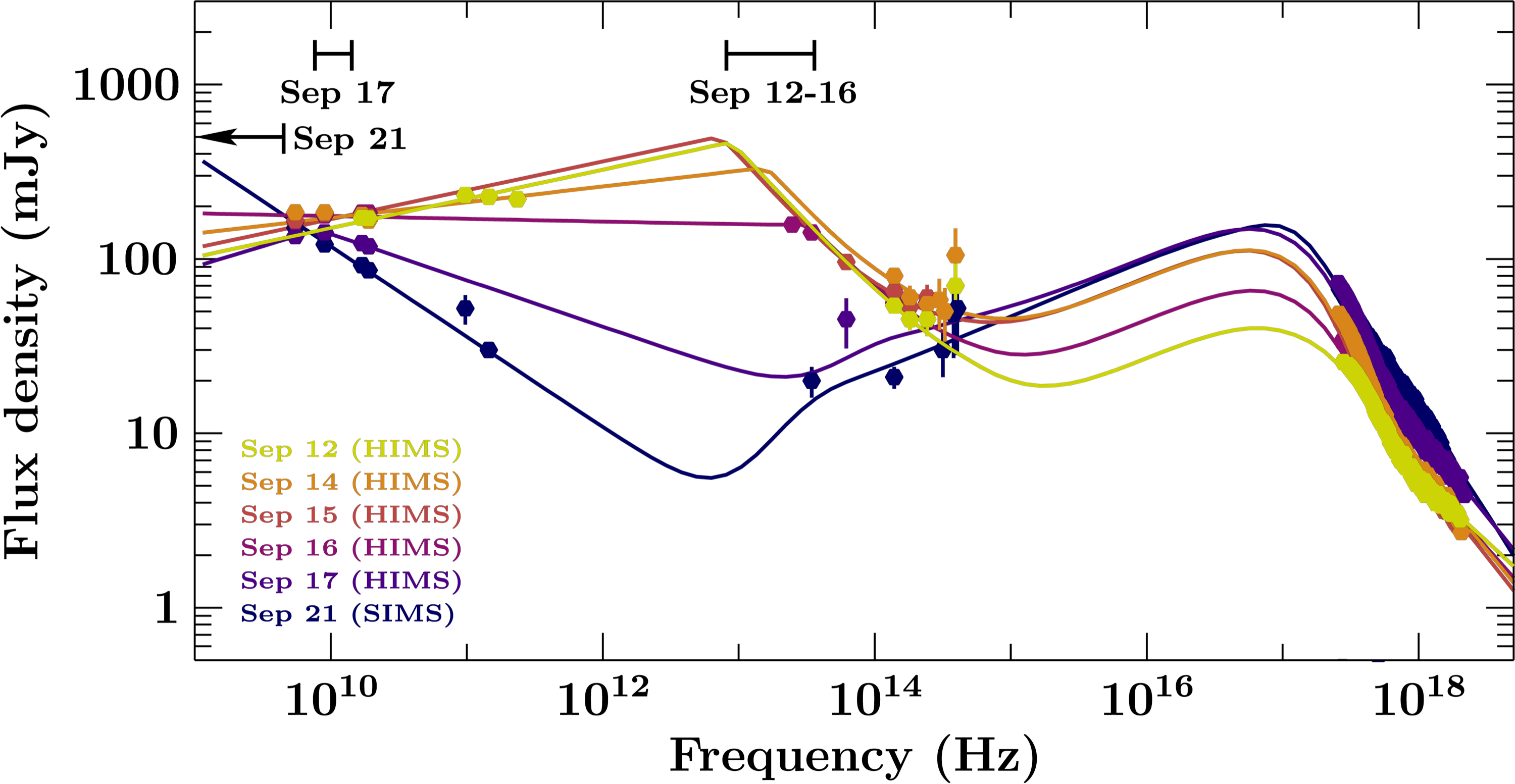}
\caption{Spectral change of MAXI~J1535--571 during its hard-to-soft transition during the 2017 outburst. The colors symbolize the dates, with HIMS for hard-intermediate state and SIMS for soft-intermediate state. Data points are the observations, solid lines represent a broad-band model. The horizontal bars show the location of the jet's spectral break $\nu_{\rm break}$. Adapted from \citet{2020MNRAS.498.5772R}.}
\label{fig:Russell20}
\end{figure}

In turn, `transient jets' are observed when the source transitions to the soft state, during the hard-to-soft transition. During this phase, the jet spectral break moves to lower frequencies \citep[e.g.,][]{2015ApJ...814..139K}, and the optically thin part of the synchrotron spectrum ($\alpha \approx -0.5$), which lies above the break, comes to dominate over an increasingly broad frequency range -- often interpreted as a signature of jet expansion and a corresponding decrease in internal density. This evolution is shown in Fig.\,\ref{fig:Russell20} in the case of MAXI~J1535--571 during its 2017 outburst, where the jet spectral break decreases from $\nu \approx 10^{13}$\,Hz at the beginning of the transition on Sep 12, 2017, down to $\nu \lesssim 10^{10}$\,Hz after the transition (Sep 21), before being quenched in the soft-state. The jets observed during this transition are only observed for a day or two and are thus called `transient'. Note that during the soft-to-hard transition these transient jets are not observed \citep{FBG2004, 2004ApJ...617.1272C, 2023MNRAS.522...70W}. These jets , or more probably blobs, are also relativistic \citep[e.g.,][]{2023MNRAS.522...70W,2025ApJ...984L..53W} and have been observed (much) further from the source than compact jets, i.e., up to $10^{18}$\,cm \citep{2021MNRAS.504..444C, Carotenuto2024}.

Finally, there are also cases when no radio emission is observed. This is the case during both quiescence \citep[see, e.g.,][]{2006MNRAS.370.1351G, 2014MNRAS.445..290G, 2013MNRAS.428.2500C, 2020MNRAS.493L.132T, 2017ApJ...834..104P, 2021MNRAS.503.3784P} and the soft-state \citep{2000A&A...359..251C, FHB2009, 2011ApJ...739L..19R}. During quiescence, it is still unclear whether jets are present or not, as the X-ray emission (and thus the accretion power) is expected to be too low for the jets’ power to produce detectable radio fluxes \citep[see however, e.g.][]{2005MNRAS.356.1017G}, and jets have been observed in some cases \citep{2006MNRAS.370.1351G, 2011ApJ...739L..18M}. During the soft-state, the accretion power is as strong as during the hard-state, and yet the radio is orders of magnitude lower (and below detection limits). It is thus commonly accepted that the jets have disappeared during this phase \citep[see, however,][]{2017MNRAS.466.4272D, 2020ApJ...894L..18Z}.

\subsubsection{Theory and important questions}

Both transient and compact jets are very powerful and clearly coupled to the accretion process \citep{FBG2004}. However, the reasons behind the existence of these two different flavors remain elusive and seldom studied, with ideas ranging from a difference in the structure of the inner (hot) accretion flow itself (see, e.g., \citealt{2023PASJ...75..677I}), to a difference in the matter content (baryonic vs leptonic, \citealt{2024ApJ...967L...7Z}; see also \citealt{2000ApJ...534..109S} in the context of AGN).
The question of jet composition remains a core issue in ejection physics \citep[e.g.,][]{2000ApJ...534..109S, 2021MNRAS.504.3862T, 2022ApJ...928L...9Z, 2022ApJ...935L...4Z}. While the observed synchrotron radiation requires relativistic electrons (or positrons), some estimates of the jet kinetic power suggest a baryonic contribution \citep[see however][]{2021MNRAS.504.3862T}. Resolving this question will require broadband and polarimetric observations, particularly in the radio and $\gamma$-ray bands.

Another key issue is the mechanism responsible for jet production. Two main processes are typically considered: (i) the \citeauthor{BZ77} (\citeyear{BZ77}, hereafter BZ) mechanism, which extracts rotational energy from the black hole via magnetic fields threading the event horizon; and (ii) the \citeauthor{BP82} (\citeyear{BP82}, hereafter BP) mechanism, which taps the accretion flow by launching jets along magnetic field lines anchored in the disk \citep[see also][]{1995A&A...295..807F, 1997A&A...319..340F, BB99}.
At first glance, the BZ mechanism appears better suited to explain the high jet powers and velocities observed in BHXB \citep{1999ApJ...512..100L, 2001Sci...291...84M}. However, if BZ dominates, one expects a strong correlation between black hole spin $a$ and jet power $P_{\rm jets} \propto a^2$. While such a link has been proposed for transient jets \citep{2014SSRv..183..295M}, it is not observed in steady, compact jets \citep{2010MNRAS.406.1425F, 2016ASSL..440...99M}, though spin estimates remain uncertain \citep{2026NewAR.10201746Z}.
More broadly, the ubiquity of jets across accreting systems (including protostars and neutron stars) challenges a purely spin-powered scenario. In addition, the suppression of jets in the soft state is not naturally explained by BZ alone, unless one assumes that efficient BZ jets require geometrically thick flows \citep[as in, e.g.,][]{2015ASSL..414...45T}. This state dependence instead points to a stronger connection with accretion flow geometry, favoring BP-like processes, although both mechanisms may operate under different conditions \citep{2006A&A...447..813F, 2014SSRv..183..295M}, possibly with BP powering the mildly relativistic compact jets and BZ powering the highly relativistic transient ones \citep[e.g.,][]{2012MNRAS.419L..69N, 2013EPJWC..6101005F}. We therefore regard this as an open, and critical, question.

\section{Variability and timing} \label{sec:timing}

Variability measurements provide fundamental constraints on the nature of astrophysical objects by probing spatial scales that are inaccessible to direct imaging or continuum emission. Rapid variability is one of the most powerful tools for understanding emission mechanisms and geometries, providing direct insight into the physical processes operating close to compact objects.

\subsection{Introduction and context}

Variability is most naturally studied in Fourier space. The power spectral density (PSD), defined as the squared modulus of the Fourier transform of the light curve, quantifies the variability power as a function of temporal frequency and is the primary tool for characterizing both aperiodic variability and quasi-periodic oscillations (QPOs), discussed below. Cross-spectral techniques extend this framework to pairs of light curves in different energy bands or wavelengths: the cross-spectrum yields the phase lag (often interpreted as time lag) between bands as a function of Fourier frequency, while the coherence measures the degree to which variability in one band is linearly correlated with that in another \citep{Vaughan1997}. These tools have been applied extensively in the X-ray band, where \textit{RXTE} provided the long baseline and high count rates needed to measure PSDs and lags across many sources and accretion states \citep{vanderklis2006}, and more recently with \textit{NICER}, whose low-energy response and high throughput have opened new windows on soft X-ray variability and reverberation \citep[e.g.,][]{2021A&A...654A..14D, Wang2022}; see also \citet{Uttley2014} for previous work with \textit{XMM-Newton}. At longer wavelengths, where count rates are lower and detector read-out speeds have historically been limiting factors, these same Fourier methods are increasingly being applied to simultaneous multi-wavelength data sets, enabling direct measurements of the X-ray/optical and X-ray/radio cross-spectra that probe the coupling between the corona, disk, and jet.

Quasi-simultaneous slow (day-timescale) variability has been well studied in X-ray binaries for several decades. Rapid optical variability studies were moderately common in the early part of the satellite astronomy era, when optical telescopes were often equipped with photomultipliers \citep{Motch,Imamura}. As charge-coupled devices (CCDs) became the dominant photometric detectors, this work largely ceased until the development of higher-speed solid-state detectors; modern CCDs can now read out fast enough for most fast optical timing applications (e.g., ULTRACAM, \citealt{ULTRACAM}, and HiPERCAM, \citealt{HIPERCAM}). Rapid radio timing was long limited not by detector speed, but by sensitivity, particularly for single-dish telescopes, and by the need for frequent calibration of interferometric arrays. Improvements in bandwidth on the VLA, together with the development of sub-array observing modes that permit continuous observations, have now enabled rapid radio timing measurements.

The most sensitive timing studies of BHXBs have been carried out with satellites in low-Earth orbit. Their typical duty cycle of approximately $60\%$, consisting of about 50 minutes of uninterrupted observing followed by roughly 40 minutes of Earth occultation, leaves poor sensitivity to variability on timescales between a few tens of seconds and a few days. Among satellites in high-Earth orbit, only \textit{EXOSAT} and \textit{INTEGRAL} had combined large collecting areas (i.e., $> 1000\,{\rm cm}^2$) with the ability to observe bright sources without severe pile-up. Consequently, studies of BHXB variability have predominantly revealed oscillations with periods of about 10 seconds and shorter, together with a smaller number of discoveries of slower oscillations and periodic or quasi-periodic signals on timescales comparable to or exceeding the binary orbital period. Because of these instrumental selection effects, however, it remains unclear how common oscillations on intermediate timescales actually are.

\subsection{X-rays}

Much has been learned from measurements of X-ray binaries made solely in X-rays.  It was recognized very early on that strong variability is often seen from accreting black holes in hard spectral states \citep{oda}, and subsequently that the variability can be stochastic, rather than periodic \citep{Terrell}. Over time, it has become clear that accreting stellar mass black holes show strong variability only when their X-ray spectra include a hard component (\citealt{2001MNRAS.321..759C}). In turn, variability tends to be very weak during disk-dominated (soft) states, i.e., when the spectra corresponds to the \citet{SS73} prediction. This represents probably the most substantial mystery regarding the connection between the data and models for the soft states, as the standard \citep{SS73} model is found to be thermally unstable \citep{LightmanEardley}; see Sect.~\ref{sec:softXrays}.

\subsubsection{Aperiodic variability}

Early variability observations of X-ray binaries showed power spectra in hard states that were well-modeled by twice-broken power laws \citep{BelloniHasinger}.  The typical break frequencies in the power laws in bright hard states are $\nu \sim 10^{-2}-10^{-1}$Hz. If the PSD is expressed as $P_\nu \propto \nu^\alpha$, then $\alpha$ is typically approximately zero below the lower break frequency and typically close to $-1$ above the break frequency, and the normalization of the power spectrum above the break is typically nearly constant if the data are presented in fractional rms units \citep{BelloniHasinger}. At higher frequencies, in the best data sets, an additional break is seen \citep{BelloniHasinger}. Some systems tend to show more rapid variability at higher energies \citep{MCP2000,Lin2000}. This rules out diffusive processes like light travel times due to repeated Compton scatterings as a mechanism for making the hard X-rays typically lag the soft X-rays.

With higher quality data from \textit{RXTE}, it became clear that the power spectra show ``wiggles'' that are not well fitted by the broken power law model, leading to the use of multiple Lorentzians to fit the data \citep{Nowak2000}. It was later shown that similar modeling also works well for neutron stars in low luminosity states \citep{vanStraaten2001, Belloni2002}, and, largely speaking, the components move together in frequency space, whether they have high or low quality factor values for their oscillations \citep{WijnandsBreaks, Belloni2002, 2008ApJ...675.1407K, 2020MNRAS.496.5262V}.

The variability of X-ray binaries in hard states shows multiple lines of evidence for nonlinearity. In short data segments, the absolute rms amplitude scales linearly with brightness \citep{UttleyMcHardyRMS}, and the flux distributions are well-modeled by log-normal distributions \citep{UttletNonlinear}, indicative of processes in which perturbations are multiplied together rather than added independently.  Light curves are also not time reversible \citep{BispectrumMaccaroneCoppi}, showing short timescale variability that typically rises more slowly than it falls off.

Short-term aperiodic variability has long been used to probe the characteristic scales of the hot Comptonizing medium–cold reflector (accretion disk) system. One approach is to study spectral variations on different timescales, or, equivalently, at different Fourier frequencies. In this method, the variable component of the count rate is extracted over a range of Fourier-frequency and photon-energy bins. These variable parts of the count rates are then converted into energy spectra corresponding to slower or faster variations. This forms the basis of the frequency-resolved (or Fourier-resolved) spectroscopy technique.

\begin{figure}[h!]
    \centering
    \includegraphics[width=\linewidth]{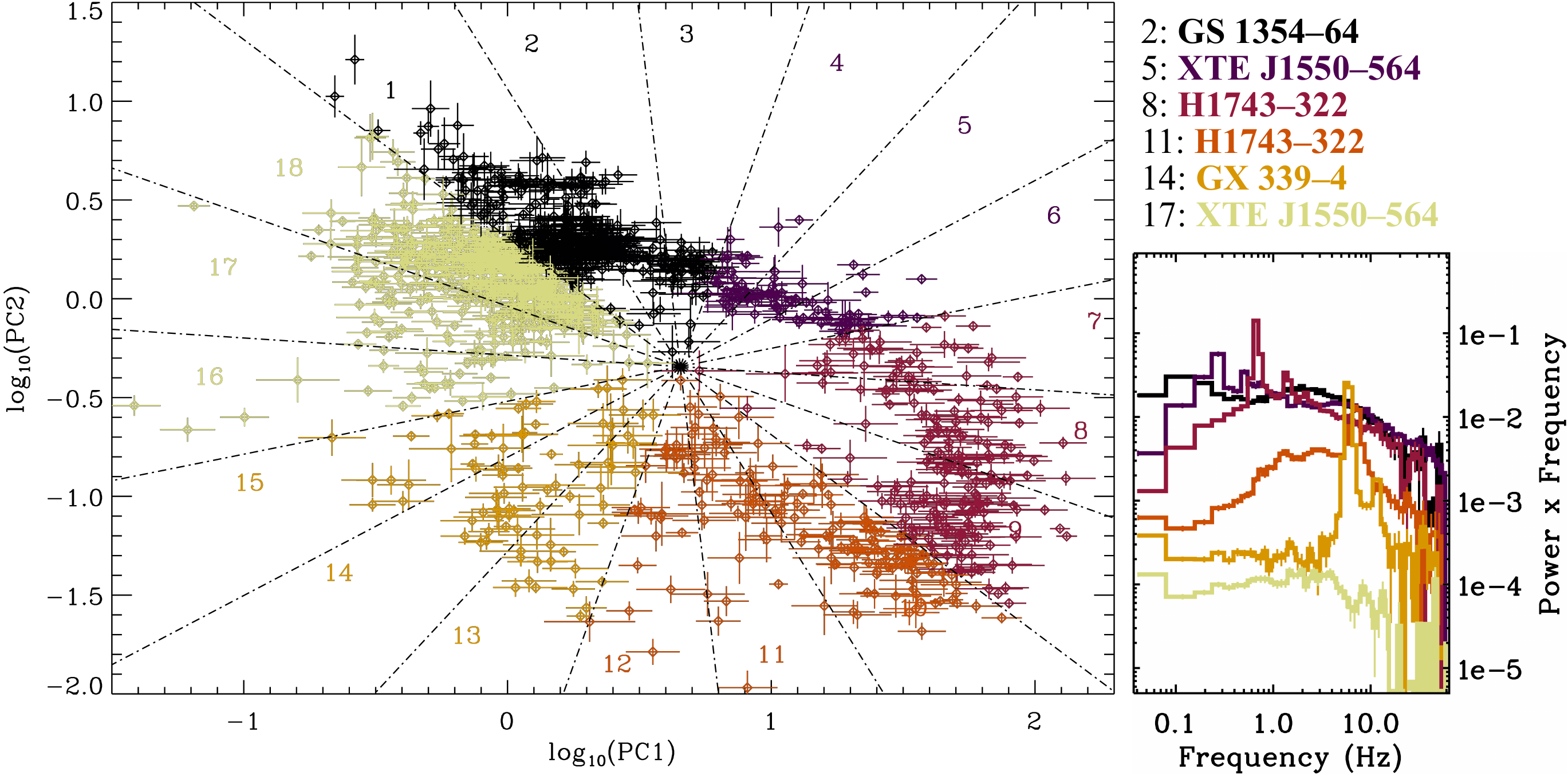}
    \caption{Left: Position in the PC1--PC2 plane, derived from a principal component decomposition of the power spectral shape, for a sample of black hole X-ray binaries; oblique (dot-dashed) lines separate empirically defined variability states. Right: Six representative power spectra (colored to match their clock position), spanning hard (black) to soft (yellow) states, illustrating the corresponding drop in broadband variability power as sources move around the diagram. Adapted from \citet{2015MNRAS.448.3339H}.}
    \label{fig:Heil2015}
\end{figure}

We illustrate in Figure~\ref{fig:Heil2015} this state-dependent variability using power-colors, where the shape of each source's PSD is decomposed into two principal components, or colors: power color 1 (PC1) and power color 2 (PC2), see \citet{2015MNRAS.448.3339H}. In a power color diagram, the position is tracked via  an angle called the `hue', which tracks the spectral/timing state in a way that is largely source-independent: different sources trace out a common, quasi-continuous sequence of positions as they evolve through an outburst. This diagram should be used in addition to the HID (Fig.\,\ref{fig:Zdz02-04}) to describe a source evolving during an outburst. \citet{2015MNRAS.448.3339H} divided the plane in 18 positions, and we show six representative PSD drawn from different positions and sources, spanning hard (black) to soft (yellow) states. The hard-state PSD (cases 2 and 5) show strong, broadband variability, with power concentrated around $\sim 0.1$--$1$\,Hz and fractional rms of several tens of percent. They also show narrow components, called quasi-periodic oscillations (QPO), see Sect.~\ref{sec:lfqpos}. As sources move toward softer states (cases 8, 11, and 14), the total variability power drops by an order of magnitude or more, and by the softest state sampled here (case 17), the PSD is essentially featureless and more than two orders of magnitude weaker than in the hard state. This progressive suppression of broadband noise, independent of the source producing it, exemplifies why PSD shape alone can serve as a state diagnostic complementary to spectral hardness.

The technique was first developed to compare the characteristic variability frequencies of the reflection continuum and fluorescent iron line with those of the underlying primary continuum, thereby placing constraints on the distance between the illuminating source and the reflector \citep{1999A&A...347L..23R, 2001A&A...380..520R}. These studies assumed filtering of variability by light-travel delays, and with this assumption showed that this distance decreases by a factor of $\sim 10$--100 as accreting sources transition from the hard to the soft spectral state \citep{2000MNRAS.316..923G,2021MNRAS.507.2744A}. When applied to broadband X-ray spectra, frequency-resolved spectroscopy further revealed the simultaneous presence of a softer, slowly varying spectral component alongside a harder, rapidly varying component \citep{2018MNRAS.480..751A}.

\begin{figure}[h!]
    \centering
    \includegraphics[width=.49\linewidth]{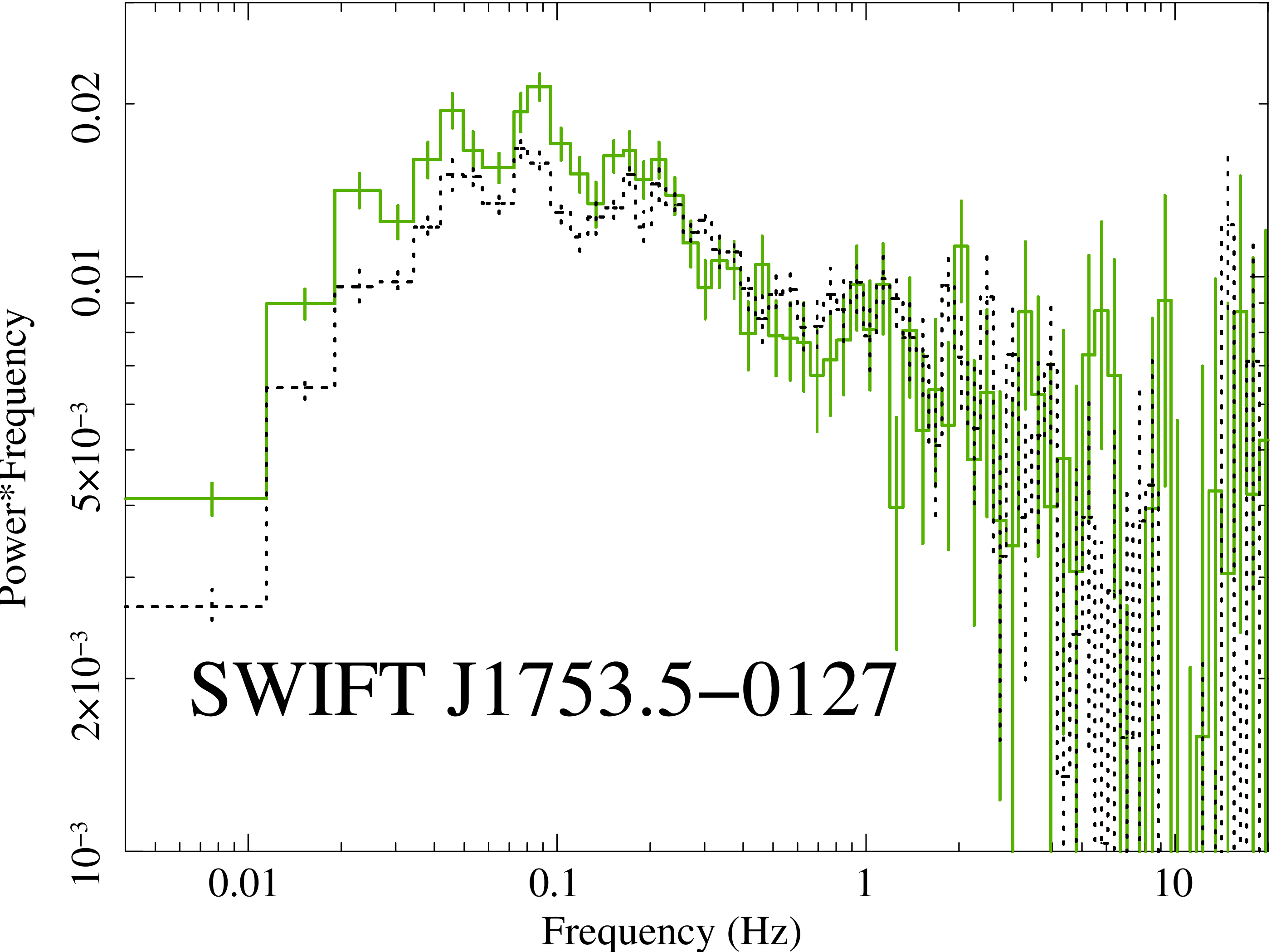} \includegraphics[width=.49\linewidth]{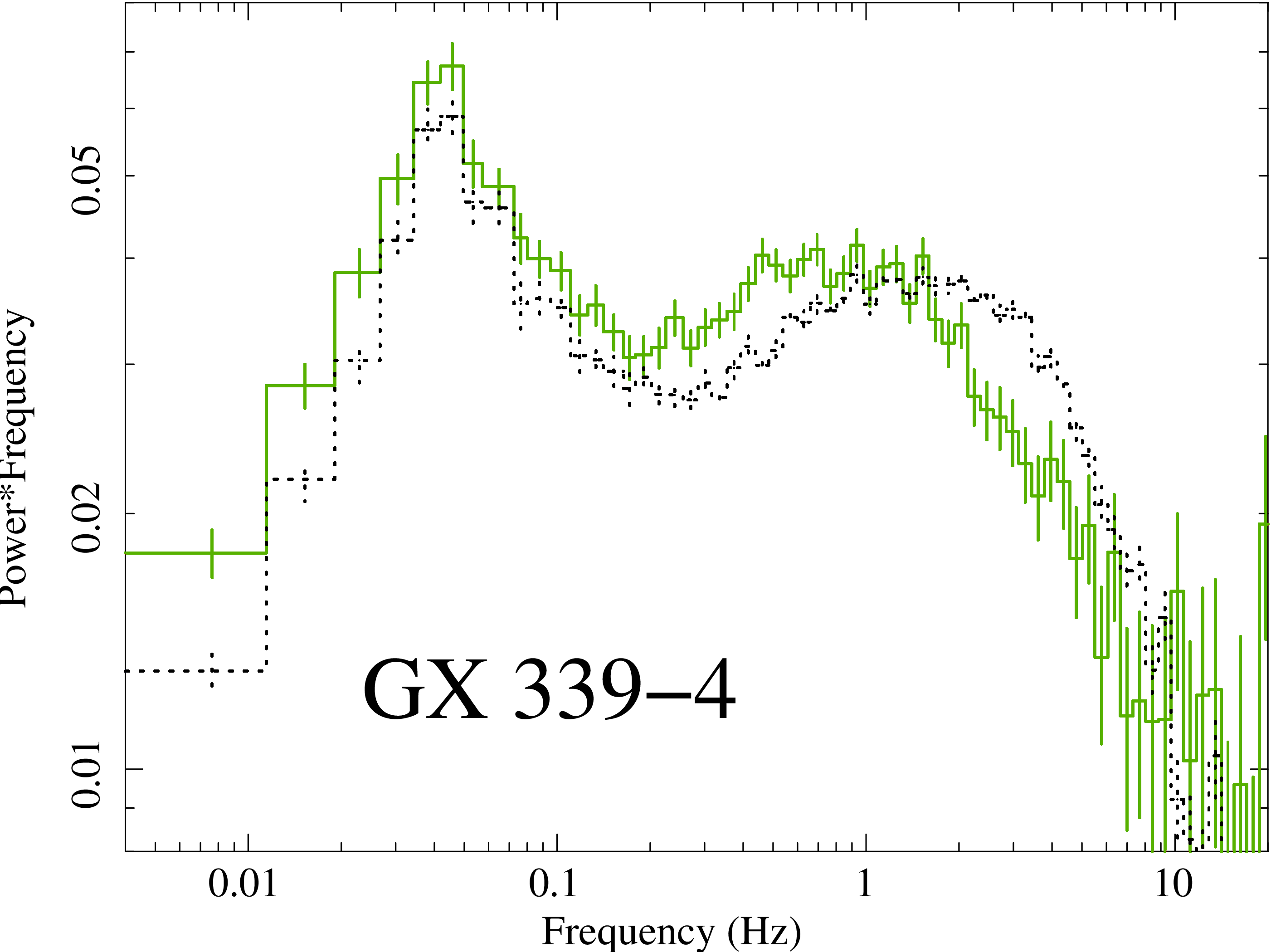}
    \caption{Power spectra from \textit{XMM-Newton} for two hard state X-ray binaries. The solid green curves are the $0.5-1.0$\,keV energy band, coming from the thermal component (disk), while the dashed curves are the $2-10$\,keV, coming from the power law component (hot flow). Adapted from \citet{WilkinsonUttley}.}
    \label{fig:wilkinson}
\end{figure}

Since the end of the \textit{RXTE} mission, studies of aperiodic rapid variability have mostly focused on studies of time lags due to reflection (see Sect.~\ref{sec:timelags} and \ref{sec:reflection}), and studies of the soft X-ray variability, a topic opened up by the launches of \textit{XMM-Newton} and \textit{NICER}.  The accretion disks of hard state X-ray binaries were undetectable in the \textit{RXTE} energy range, and the lack of variability in the soft states' accretion disks led to a lore that thermal disks do not vary.  It was thus a surprise when the bands with the thermal components in hard states showed {\it stronger} variability than harder X-ray bands (\citealt{WilkinsonUttley}, see also \citealt{2015MNRAS.454.2360D}).  The power spectra of the thermal emission bands look broadly similar to those in the hard X-ray component. The high frequency power spectrum cuts off more sharply in the soft, thermal band than in the non-thermal band, but the shapes are quite similar, and there is more power at lower frequencies in the soft component (see Fig.\,\ref{fig:wilkinson}).

\subsubsection{Orbital and longer modulations}

In many BHXBs, a modulation is seen on the orbital period at X-ray and other wavelengths.  This is extremely common in systems with high donor masses, where binaries are often eccentric, and where the stellar winds from the donor star can lead to (sometimes substantial) modulation of the foreground absorption (see, e.g., \citealt{FornasiniReview} and references within).  Some LMXBs also show X-ray orbital modulation, usually because they are eclipsing \citep{WhiteEclipse}, or have ``dips'' because their inclination angles are so large that their flared outer disks often obscure the inner X-ray emitting region \citep{DippersWalter}.  Evidence for the orbital periods in the optical band are much more common, and result from a mixture of spectroscopic evidence for an orbit and ellipsoidal modulations (e.g., \citealt{1986ApJ...308..110M}) and heating of the inner faces of the donor stars by the X-ray sources \citep{1987ApJ...312..739T}.

Some X-ray binaries also show superhump modulations \citep{DOD1996,HaswellSuperhumps}.  These are thought to be on a beat frequency between the orbital period, and a much slower precession period of the outer accretion disk. As they are generated in the outer accretion disks of X-ray binaries, they appear in the optical and IR bands, rather than in the X-ray band. They are very well studied in CVs, where they allow to estimate the orbital period to within a few percent from photometric observations in outburst \citep{PattersonSuperhumps}, and hence allow orbital period estimation in some systems where it is otherwise quite challenging, such as ultracompact binaries \citep{PichardoMarcano}. They are seen when mass ratios are less than about 0.3, as something which should apply to essentially all black hole LMXB \citep{PattersonSuperhumps, HaswellSuperhumps}, although the longer typical orbital periods for X-ray binaries than for CVs makes detecting them challenging. The period excesses, i.e., the differences between the superhump and orbital periods, for X-ray binaries appear to be mostly smaller than for CVs with the same mass ratios \citep{KosenkovSuperhumps}. Note that this is based on a small sample size, and the most recent black hole superhumper (MAXI~J1820+070) shows a larger period excess than expected from the CV trend line \citep{TorresMAXI1820,Thomas1820}.

Superorbital periodicities are claimed in many HMXBs as well as some LMXBs.  In many cases, these periods are long enough that they are hard to distinguish from red noise, but in other cases, they are well-established to be real.  A superorbital period of about 35 days was seen in the relatively early era of X-ray astronomy from neutron star X-ray binary \mbox{Her~X-1} \citep{Giacconi1973}, and a larger sample of such superorbital periods have been seen \citep[and references within]{MaloneyBegelman}. 
The first detection of X-ray superorbital variability in a BHXB was made in Cyg X-1 using data from Vela 5B satellite \citep[with a $\sim\,300$~day period][]{1983ApJ...270..233P}.
Subsequently, this variability was confirmed in soft and hard X-rays, optical polarimetric and radio data, yet in some cases the period was found to be half of the original period \citep[e.g.,][]{1983ApJ...271L..65K,2007MNRAS.381..723I}.
The superorbital periods are typically $\sim 30$ times the orbital periods, with some substantial scatter in the ratio.  In many cases, these periods are transient, lasting only a few cycles, making lists of systems with superorbital periods very susceptible to contamination from red noise.  Theoretical models for these will be discussed in Ogilvie et al. (submitted to SSRv).

\subsubsection{Low-frequency quasi-periodic oscillations} \label{sec:lfqpos}

Of particular interests are quasi-periodic oscillations, defined as narrow peaks in the PSD. They clearly have characteristic frequencies at which there is more power than at adjacent frequencies, while they are also clearly not strictly periodic, coherent phenomena. The review of \citet{vanderklisreview2000} provides an excellent introduction to the basics and phenomenology of QPOs, and the more recent review of \citet{IngramMotta} gives an up-to-date account. We summarize the observational aspects of the problem briefly here for completeness, but refer the reader to that review for more details on both observations and theoretical models.

Three classes of oscillations are reported at low-frequencies, i.e, around 1\,Hz. These have creatively been named Type A, Type B and Type C QPOs (\citealt{Belloni2002, Casella2005}; see \citealt{2026A&A...710A.387M} for a tentative unification).  The Type C QPOs are the strongest, and most frequently seen. Clear evidence exists that the Type C QPOs have a geometric origin,\footnote{In this context, the term `geometric' refers to QPOs being caused by large-scale structural or spatial dynamics in the accretion flow, rather than purely energetic/radiative processes.} as there are inclination angle dependencies for both the amplitude of the modulations \citep[][]{Motta2015, 2026MNRAS.549g1004V}, phase lags between energy bands \citep{vandenEijnden2017}, and the bispectral properties of the modulations \citep{Arur2020}.  The geometric model that has drawn the most attention is the Lense-Thirring solid-body precession model, in which there is a misalignment between the black hole's spin axis and the outer accretion disk that leads to disk warping and frame-dragging \citep{2026SSRv..222...25F}, and in turn, to a precession \citep{Fragile07, IngramDoneFragile}. Despite the similarity in the names, this model is different from the relativistic precession model \citep{StellaVietri}; see \citet{Motta18} for a discussion.  Perhaps the biggest difficulty in explaining those QPOs lies in their very low observed frequencies compared to the dynamical frequencies of the system \citep{2020A&A...640A..18M}. Other models, such as spiral arms in the accretion disk \citep{Varniere}, predict QPOs of about the correct timescale. 
Neither model is well-enough developed theoretically to make predictions that match the full richness of the data. 

A different, more phenomenological approach is offered by the time-dependent Comptonization model \textsc{vKompth} \citep{2020MNRAS.492.1399K, 2022MNRAS.515.2099B}, which treats the QPO as a small, quasi-periodic oscillation of the emitting regions around its time-averaged solution. Rather than attempting to explain the QPO mechanism from first principles, \textsc{vKompth} is fit jointly to the time-averaged spectrum and to the energy-dependent rms and phase-lag spectra of the QPO itself. This method has been applied to a number of sources, including MAXI~J1348--630 \citep{2021MNRAS.501.3173G}, MAXI~J1535--571 \citep{2022MNRAS.512.2686Z}, and GRS~1915+105 \citep{Mendez2022}. This approach has the advantage of directly linking the observed lags and rms amplitude to physical corona parameters (size, temperature, feedback fraction), yielding geometric constraints. It has, however, two main limitations so far. First, the corona is modeled as a uniform sphere, an oversimplification that likely limits the reliability of the inferred geometric parameters. Second, the model is agnostic about the underlying trigger of the oscillation itself, although a possible physical origin has recently been proposed where a disk-corona radiative feedback loop naturally produces limit-cycle oscillations at the QPO frequency \citep{Mastichiadis22, Mastichiadis26}.

The Type A \& B QPOs remain poorly constrained by current models. They are weaker and less frequent than Type C, appearing predominantly near spectral state transitions, and show the opposite rms vs. inclination-angle dependence \citep{Motta2015}. There have been hundreds of Type B and Type C detections, and a few cases of simultaneous B+C detections, see \citet{2026A&A...710A.387M} for a tentative unification and a review of each individual case. There are some suggestions that the Type B QPOs are associated with jet ejections due to frequent quasi-simultaneous appearance of both phenomena \citep{2008MNRAS.383.1089S, Motta2012}. However, in at least one case, the jet ejection has been seen prior to the appearance of Type B QPOs \citep{FHB2009, Carotenuto2024}, challenging this view. Moreover, because some X-ray binaries jitter back and forth on short timescales near the state transition, making multiple state transitions in the same outburst, it can be challenging to make definitive determinations about the order of these phenomena \citep{Carotenuto2024}.

It is worth noting that with next-generation all-sky monitors, as proposed for missions such as \textit{eXTP} and \textit{STROBE-X}, it may be possible to monitor low-frequency QPOs with wide-field instruments. The signal-to-noise for aperiodic variability searches is given by \citep{2019RNAAS...3..116M} as
\begin{eqnarray}&&{({S}/{N})}_{\mathrm{var}}=\displaystyle \frac{1}{2}{({S}/{N}{)}_{\det }^{2}{r}^{{\prime} 2}(\lambda T)}^{-0.5},
\end{eqnarray}
where $(S/N)_{\mathrm{var}}$ is the signal to noise ratio of a variability feature, $(S/N)_{\mathrm{det}}$ is the signal-to-noise for detection of the source, ${r}^{\prime}$ is the intrinsic fractional rms amplitude, $T$ is the exposure time, and $\lambda$ is the FWHM of the feature in the power spectrum.  Thus, for an X-ray monitor with a few mCrab daily sensitivity, an outburst level of about 1\,Crab, and a 5\% rms amplitude of an oscillation with 1\,Hz frequency width, detection of QPOs should be possible.  Type C QPOs will be straightforward, and Type B QPOs will be detectable much of the time.  Because Type A QPOs are shorter-lived, weaker, and broader, they will be detectable only under exceptionally fortunate circumstances -- these have only rarely been detected even with pointed instruments dedicated to timing measurements.

\subsubsection{High frequency QPOs}

In a small number of BHXBs, QPOs have been seen at frequencies above 20\,Hz.  These likely hold clues about the nature of the innermost accretion flow, where effects of the black hole spin are most pronounced.  The high frequency QPOs to date have all been detected with \textit{RXTE}, with some being seen only in very high energy bands.  The first QPOs seen above 20 Hz in accreting black holes were from GRS~1915+105 \citep{Morgan1997}, at 67\,Hz.  Later, QPOs have been seen in numerous observations from GRO~J1655--40 at 300 and 450\,Hz \citep{Remillard1999,Strohmayer2001}, and XTE~J1550--564 at about 180 and 270\,Hz \citep{Homan2001}.  Other systems have shown more tentative evidence for high frequency QPOs \citep{BelloniHFQPO}.  Whether the 3:2 ratios between the frequencies of some of the aforementioned QPOs is a coincidence, or is highly constraining for models, is challenging to say with current data samples.  There is also some evidence that there may be an inclination angle dependence of the amplitude for these QPOs, something also challenging to determine from the meager sample of detections.

The emission spectra of these QPOs, especially the higher frequency ones in the 3:2 pairings, is extremely hard -- e.g., the 450\,Hz QPO in GRO~J1655--40 was originally found only in a light curve in the band $27$--$40$\,keV \citep{Strohmayer2001}. New softer band timing missions have opened up new avenues in X-ray variability studies, but are much less sensitive to high frequency QPOs than \textit{RXTE} was. Notably, despite its hard-X-ray sensitivity, Insight-HXMT has not reported any HFQPO detections (in black hole systems) to date, suggesting that hard-band sensitivity alone is not sufficient: the relevant intermediate states in which HFQPOs are found may themselves be only moderately hard compared to the canonical hard state, and large effective area may be at least as important as raw hard-band reach for detecting these weak signals. Further progress may thus require a large-area timing mission with good hard-band sensitivity, such as the LAD instrument proposed for \textit{LOFT} or \textit{STROBE-X}.

\subsubsection{Time lags}
\label{sec:timelags}

From relatively early on, it was determined that, in hard X-ray bands, the harder X-rays arrive, on average, later than softer X-rays during the hard state \citep{Nolan1981, 1989Natur.342..773M}. For some time, interpretations focused on light travel times in Comptonizing media (\citealt{Cui1997}, see however \citealt{Nowak1999}), in part motivated by the ln\,E dependence of the lags, which is a natural prediction of a model in which lags are derived from increased light travel times due to repeated Compton scatterings \citep{Kazanas97}. However, the required sizes for the hot-flow/corona are far above what is expected \citep{Kylafis2008, UttleyMalzac}. The lags show a frequency dependence that is typically approximately $f^{-1}$--$f^{-0.7}$ (or, equivalently, nearly constant phase lags, \citealt{1988Natur.336..450M}), and this means that the peaks of the cross-correlation functions are generally consistent with zero lag, with the cross-correlation functions being asymmetric. 
We caution that these are, in essentially all cases, \emph{phase} lags $\phi(\nu)$ measured from the cross-spectrum, converted into a time lag via $\tau(\nu)=\phi(\nu)/2\pi\nu$; this conversion implicitly assumes that a single, coherent time delay operates at all frequencies. In practice, the measured phase can also result from other physical processes, which are also known to produce phase lags even without any delays \citep[see, e.g.,][]{RM26}. The direct identification of a `time lag' with a physical light-crossing or propagation delay should therefore not always be taken at face value, and depends on the specific model invoked to explain it.
As it became clear that the characteristic variability timescales in hard X-rays were {\it shorter}, rather than longer than in soft X-rays \citep{MCP2000, Lin2000}, alternative models gained attention, largely involving actual evolution of the source spectrum \citep[pivoting,][]{PoutanenFabian1999, 2004A&A...414..795K}, or inward propagation of disturbances \citep[or fluctuations,][]{Lyubarskii1997, 2001MNRAS.327..799K, Arevalo2006, Ingram2013}. The inwards propagation model has gained traction, largely because it also explains the observed linear rms-flux relation as variations in the accretion rate from different regions of the accretion flow combining multiplicatively \citep{UttleyMcHardyRMS}.

At the highest temporal frequencies, the lag reverses: soft X-rays arrive, on average, after hard X-rays. This is typically interpreted as reverberation: X-rays that irradiate the disk to be reprocessed and re-emitted take a longer path to the observer than those that are observed directly \citep{Uttley2014, Bambi2021}. Since the emergent spectrum of these `reflected' X-rays includes a soft X-ray excess caused by partial thermalization of the irradiating flux, light-crossing delays between direct and reflected emission can cause the observed soft lags. It is the combination of high time resolution, high collecting area in the soft X-ray band, and good spectral resolution from \textit{XMM-Newton} and \textit{NICER} that has led in recent years to the detections of these soft lags \citep[e.g.,][]{2021A&A...654A..14D, Wang2022}. They were first detected in AGN \citep{Fabian2009}, with X-ray binaries proving more challenging due to the shorter delay times associated with smaller black holes. The first detection of soft lags in an X-ray binary was by \citet{Uttley2011} using data from GX~339--4, who found that, at low temporal frequencies, the soft thermal component leads the hard power-law, whereas at high temporal frequencies, it lags it. This was interpreted as the disk driving low-frequency variability in the hot inner flow (or corona), while the disk's most rapid variability is primarily caused by variable heating from irradiation.

\begin{figure}[h!]
    \centering
    \includegraphics[width=1.0\linewidth]{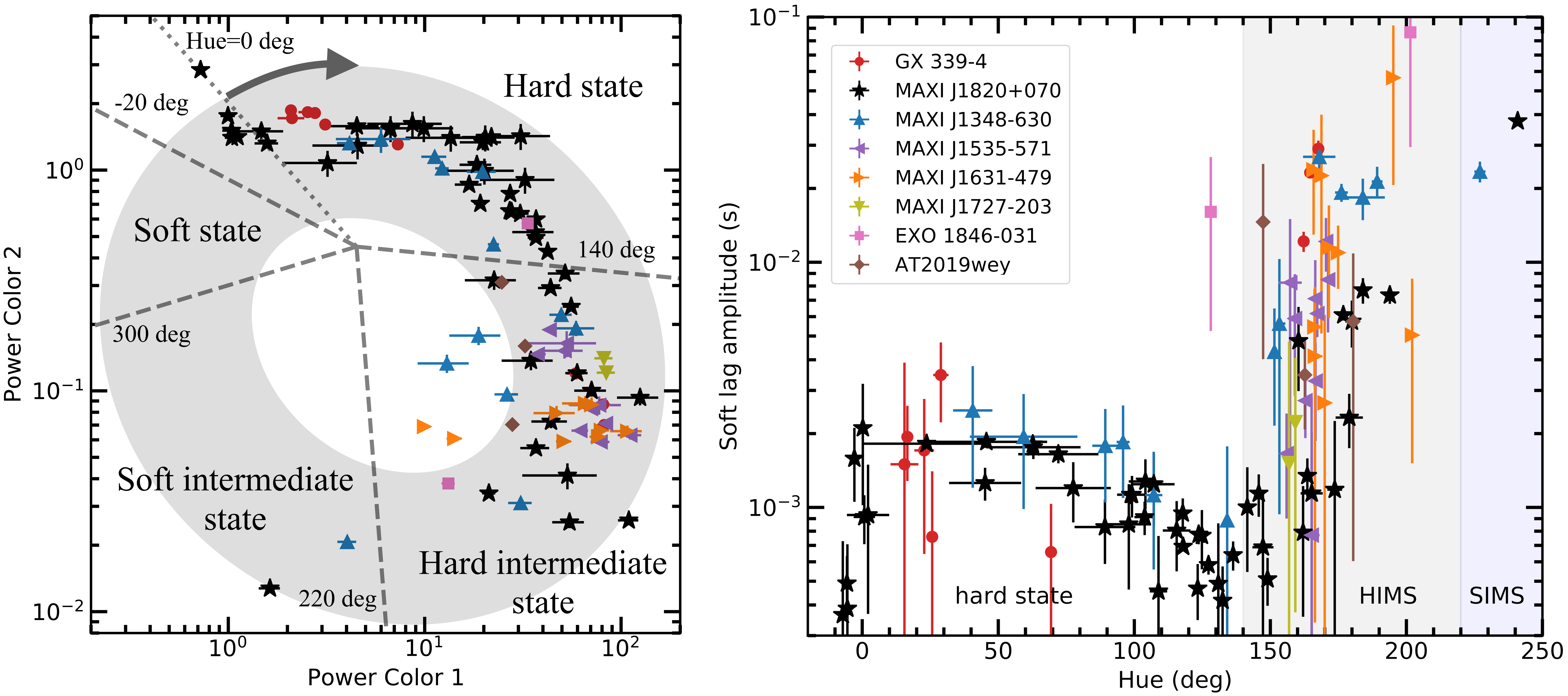}
    \caption{Left: Selected observations of eight sources shown in a power color diagram \citep[see Fig.\,\ref{fig:Heil2015} and][]{2015MNRAS.448.3339H}. Right: Measured soft X-ray lag amplitude for each observation and its associated state. Adapted from \citet{Wang2022}.}
    \label{fig:softlags}
\end{figure}

The emergent reflection spectrum, discussed in more detail in Sect.~\ref{sec:reflection}, also includes a prominent iron K$\alpha$ emission line at $\sim 6.4$\,keV. The iron line provides perhaps a more reliable diagnostic of reverberation than soft lags, given the physical complexity of the soft X-ray band. Iron line reverberation features were, again, first detected in AGN, again due to the longer lags associated with larger objects. Such features have been detected in at least $25$ AGN to date \citep{Kara2016}. For X-ray binaries, hints of an iron line lag were seen first with \textit{XMM-Newton} data \citep[][see however \citealt{2001MNRAS.327..799K}]{DeMarco2017}, but first began to be studied in detail with \textit{NICER} \citep{Kara2019}. A systematic study of the \textit{NICER} lags \citep{Wang2022} shows soft lags strongly increasing in duration as a source transitions from hard states to softer states, for a range of sources (see Fig.\,\ref{fig:softlags}).  There is some debate about whether this should be interpreted as evidence for a large, vertically extended corona in the transition states \citep{Wang2022}, the results of complicated feedback processes \citep{UttleyMalzac}, or a consequence of physical effects that cause a difference in the power-spectral filtering in the two energy bands \citep{RM26}. X-ray reverberation modeling is discussed in more detail in Sect.~\ref{sec:reflection}.

\subsection{Multi-wavelength correlated timing measurements} \label{sec:mwtiming}

A few fast optical timing experiments on X-ray binaries were performed in the 1970s and 1980s, when photomultipliers were commonly used \citep{Motch}, but only with the development of fast solid state detectors did this topic really take off \citep{Phelan2008}. Timing observations hold the key to disentangling the different possible emission components (irradiated disk, jet, hot accretion flow), as the temporal signatures of the different components can be quite different in circumstances where the spectral signatures are quite similar. 

Early optical work focused on using time lags to estimate the size scale of the accretion disk, and to search for temporal evidence of reflection from the companion star \citep{hynes1998}, and found that, at least in some cases, the details of the time delay indicated a level of disk flaring sufficient to shield the donor star from the X-rays arising in the inner disk regions \citep{obrien2002}.  
The discovery of a more complicated structure of the cross-correlation function (CCF) between X-ray, UV and optical emission in XTE~J1118+480 \citep{Kanbach2001,Hynes2003}, showed that there was some emission component other than thermal reprocessing of the X-rays.  Two features stand out in this analysis: (i) the auto-correlation function is narrower in the optical than in X-rays, indicating that the characteristic timescale for the optical-UV emission is faster than that for X-rays; (ii) the CCF shows both a positively lagged (i.e. optical-UV lags X-rays) component with positive correlation and a negatively lagged component with an anti-correlation.  Neither of these features can be explained in the context of thermal reprocessing. 
Potential sources of such anti-correlation include synchrotron emission from the jet \citep[either sharing a common energy reservoir with the X-ray-emitting inner flow or being powered by internal shocks;][]{Malzac2004,Malzac2013}, emission from the hot accretion flow \citep{Veledina2011}, or a combination of the two.

\subsubsection{Steady states}

The first rapid simultaneous IR and X-ray variability in GX~339--4 \citep{Casella2010} showed only positive correlations with IR lagging the X-rays.  Furthermore, the short time lag (0.1 seconds) and the high brightness temperature ($2.5\times10^6$ K) for the IR emission could be explained only by a relativistic jet.  
Subsequent studies revealed a rich diversity in the shapes of the IR-Optical--X-ray CCFs across different sources, ranging from cases dominated by an anti-correlation component \citep[in some instances occurring at positive lags,][]{Durant2008,Durant2011,Paice2019,Vincentelli2021} to those characterized by a narrow positive peak \citep{Gandhi2010,Vincentelli2025} or a broader positive correlation \citep{2005ASPC..330..237H}. 
In many systems, however, these features appear to coexist. 
Follow-up monitoring further demonstrated that the CCF morphology evolves with spectral state \citep{2005ASPC..330..237H,Paice2021}. 
Nevertheless, nearly all high-time-resolution observations report the presence of both a dip (anti-correlation) and a narrow positive peak in the CCF, suggesting that the optical emission arises from the superposition of multiple spectral components.

As more observations of more sources have been obtained, a broader range of phenomenology has developed, with QPOs being seen in IR and optical wavelengths \citep{Durant2009,VeledinaSwiftQPO,Kalamkar,Paice2021,Vincentelli2025}. In fact, there are several cases where QPOs have been seen at long wavelengths, but not seen in the corresponding X-ray data.  This is most likely because the count rates in the optical and IR bands are much larger than with current X-ray facilities, and because these particular observations were done with soft X-ray timing measurements, for which the fractional amplitude of variability is much lower than in the hard X-ray band.

\begin{figure}[h!]
    \centering
    \includegraphics[width=0.63\linewidth]{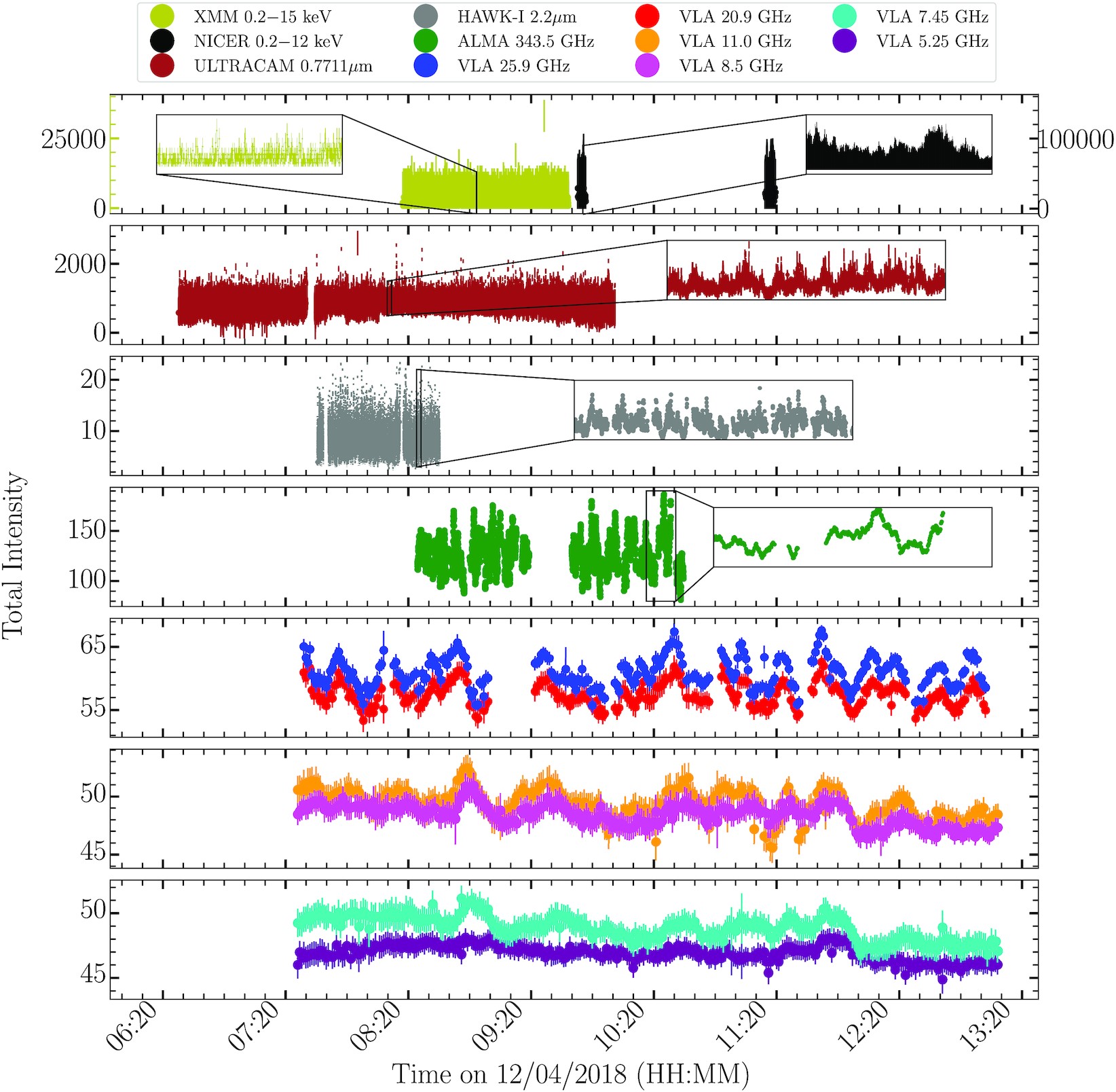}
    \includegraphics[width=0.33\linewidth]{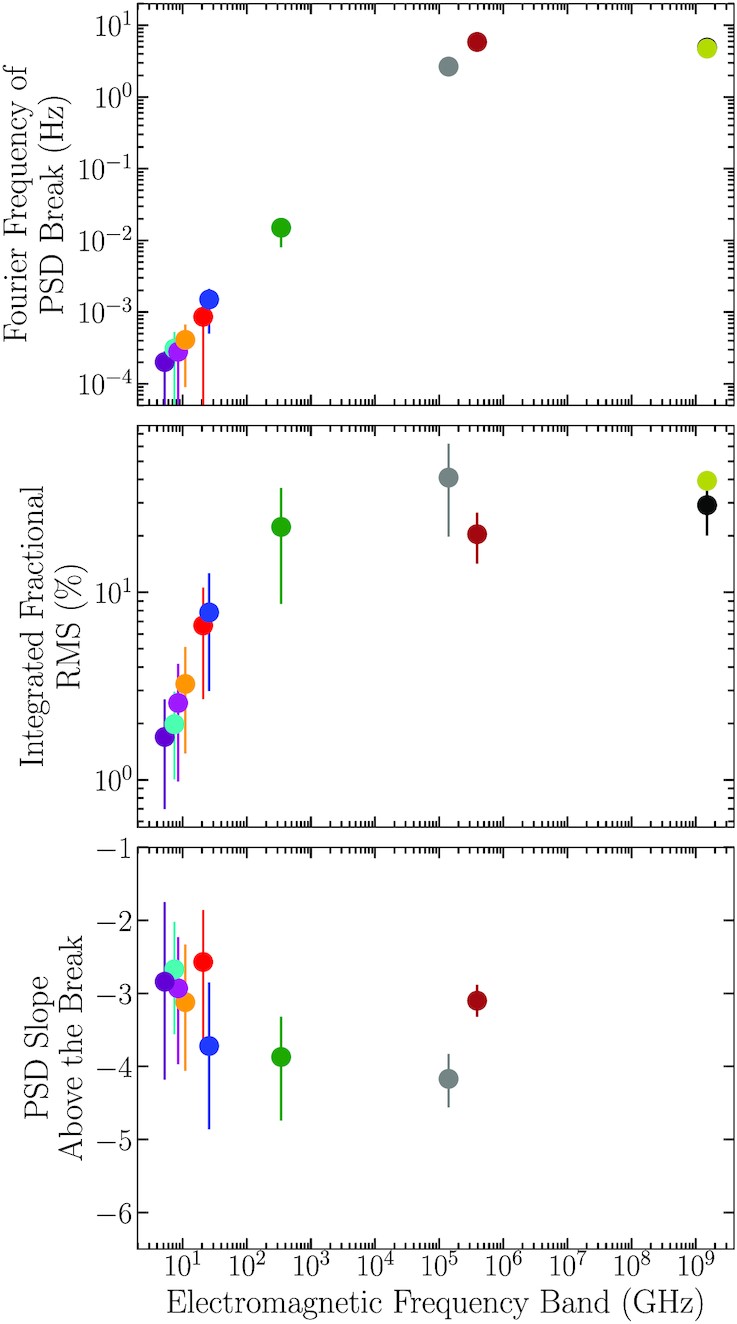}
    \caption{The results from multi-wavelength timing of MAXI~J1820+070. Left: light curves in numerous bands, arranged in increasing wavelength from top to bottom. Right: Parameters of the power spectra as a function of frequency.}
    \label{fig:atetarenko2021}
\end{figure}

More recently, radio time series analysis has been conducted for a few objects \citep{Tetarenko2019, 2021MNRAS.504.3862T, 2022ApJ...925..189Z}. The data for MAXI~J1820+070, presented in Fig.\,\ref{fig:atetarenko2021}, show characteristic break timescales consistent with being linear with wavelength over the bands from radio through optical, and with much stronger variability at the shortest wavelengths.  At least with this quality of data, the results are thus consistent with a scenario in which the size scale of the jet is linearly proportional to the wavelength (in agreement with the \citealt{BK79} model), along with a constant jet opening angle and speed. 

In Cygnus X-1, the situation may be more complex: a straightforward interpretation of the lag estimates suggests a height scaling with wavelength to the 0.4 power \citep[][see however \citealt{2022ApJ...935L...4Z}]{Tetarenko2019}. This may be indicative, instead of acceleration of the jet, or may just be due to the challenges in measuring lags between different radio bands \citep{Tetarenko2019}. It has also been suggested that free-free absorption that leads to modulation of the radio emission on the orbital period \citep{1999MNRAS.302L...1P, Szostek2007} may also affect the photospheric height versus wavelength in the radio band, leading to some distortions in the timing data that will not be seen in typical LMXBs \citep{MillerJones2021}.

Similar work has also been done for neutron stars.  Most notably, thermonuclear bursts have recently been shown to lead to enhanced radio emission in multiple sources \citep{RussellJetBurst}, interpreted as being driven by Poynting-Robertson drag driving up the accretion rate through the disk in response to the bursts.  The time lags of a few minutes, being about twice as long at 9 GHz as 5.5 GHz, suggest a jet speed of about $0.4\,c$ (Lorentz factor $1.1$), similar to the expected escape speeds of neutron stars.

\subsubsection{Non-steady states}

Strong jet ejection episodes are often seen from X-ray binaries around the time of the hard-to-soft state transition, when the steady compact jet gives way to the discrete, relativistic ejecta we termed `transient jets' in Sect.~\ref{sec:radio} \citep[][]{Vadawale2003, FBG2004}. High-cadence radio monitoring campaigns have considerably refined this picture in recent years, tracking the launching, proper motion, and long-term fading of individual ejecta in a growing number of sources, see for example MAXI~J1820+070 \citep{2020NatAs...4..697B, 2021MNRAS.505.3393W, Carotenuto2024}, MAXI~J1348--630 \citep{2021MNRAS.504..444C}, Swift~J1727.8--1613 \citep{2025MNRAS.542.1803H}, MAXI~J1535--571 and XTE~J1752--223 \citep{Carotenuto2024}.

For a few systems that sit near the Eddington luminosity, repeated jet ejection events take place. GRS~1915+105 is a well-studied case of this behavior, showing repeated, shorter-timescale jet ejections associated with particular X-ray states, as traced by simultaneous radio and X-ray light curves \citep[e.g.,][]{1997MNRAS.292..925P, 2002MNRAS.331..745K}. V404~Cyg during its 2015 outburst illustrates the same phenomenon in an even more extreme regime. Rather than the single, well-defined transient-jet episode typically seen at the hard-to-soft transition of a canonical outburst (e.g., MAXI~J1820+070; Sect.~\ref{sec:radio}), V404~Cyg produced numerous discrete, bi-polar, relativistic ejections within a single observing window \citep{TetarenkoV404}.

Even in the best studied of these episodes, V404~Cyg in 2015, there is no clear correlation between X-ray flaring and radio flaring \citep{TetarenkoV404}. It is unclear whether this is because of a real lack of correlation between episodes that lead to increased X-ray production and jet ejections, or because in these states, strong disk winds are generated that can lead to self-obscuration that make estimating the intrinsic X-ray production challenging \citep{TetarenkoV404}. Beyond their association with the broad hard-to-soft transition, transient jets have also been linked on a finer timescale to the appearance of Type~B QPOs (Sect.~\ref{sec:lfqpos}): the two phenomena are frequently seen in close temporal proximity \citep[e.g.,][]{2008MNRAS.383.1089S, IngramMotta}, although the order in which they occur is not always the same \citep{FHB2009, Carotenuto2024}, see Sect.~\ref{sec:lfqpos}.

\section{Emission lines} \label{sec:lines}

In this section, we focus particularly on low-mass X-ray binaries. The primary spectral signature of LMXBs is the variable zoo of emission lines observed across the electromagnetic spectrum, from IR through X-ray wavelengths.

\begin{figure}[h!]
\centering
\includegraphics[width=0.72\textwidth]{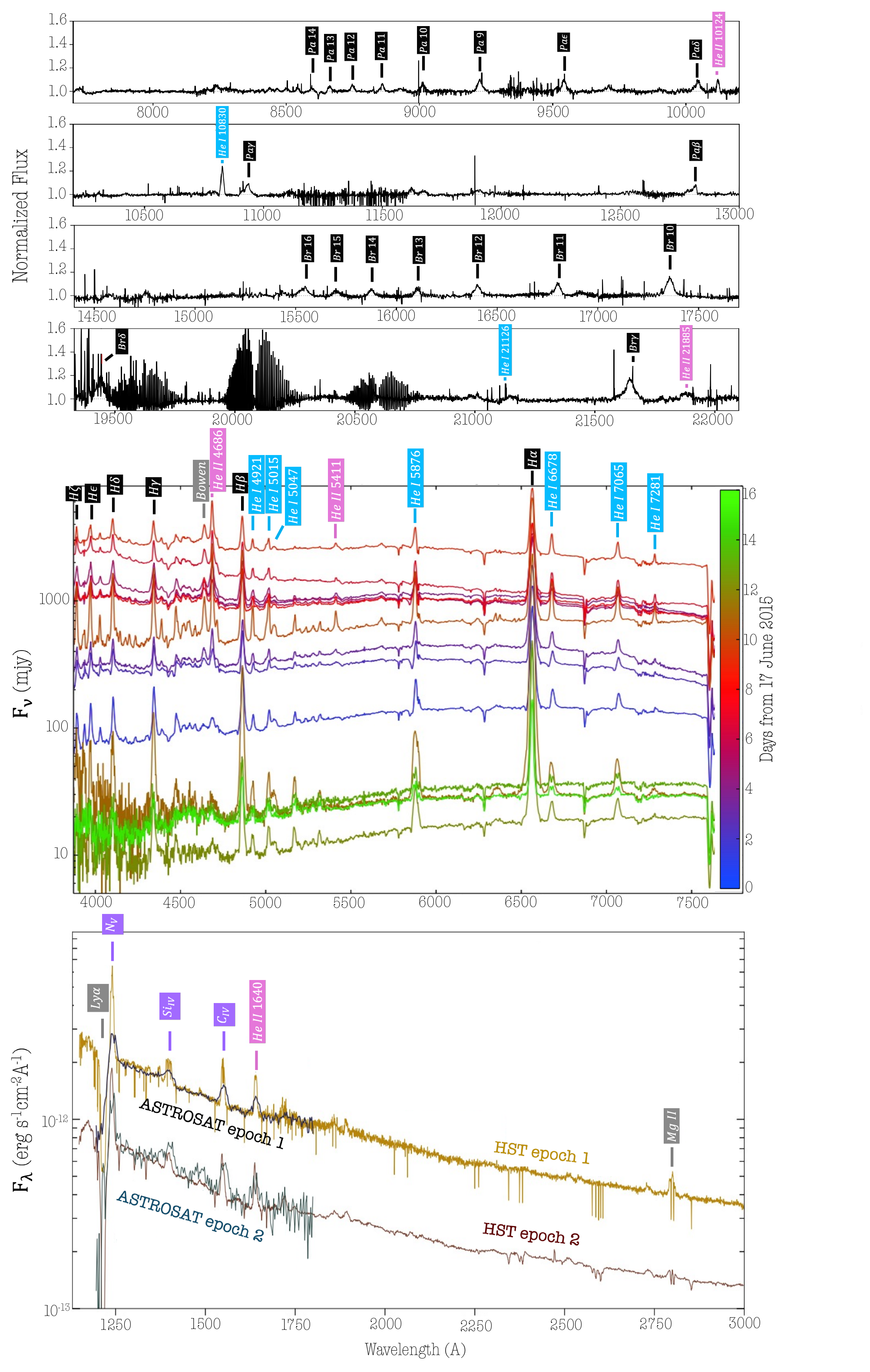}
\caption{From bottom to top: UV, optical, and near-IR spectra taken during outbursts of BH-LMXBs MAXI~J1820+070, V404~Cyg, and 4U~1543--47, respectively. Strong recombination lines in H (black), He {\sc I} (blue) and He {\sc II} (pink), as well as strong UV resonance lines (purple), are labeled. {\it UV}: Hubble Space Telescope (HST) and AstroSAT. {\it Optical}: Gran Telescopio Canarias (GTC), William Herschel Telescope (WHT), Nordic Optical Telescope (NOT), Isaac Newton Telescope (INT). {\it Near-IR}: Very Large Telescope (VLT). Figures adapted from: \citep{georganti2026, matasanchez2018, sanchezsierras2023b}.}
\label{fig:spectra}
\end{figure}

\subsection{Optical and Infrared} \label{sec:OIRlines}

\subsubsection{Disk-formed Recombination Lines}

Emission lines in Hydrogen and Helium, formed as a result of recombination (Balmer/Paschen/Brackett series H, He I, He II), are among the strongest O-IR spectral features of LMXBs in both quiescence and outburst \citep[see, e.g.,][for recent spectral studies]{sanchezsierrras2020, tetarenko2021, panizoespinar2022, matasanchez2022, tetarenko2023, sanchezsierras2023, sanchezsierras2023b, 2025MNRAS.537.1385G, ambrifi2025}. These lines, originating in the atmosphere of the accretion disk, typically display line widths on the order of hundreds to thousands of km\,s$^{-1}$ \citep{casares2015,cuneo2023}, and double-peaked profiles, due to Doppler shifts associated with the roughly Keplerian velocities in the accretion disk \citep{crawfordkraft1956,hornemarsh1986}. Single-peaked lines can also be observed, depending on the orientation at which the system is being observed and/or the distribution of emission over the disk itself. The strength, shape and appearance/disappearance of these recombination lines will vary throughout a binary orbit \citep{marsh2001,marsh2005}. See Fig.\,\ref{fig:spectra} for examples of O-IR spectra of LMXBs.

Most of the O-IR light emitted by the accretion disks in LMXBs comes from the outer disk regions ($\gtrsim$ hundreds of gravitational radii), that reprocess X-rays produced close to the compact object \citep{vanParadijsMcClintock, vanparadijs1996, Russell2006OIR}. For recombination lines to be produced in this environment, a temperature inversion must exist in the atmosphere of the disk \citep{shaviv1986,hubeny1990}. For this reason, the X-ray irradiation mechanism has long been thought to be the likely process behind the recombination line production in LMXBs.

\begin{figure}[h!]
\centering
\includegraphics[width=1.0\textwidth]{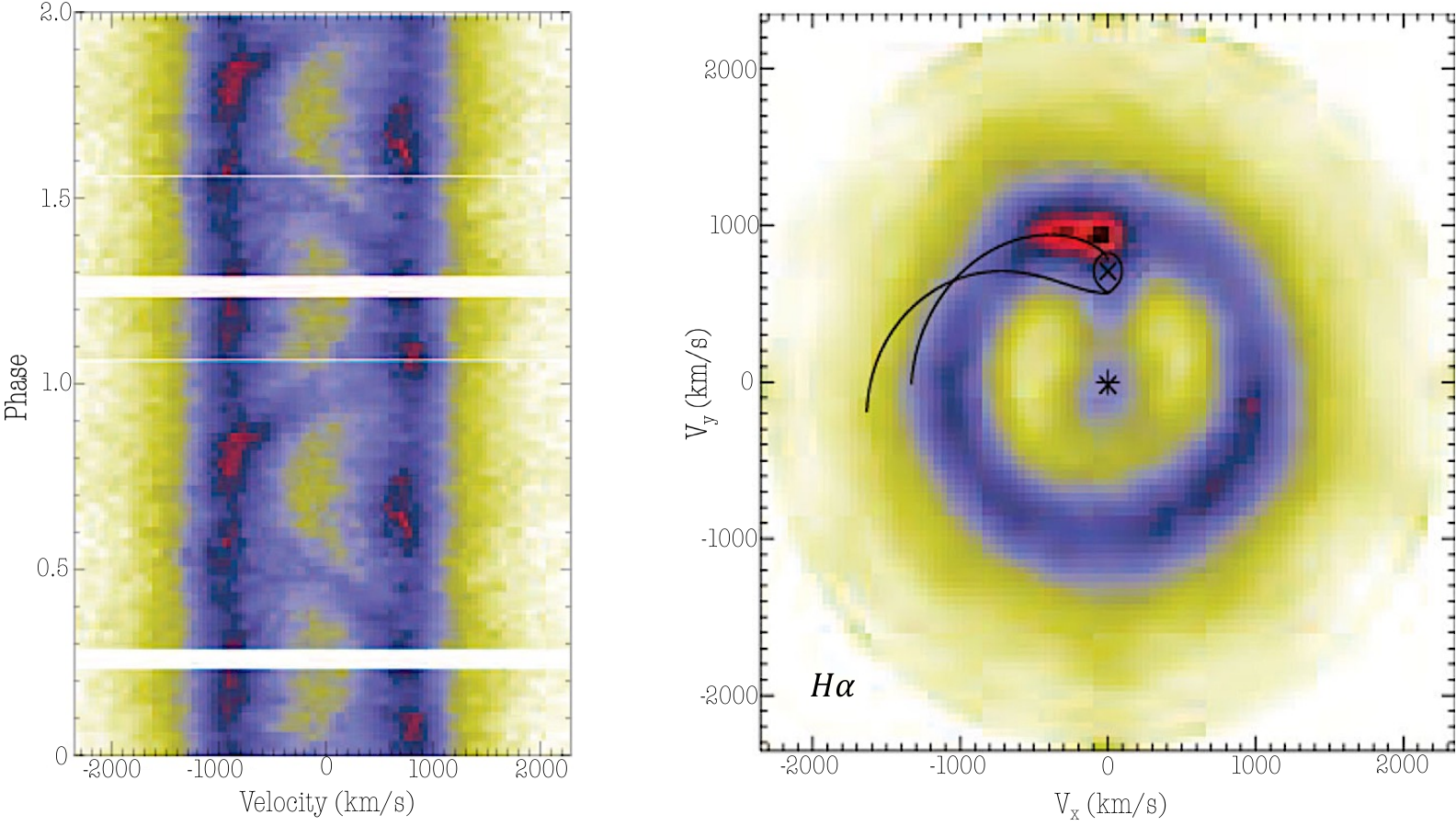}
\caption{Doppler tomography of H$\alpha$ emission, during quiescence, of BH-LMXB XTE~J1118+480. Displayed are the trailed spectrum (left) and corresponding tomogram (right), created from spectra taken with the Optical System for Imaging and low Resolution Integrated Spectroscopy (OSIRIS) on GTC. The companion star's Roche lobe, gas stream, and binary center of mass, are overplotted on the tomogram for known orbital parameters. Figure adapted from \citep{zurita2016}.}
\label{fig:dopp_quies}
\end{figure}

During outburst, X-ray irradiation is most significant in the cooler ($T_e\gtrsim10^{4}$K) outer accretion disk, making H/He recombination lines ideal tracers of its source and effect on thermal properties of the gas in the disk. This expected empirical connection existing between the line emitting regions, and physical properties of the X-ray source heating the accretion disk, has now been observed in multiple LMXBs. These studies show that the properties of the evolving high-energy X-rays heating the accretion disks in these systems are actually imprinted on the observed disk-formed H/He recombination line profiles themselves \citep[see, e.g.,][]{tetarenko2021,tetarenko2023}.

The global shape of a recombination line profile depends on the distribution of emission over the disk surface. The profile itself is built from the sum of independent contributions from each part of the disk surface, taking into account Doppler shifts due to the orbital motion of the disk material. 
Consequently, these emission lines encode within them a projection of the disk itself along the line of sight. As the spatial distribution of line emission is a tracer of the disk structure, these emission lines can, in principle, be used to effectively trace how the accretion disk gas behaves and evolves over time \citep{hornemarsh1986,marsh2001}.

A set of time and velocity dependent recombination line profiles observed through a binary orbit, can be inverted using a technique called Doppler tomography to construct images of the accretion disk on micro-arcsecond scales. Analogous to a CT-scan (sequential 2-D X-ray images taken at different angles, to construct a 3-D image of the human body), tomography uses a sequence of 1-D spectra to construct a 2-D velocity-resolved map of line emission across the disk \citep{marsh2001,marsh2005,steeghs2004}. As line emissivities depend sensitively on physical conditions in the disk \citep{marshhorne1988}, tomographic velocity-space maps (often referred to as tomograms) can effectively be used to track disk structures and physical gas properties over time.

\begin{figure}[h!]
\centering
\includegraphics[width=1.0\textwidth]{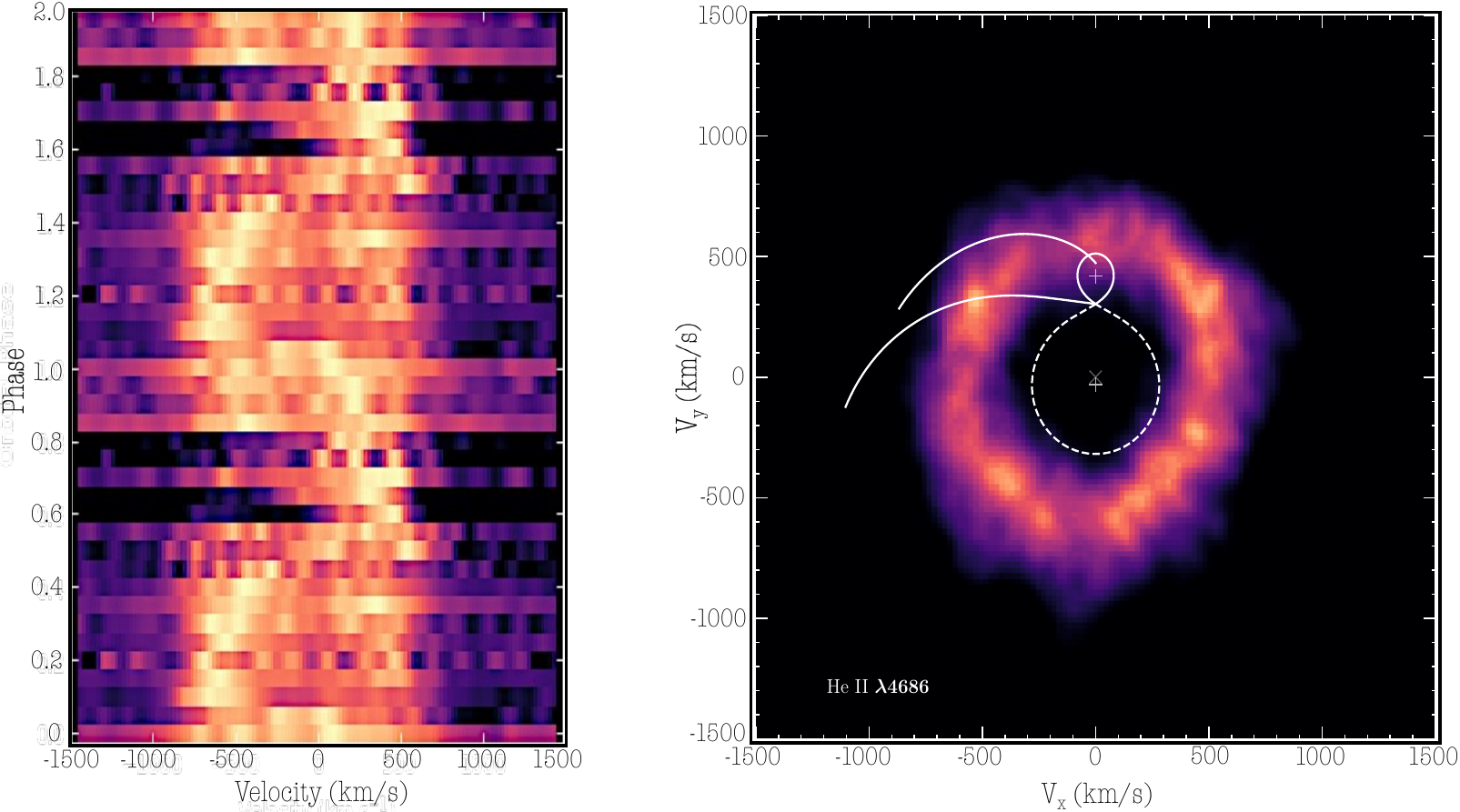}
\caption{Doppler tomography of He {\sc II} $\lambda = 4686 \,\si{\angstrom}$ emission, during the 2018 outburst of BH-LMXB MAXI~J1820+070. Displayed are the trailed spectrum (left) and corresponding tomogram (right), using spectral data from GTC/OSIRIS and the Fiber-fed RObotic Dual-beam Optical Spectrograph (FRODOspec) on the Liverpool Telescope. Overplotted on the tomogram are the Roche lobes of the compact object (dashed line) and companion star (solid line), gas stream, and center of mass of the binary, using known orbital parameters. Figure adapted from \citep{tetarenko2023}.}
\label{fig:dopp_out}
\end{figure}

To date, tomographic Doppler maps constructed from single emission lines have proven excellent at revealing global disk structure (i.e., gas streams, spiral shocks, truncation) and evolution \citep{marsh2001,marsh2005}, in both quiescent \citep[see Fig.\,\ref{fig:dopp_quies} and e.g.,][]{casares1995, DAvanzo2005, GH2010, zurita2016} and outbursting LMXBs \citep[see Fig.\,\ref{fig:dopp_out} and, e.g.,][]{hynes2001,shaw2016,tetarenko2021,tetarenko2023,killestein2023}.

In addition to tracing inflowing matter within the accretion disk, recombination lines can also be used as powerful diagnostics for detecting the presence of outflowing matter from the disk as well as deriving their physical properties.
During outburst, H/He recombination lines have been observed to display signatures of disk wind outflows, in the form of blue-shifted absorption features (P-Cygni profiles or absorption troughs), broad emission line wings, flat-top line profiles, and asymmetric line shapes \citep{bandyopadhyay1997, bandyopadhyay1999, sanchezsierrras2020, panizoespinar2022}. Although these features were originally detected using visual inspection or an excesses diagnostic diagram \citep{munozdarias2016,matasanchez2018,munozdarias2022}, more recently, machine learning algorithms have been applied to this complex task \citep{matasanchez2023}. To date, O-IR outflows have been detected in the recombination line profiles of nine LMXBs \citep{munozdarias2016,munozdarias2018,munozdarias2019,cuneo2020,munozdarias2020,panizoespinar2021,matasanchez2022,panizoespinar2022,sanchezsierras2023,matasanchez2024}. For a detailed discussion of accretion disk-wind outflows, see \citet{2026SSRv..222...39M}.

During quiescence, the O-IR spectrum of LMXBs is dominated by the companion star, with the addition of broad, disk-formed recombination lines superimposed on the continuum \citep{charlescoe2006}. The recombination lines are much stronger (i.e., larger equivalent width), compared to the absorption lines from the companion star, in quiescence \citep{fender2009lines}. As such, it is possible to extract fundamental binary orbital parameters (previously only accessible via classic dynamical studies) encoded within the shape of the line profile itself in quiescent spectra.

\begin{figure}[h!]
\centering
\includegraphics[width=0.85\textwidth]{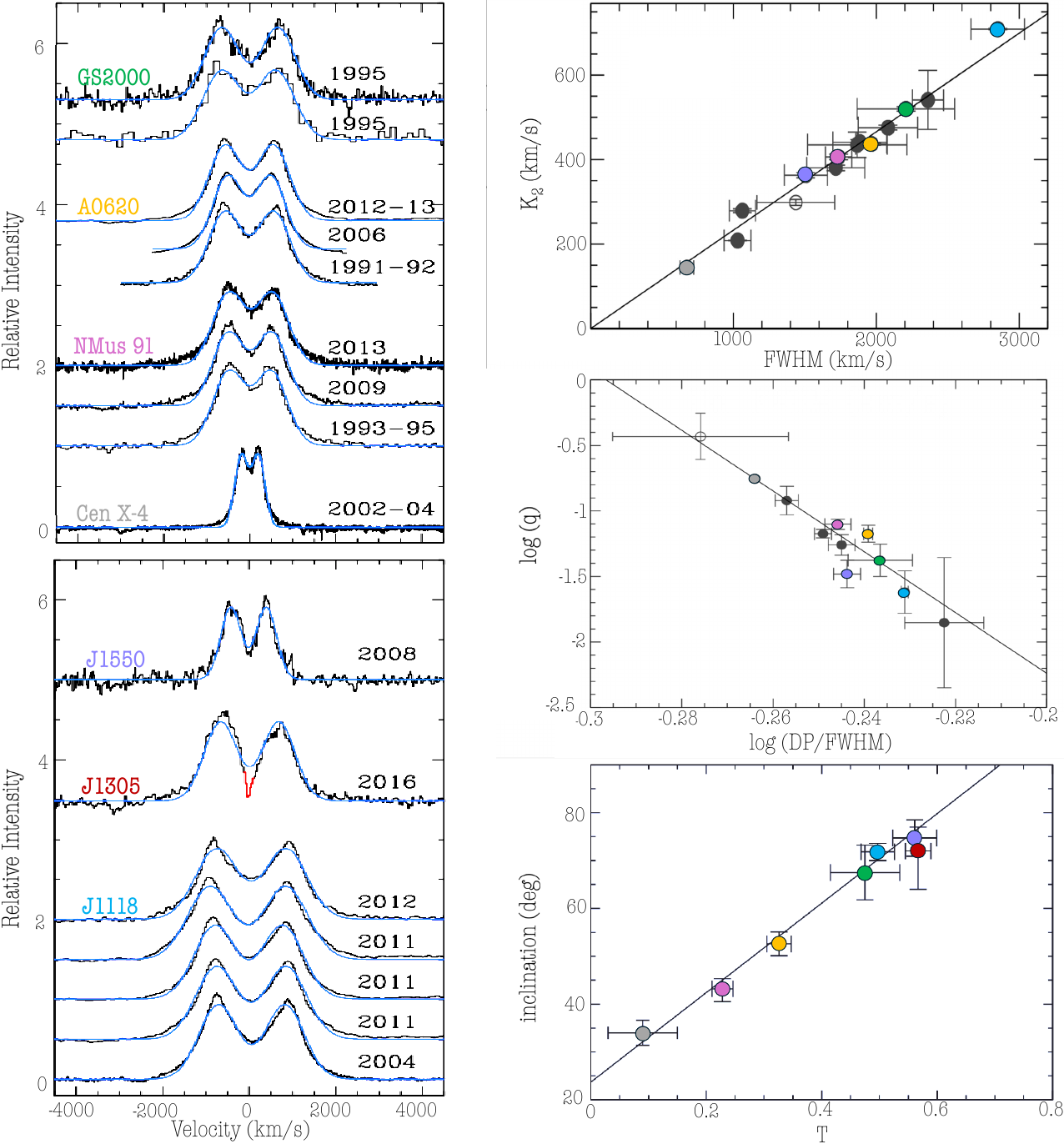}
\caption{Left: Example orbital averaged H$\alpha$ line profiles, taken during quiescence of seven LMXBs. The best-fit 2-Gaussian model is overplotted in blue. Right: Empirical correlations between the (top) FWHM of the H$\alpha$ line and companion star's velocity ($K_2$), (middle) ratio of the double-peak seperation and FWHM of the H$\alpha$ line profile (DP/FWHM) and the binary mass ratio ($q$), and (bottom) depth of the double peak trough in the H$\alpha$ line ($T$) and binary inclination ($i$). Colors correspond to the calibration sources in the left panel. The spectral library used here includes data from the VLT, WHT, GTC, Keck, and Magellan telescopes. Figures adapted from \citep{casares2015,casares2016,casares2022}.}
\label{fig:casares_corrs}
\end{figure}

As mentioned in the introduction, using multiple existing empirical correlations (see Fig.\,\ref{fig:casares_corrs}), it is possible to derive key binary orbital parameters by analyzing the shape of the line profile in quiescence. That is, full width half maximum (FWHM), double peak separation (DP), and trough depth of the double peak profile (T). With the first correlation alone, one can derive the semi-amplitude of the donor star's radial velocity curve. Combining knowledge of the orbital period, with the remaining two correlations, it is possible to attain compact object mass \citep{casares2015,cuneo2023}, the mass ratio \citep{casares2016} and the inclination \citep{casares2022}, of the binary.

Most LMXBs are located along the Galactic Plane, where interstellar extinction is high. Furthermore, the companion stars in these systems are typically low-mass and intrinsically faint enough that their apparent brightness often falls beyond the ability of even the largest available O-IR telescopes \citep{tetarenko2016,corralsantana2016}. As a result, there exists an extreme bias to not only detecting but also determining the compact object type in only the brightest and closest LMXBs. For this reason, of the estimated $\sim10^3-10^4$ LMXBs expected to exist in the Galaxy \citep{romani1998,yungelson2006,kiel2006}, only a few hundred have been discovered so far, with $\lesssim50\%$ of this population having known compact objects: $\sim 20$ confirmed black holes (see Table~\ref{tab:BHXBs}) and around $150$ confirmed neutron stars (see Baglio et al., submitted to SSRv).
Thus, not only does the novel technique (using the FWHM) have the ability to access binary systems up to 2.5 magnitudes fainter than classic dynamical studies \citep[see, e.g.][]{torres2021,casares2023,yanesrizo2024}, but it also provides an alternative method to search for currently undiscovered LMXBs in the Galaxy, outside of their bright outburst intervals \citep{casares2018,casares2018b}.

\subsubsection{Bowen Fluorescence from Irradiated Companion Stars}

\begin{figure}[h!]
\centering
\includegraphics[width=.8\textwidth]{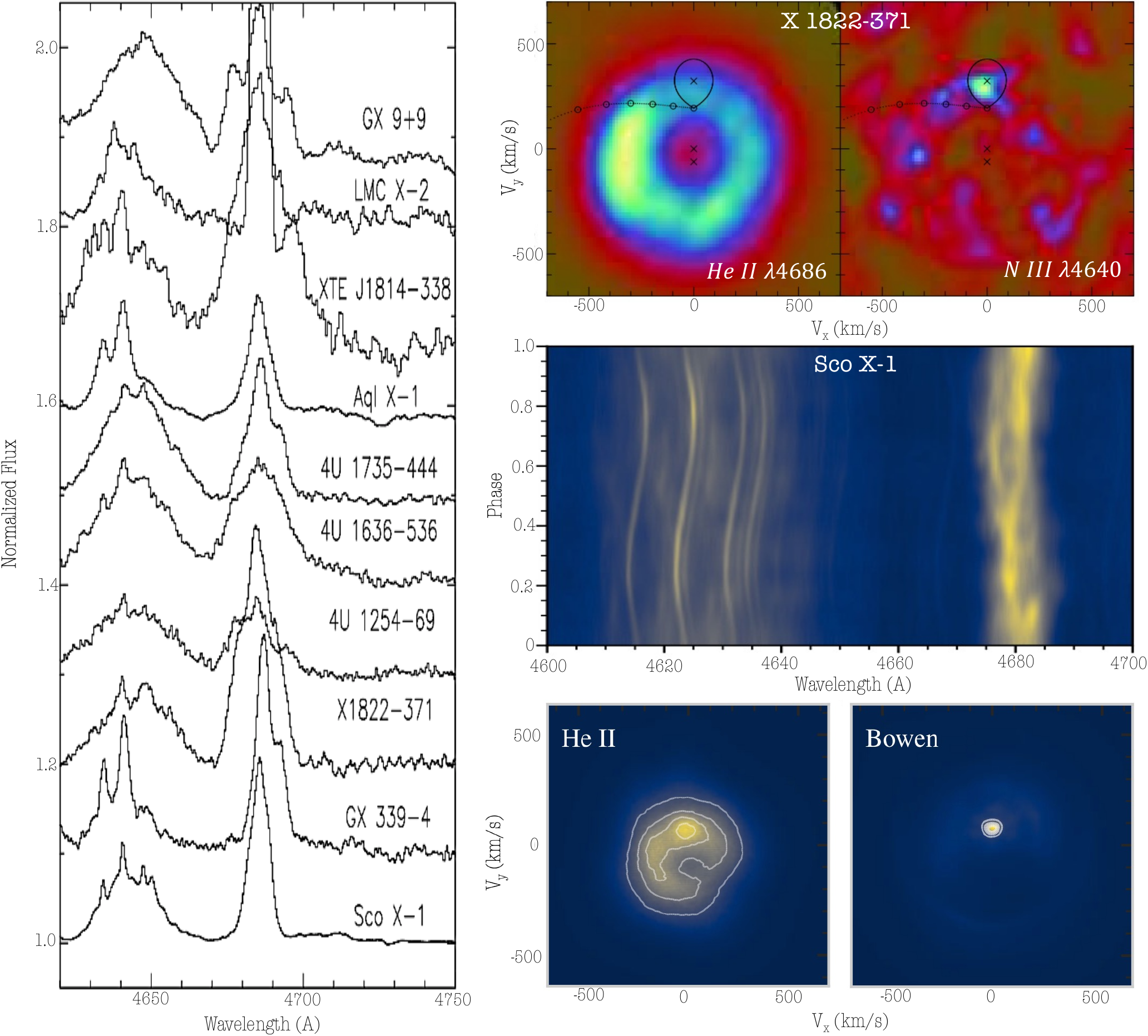}
\caption{Left: Averaged spectra of the Bowen region, in the companion star rest frame, for a sample of LMXBs (including some harboring Neutron Stars). Top-right: Doppler tomograms of He{\sc II} $(\lambda = 4686\,\si{\angstrom})$ and N{\sc III} $(\lambda = 4640\,\si{\angstrom})$ for LMXB X1822-371, created from data taken with the RGO spectrograph on the AAT, and the New Technology Telescope (NTT). Roche lobe of the companion, gas stream, and compact object position, are overplotted. Bottom-right: Trailed spectrum, showing the presence and movement over phase, as well as the corresponding Doppler tomograms, of He{\sc II} $(\lambda = 4686\,\si{\angstrom})$ and Bowen blend emission in LMXB Sco~X-1. Includes data taken with the Ultraviolet and Visual Echelle Spectrograph (UVES) on VLT and WHT/ISIS.
Figures adapted from \citep{cornelisse2008, casares2003, killestein2023}. }
\label{fig:bowen}
\end{figure}

In the optical regime, in addition to recombination lines, high excitation emission components in the wavelength range $\lambda = 4630 - 4660\,\si{\angstrom}$ are also observed \citep{Schachter1989Bowen, steeghs2002,hynes2004} in LMXBs. This collection of lines, often referred to as the Bowen blend \citep[][Fig.\,\ref{fig:bowen}]{hynes2003b}, include: N{\sc iii} ($\lambda = 4634$, $4641$, and $4642\,\si{\angstrom}$), and C{\sc iii} ($4647$, $4650$, and $4651$). The N{\sc iii} lines are produced as a result of UV fluorescence, through cascade recombination (via He{\sc ii} Ly$\alpha$ seed photons), and the C{\sc iii} lines come from photo-ionization and subsequent recombination \citep{cornelisse2008}.

The Bowen blend is known to come from the irradiated surface of the companion star. This is evidenced by (i) a sinusoidal radial velocity curve (indicative of a fixed structure in the frame of the binary), (ii) a very narrow (FWHM of hundreds of km\,s$^{-1}$ or less) profile, and (iii) the fact that they move in anti-phase with the compact object, as traced by the wings of the He{\sc ii} $\lambda = 4686\,\si{\angstrom}$ line profile \citep{hynes2003b,casares2004,cornelisse2008}. Furthermore, with the use of the Doppler tomography technique, it is possible to reconstruct the distribution of the Bowen blend in velocity space. In all cases, this results in a compact spot, consistent with the phasing/velocity of the companion star \citep{steeghs2002, casares2003, barnes2007, cornelisee2007, cornelisse2008, elebert2009, cornelisse2009, wang2017, wang2018, jimenezibarra2018, brauer2018}. See Fig.\,\ref{fig:bowen} for examples of this behavior, though with some harboring Neutron Stars.

Companion stars in LMXBs are typically $\sim10^6$ times fainter than the optically-emitting accretion disk in outburst, which is dominated by X-ray reprocessing.
Bowen fluorescence lines therefore offer an alternative avenue for binary parameter estimation in LMXBs, during outburst intervals when the system is bright \citep{steeghs2002, hynes2003b, casares2003, casares2006, barnes2007, cornelisee2007, elebert2009, cornelisse2009, cornelisee2012, galloway2013, matasanchez2015, wang2017, wang2018, jimenezibarra2018, brauer2018}. Though it is important to note that this method is subject to a number of systematic uncertainties as a result of the Bowen lines not being symmetric about the center of the star \citep[see for example the case of GX~339--4,][]{hynes2003b, heida2017}.

Additionally, using a unique technique called echo-tomography, it is also possible to use Bowen blend emission to map the distribution of X-ray reprocessing sites in LMXBs \citep{obrien2002,casares2005,munozdarias2006}. In LMXBs, optical emission is observed to lag (in time) behind X-ray emission, where the degree of delay depends on both the binary location of the X-ray reprocessed region, as well as the geometry of the binary itself. Echo-tomography (sometimes called echo mapping) is is a type of indirect imaging, that exploits this correlation between X-ray and optical variability in LMXBs.

By using the observed time delays between the X-ray (i.e., the driving mechanism behind X-ray irradiation) and the optical (i.e., resulting X-ray reprocessed ``echoes'') light-curves, Doppler shifts (time-delays as a function of phase), and photoionization physics, it is possible to map the geometry, kinematics, and physical conditions in the reprocessed regions within the binary \citep{obrien2002,horne2003}.
Echo-tomography studies, using X-ray and broad-band optical light-curves have since been performed on many LMXBs \citep[e.g.,][]{hynes1998,obrien2002,obrien2004,2005ASPC..330..237H,hynes2009}. It is clear from these studies that the reprocessed flux is dominated by the accretion disk.

Bowen echo-tomography exploits emission-line reprocessing rather than broad-band photometry, by using the orbital phase variable time delays between an X-ray light-curve and an optical light-curve, obtained using narrow-band filters centered around the Bowen/He {\sc ii} region \citep{casares2005,munozdarias2006}. This technique, which has since been used on multiple LMXBs \citep{casares2005, munozdarias2006, munozdarias2007, munozdarias2008}, shows clear evidence for the Bowen/HeII light-curves lagging behind the X-rays, with time delays on the order of tens of seconds. Given that binary separations in LMXBs are $\sim$ light-seconds to light-minutes, typical light travel times in these binaries are expected to be in this range. Thus, these observed time delays are consistent with X-ray reprocessing on the face of the companion star. Further, when compared to the continuum, Bowen fluorescence happens with a longer lag. A clear indication that the Bowen blend comes more from the star and the continuum originates more from the disk \citep{horne2003, 2005ASPC..330..237H, casares2005, munozdarias2006}.

\subsection{Ultraviolet: Resonance Lines}

The UV regime contains a number of prominent lines that can provide important diagnostic information, including for example recombination lines like He{\sc ii} $\lambda = 1640\,\si{\angstrom}$, that allows one to probe the extreme UV (EUV) band, which is usually unobservable. However, the most dominant features in the UV regime are the strong resonance lines observed \citep{haswell2002, froning2011, froning2014, georganti2026}. See Fig.\,\ref{fig:spectra} for an example of UV spectra of a BH-LMXB.

The UV regime contains strong, resonance line transitions in many key elements (e.g., C{\sc iv} $\lambda = 1550\,\si{\angstrom}$, Si{\sc iv} $1400\,\si{\angstrom}$, N{\sc v} $1240\,\si{\angstrom}$). These lines, which often display the characteristic double-peaked profile, act as strong tracers for outflows from the accretion disk \citep{georganti2026}. However, for LMXBs, this regime is under-studied. UV studies of LMXBs (most of which are located in the Galactic Plane) are greatly complicated by high interstellar extinction \citep{bahramian2023}, and as a result, only a few systems have spectral studies in this regime \citep{haswell2002, hynes2005b, froning2011, froning2014, georganti2026}. 

Interestingly, while orbital phase variable, double-peaked UV resonance lines (similar behavior to O-IR recombination lines) have been observed in CVs \citep[see, e.g. ][]{knigge1994}, no clear evidence for such behavior has yet been observed in LMXBs \citep{fijma2023}. However, some variation in resonance line strength and shape has been observed between hard and soft accretion states \citep[see, e.g. ][]{georganti2026}.
Furthermore, echo-tomography studies, using X-ray and broad-band UV light-curves have been performed on a few LMXBs \citep{hynes1998, obrien2002, 2005ASPC..330..237H}.
Additionally, to date, disk wind outflow signatures in UV resonance lines (P-Cygni profiles or absorption troughs) have been observed in a handful of LMXBs \citep{boroson2001, ioannou2003, bayless2010, castrosegura2022, fijma2023}.

In addition to probing the physical conditions in photo-ionized gas, UV resonance lines can also be used as an effective tool to study the evolutionary history of the binary system. The abundance/depletion of these elements found in the accreted material has actually been linked to the evolutionary stage of the companion star feeding the disk in the system \citep{haswell2002, froning2011, froning2014, castrosegura2024}.

\subsection{X-ray Emission Lines}
\label{sec:reflection}

\begin{figure}[h!]
\centering
\includegraphics[width=0.7\textwidth]{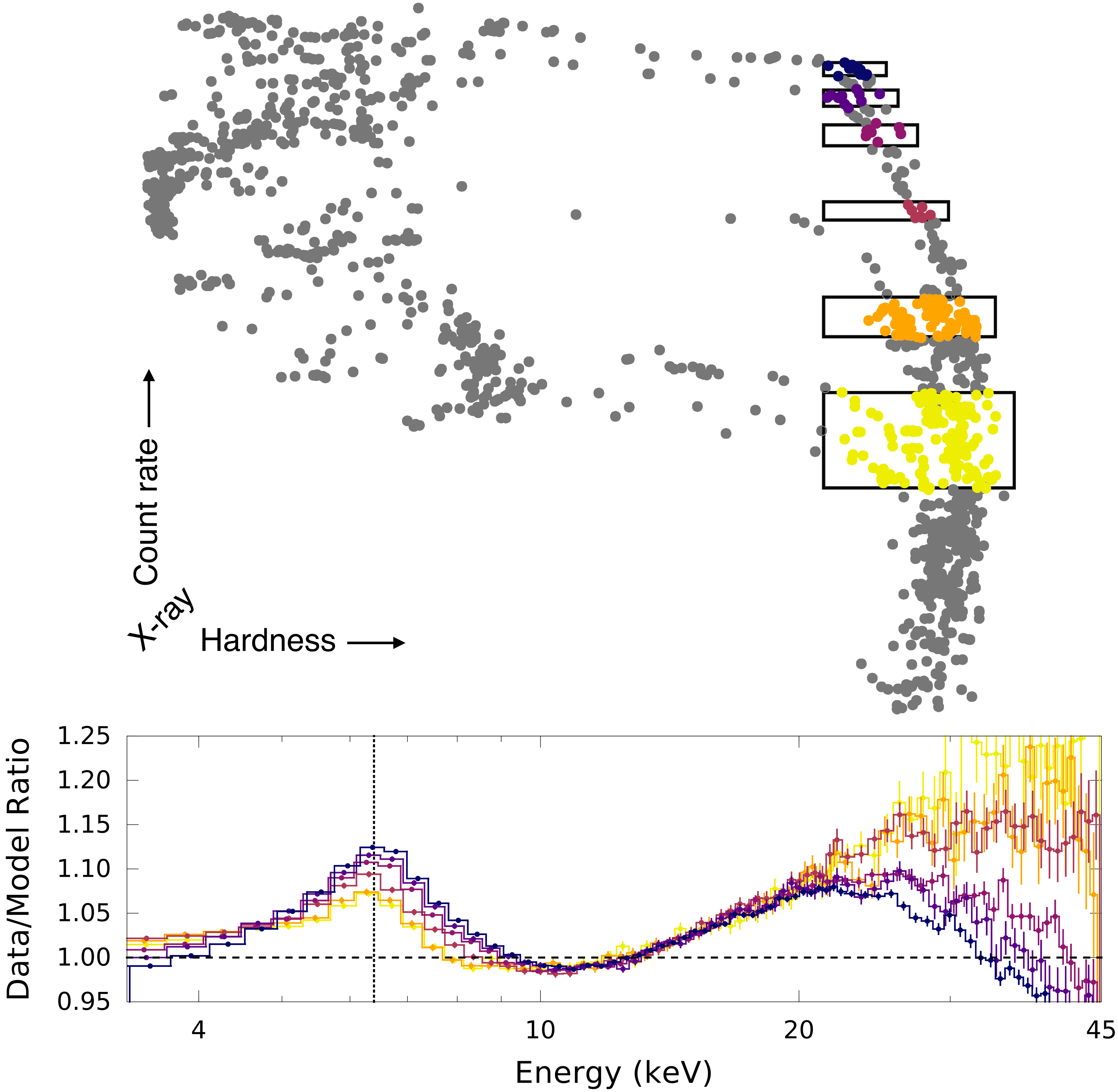}
\caption{Example of average spectra during the rise and decay phases of GX~339--4 using the whole \textit{RXTE}/\textit{PCA} archive. The top panel shows the hardness intensity diagram, with the selected spectra in each box. The bottom panel shows the resulting reflection spectra, assuming the continuum is a simple \textsc{Tbabs*powerlaw}, using a fixed hydrogen column density and a powerlaw that varies slope between 1.52 and 1.75 \citep[][Table~1]{Garcia15}. These figures are adapted from \citet{Garcia15}.}
\label{fig:Garcia15}
\end{figure}

Black hole X-ray binaries also show prominent X-ray emission lines, in particular the Fe~K$\alpha$ fluorescence complex near $6.4$\,keV \citep{1989MNRAS.238..729F, 1991ApJ...376...90L, 2014SSRv..183..277R}. These arise when hard X-rays from the hot-flow/corona (Sect.~\ref{sec:hardXrays}) irradiate the optically thick disk, producing fluorescent lines, absorption edges, a Compton hump peaking around $20$--$40$\,keV, and a partially-thermalized bump peaking in the soft X-rays \citep{Garcia13, Garcia15}. Iron dominates because of its high abundance and large fluorescence yield. Moreover, line energies shift from $6.4$\,keV (Fe~\textsc{xxv}) up to $6.97$\,keV (Fe~\textsc{xxvi}) as the disk ionization increases, defined as $\xi = 4 \pi F_X /n $ where $F_X$ is the irradiation field and $n$ is the disk density \citep{Garcia13}. We show examples of X-ray reflection spectra on Fig.\,\ref{fig:Garcia15}, where the iron line ($\sim$ 6.5\,keV) and Compton hump ($\gtrsim 10$\,keV) are clearly visible.

Among the whole reflection spectrum, the Fe~K profile is of particular interest as it encodes strong‑gravity disk geometry. Each annulus contributes at energies set by orbital motion, gravitational redshift, and light bending, yielding a broadened, skewed line with a blue horn from the approaching side and an extended red wing from the receding, strongly red-shifted inner disk. The amount of broadening constrains the inner radius and, under the usual assumption that the disk reaches the ISCO, the black hole spin \citep{Bambi2021, Reynolds21}. In some black-hole X-ray binaries, however, the Fe~K complex is better described by more than one reflection component: a broad, relativistically blurred component associated with the inner disk, together with a narrower component produced by less strongly relativistic reprocessing at larger disk radii. MAXI~J1820+070 provides a particularly clear example, with broadband X-ray spectra requiring two reflection components that can be associated with distinct regions of the accretion flow \citep[e.g.,][]{2021ApJ...909L...9Z, 2021A&A...654A..14D, 2021A&A...656A..63M}. In this picture, the narrow component is most naturally associated with the outer disk, potentially including a warped or flared region, although contributions from other distant material cannot in general be excluded. The importance of such distant reprocessing has recently become clearer with the high spectral resolution of \textit{XRISM}/Resolve: a narrow Fe~K$\alpha$ component consistent with reflection at $\gtrsim 1000\,R_g$ has been reported alongside the relativistic line in Cyg~X-1 \citep{2025ApJ...995L..12D}, although this interpretation has been questioned by \citet{2026ApJ..1005L..78Z}. Multiple narrow, stratified components spanning the disk and possible disk winds have since been identified in MAXI~J1744--294 \citep{Parra2026a, Parra2026b}. Conversely, \textit{XRISM} observations of GS~1354--64 show a clean, predominantly broad Fe~K profile with no significant narrow component \citep{2026arXiv260306883L}, demonstrating that distant reprocessing need not produce a detectable Fe~K feature in every black-hole X-ray binary.

State‑of‑the‑art models such as \textsc{relxill} \citep[see also \textsc{reflkerr},][]{2019MNRAS.485.2942N} combine relativistic transfer with angle‑ and ionization‑dependent rest‑frame reflection \citep[\textsc{xillver};][]{Garcia14, Dauser14}. They solve for ionization balance and radiative transfer in an illuminated slab, and allow different disk densities and coronal geometries (e.g., lamppost, off‑axis, extended, ring-like). Their application has allowed to estimate spin, inclination, ionization, and abundance measurements for many BHXBs and AGN (e.g., \citealt{2008MNRAS.387.1489R}, \citealt{Garcia15}, \citealt{2016ApJ...821L...6P}, \citealt{2021ApJ...908..117Z}; see \citealt{Reynolds21} for a review and \citealt{2026ApJ..1005L..78Z} for a comparison of different models).

\begin{figure}[h!]
\centering
\includegraphics[width=0.55\linewidth]{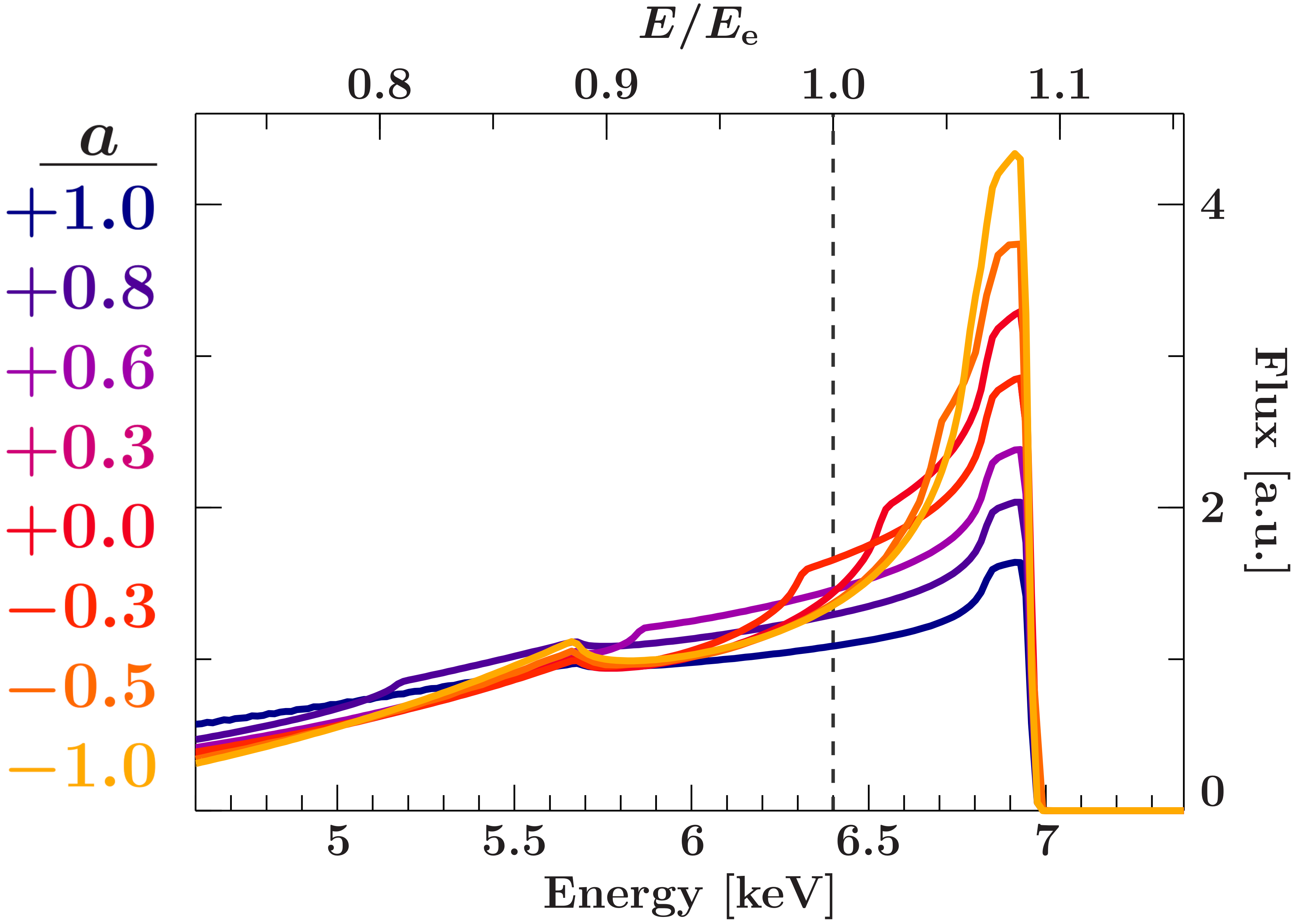}
\includegraphics[width=1.0\textwidth]{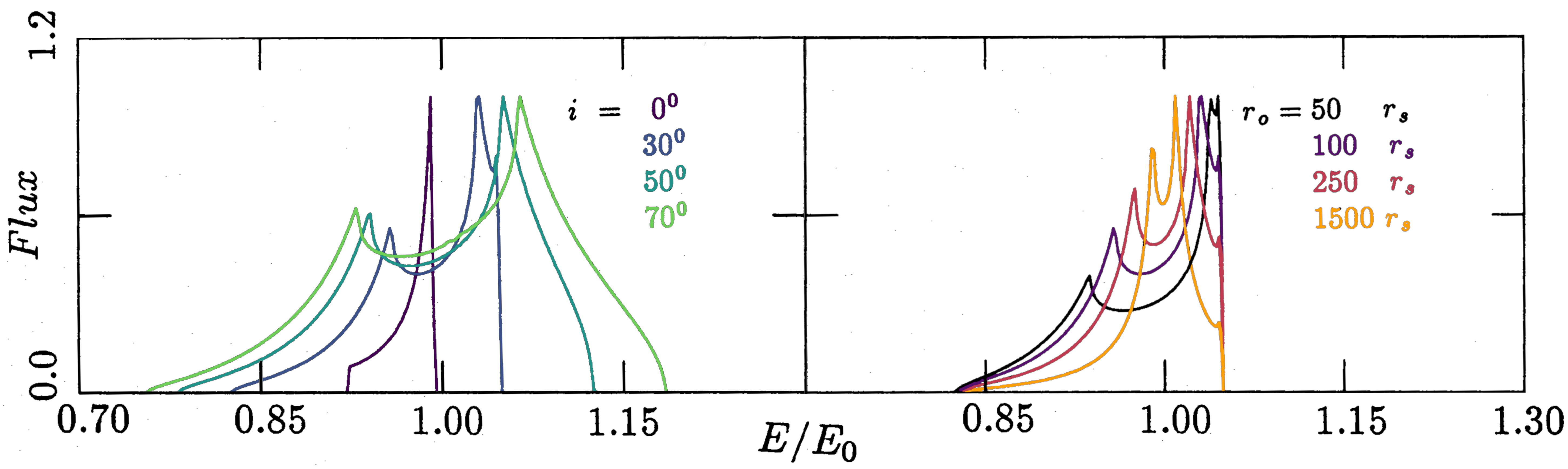}
\caption{Top: Impact on the black hole spin on the shape of the Fe~K profile, adapted from \citet{Dauser16}. Bottom: Impact of the disk inclination (left) and outer radius (right), as calculated in the seminal paper of \citet{1989MNRAS.238..729F}. The emissivity profiles and source geometries assumed are detailed in the original works.}
\label{fig:Dauser16}
\end{figure}

However, several systematic issues complicate inference \citep[e.g.][]{2017ApJ...851...57C, 2026NewAR.10201746Z}. Angle-averaged reflection can bias Fe abundance and other parameters \citep{Garcia13,Garcia14}. We show in Fig.\,\ref{fig:Dauser16} the impact of spin, outer radius, and inclination on the line profile, illustrating the complex and potentially degenerate ways in which individual parameters shape the observed signal. Emission from the plunging region, coronal Comptonization of reflection, high disk density, and returning radiation can also all modify the Fe~K profile and Compton hump, mimicking higher spin or super‑solar iron abundance if neglected \citep[e.g.][]{Steiner2017, Bambi2021, 2025PhRvD.112l3030S}. Simplified emissivity prescriptions and assumed corona geometries (especially strict lamppost models) can misestimate spin and other model parameters, particularly when the true illumination is complex or off‑axis \citep{2024ApJ...965...66M, 2025PhRvD.112l3030S, 2025ApJ...984..173F}, see also, e.g., \citet{2026ApJ..1004...68S} for the impact of the disk geometry itself. Reflection modeling also depends on the assumed shape of the primary continuum that is often taken to be a single power law or Comptonization component, despite evidence for at least two Comptonization components in some hard-state spectra \citep[e.g.,][]{2017MNRAS.472.4220B, 2021ApJ...914L...5Z, 2022MNRAS.511..536K, 2026ApJ...998...36S}. Moreover, similar degeneracies are expected to arise from disk winds, that can produce narrow emission lines through scattering in the wind material \citep[e.g.,][]{2023MNRAS.523.3441T, 2025ApJ...994..258T} that are difficult to distinguish from narrow reflection features. In practice, should both contributions (winds and reflection) exist, the current data do not allow us to robustly disentangle them \citep[][although this study was performed for AGN]{Parker2022}. Finally, post-processing GRMHD simulations \citep[see, e.g.,][]{2013ApJ...769..156S, 2016ApJ...826...52K} directly challenge several assumptions underlying these toy models. Recently, \citet{Nagele26a} find that the disk temperature is highly inhomogeneous even in its `cold' regions, while \citet{Nagele26b} show that Fe~K$\alpha$ photons originate mainly from $\sim10\,R_g$ rather than near the ISCO, yet still produce broad profiles via Compton broadening and multiple ionization states. This demonstrates that an observed broad line does not necessarily imply high spin.

Reflection fitting has also been combined with X-ray reverberation mapping, where soft lags are interpreted as light-travel delays between the corona and the reflecting disk \citep[e.g.,][]{Fabian2009, DeMarco2017, Kara2019, Wang2022}. However, the spectral and timing results can be mutually inconsistent: in MAXI~J1820+070, the reflection fits require the disk to remain at the ISCO throughout the hard state, while the reverberation lag increases dramatically, demanding large changes in the coronal geometry \citep{Kara2019}. Indeed, it is not possible to jointly fit the X-ray spectrum and the time lags of MAXI~J1820+070 with a simple lamppost corona model \citep{Wang2021}. A joint fit is possible assuming a large, vertically extended corona \citep{Lucchini23}, but we will see in Sect.~\ref{sec:polar} that this contradicts polarization measurements, strongly supporting alternative explanations \citep{2021A&A...654A..14D, 2021MNRAS.507.2744A, 2022ApJ...928...11Z}.

It is likely that the inconsistency is worsened by the soft X-ray signal, since this is where the data quality is highest and the theory is most uncertain. Whereas iron line lag features are very likely driven by reverberation, the soft lags, at least in some states, may be dominated by other processes. For example, the response of the reflection spectrum to spectral pivoting of the Comptonization continuum \citep{2021MNRAS.507.2744A, UttleyMalzac} may be more complex than the prescriptions used in current reverberation models can capture, or the process of thermalisation itself may take a comparable time to the light crossing delays \citep{Salvesen2022}. Joint spectral–timing modeling with reverberation codes such as \textsc{reltrans} add geometric constraints by fitting both time‑averaged spectra and Fourier‑dependent lags, including ionization‑dependent lag contributions \citep{Ingram19, Gullo21}. Such an analysis can also be used to estimate black hole mass, and reasonable values have been returned for Cygnus X-1 \citep{Mastroserio2019, ONeill2026} and the AGN Ark 564 \citep{Lewin2022}. However, timing‑based and spectral inferences can still be in tension in other sources, indicating that simple geometric assumptions such as a single compact lamppost may be inadequate (see, e.g., \citealt{Gullo21, Lucchini23}; see also \citealt{2018MNRAS.480.2650C} for AGN). We stress that this tension does not necessarily call for more elaborate lamppost geometries (e.g., a double lamppost); it may instead point to lag mechanisms beyond pure light-travel time and/or non-standard radiative processes such as synchrotron emission \citep[as discussed in][]{Lucchini23}, in addition to geometries beyond the lamppost framework and non-vertically extended, which X-ray polarization measurements have independently disfavored (see Sect.~\ref{sec:polar}).

The underlying physics of reflection is well understood and firmly grounded in atomic and radiative transfer theory. However, as discussed, neither the current spectral models nor the numerical methods commonly employed to fit them are without significant caveats \citep{2026arXiv261035174L}. The inferred parameters are highly sensitive to the assumed disk inclination, which is often treated as a free parameter in the fit and need not coincide with the binary orbital inclination or the jet axis---since warps, precession, or misalignment between the inner disk and the outer binary orbit can produce different inclinations at different scales. Degeneracies between reflection and other spectral components (i.e., disk winds, warm absorbers), can further bias the results, and systematic uncertainties in the modeling of the illuminating continuum, disk density, vertical structure, and coronal geometry are still under-explored. Reflection spectroscopy remains valuable to probe the innermost regions, and, together with the continuum fitting method constitutes one of the principal approaches to measuring black hole spin \citep{Reynolds21}. Nonetheless, more work is needed to critically assess the systematics discussed above before reflection-based measurements can be considered fully robust \citep{2026NewAR.10201746Z}.

\section{Polarization} \label{sec:polar}

Linear polarization of radiation reflects a preferred orientation of the electric field oscillations in the observed electromagnetic waves. In the systems considered in this review, such a preferred direction can arise from ordered particle motion, for example, synchrotron radiation from an ordered magnetic field, or from a preferred orientation of scattering planes in Thomson or Compton scattering. The level of ordering, or axial symmetry of the system, determines the polarization degree (PD), while the orientation of the symmetry axis sets the polarization angle (PA). Consequently, already in the early era of X-ray astrophysics, polarization was recognized as an independent and sensitive diagnostic of accretion geometry.

To the best of our knowledge, the earliest polarimetric study of an accreting BH binary system was performed for Cyg X-1 by \citet{Kemp1976}, who reported an optical PD of 4--5\%. Subsequent studies revealed that most of this polarization is of interstellar origin, with the intrinsic polarization of the source at a sub-percent level \citep[e.g.,][]{Kemp1979,Nagae2009}. This intrinsic component is attributed primarily to scattering of light from the supergiant companion star, rather than to accretion processes \citep{Kravtsov2023}. The detected optical and infrared polarization associated directly with accretion processes is low, typically remaining at sub-percent level throughout the outburst and aligning with the axis of extended jet and/or jet ejections in many sources (V404~Cyg, \citealt{Kosenkov2017}; MAXI~J1820+070, \citealt{Veledina2019}; Swift~J1727.8--1613, \citealt{Nitindala2026}). However, the PD can greatly increase at the transition to quiescence \citep{DolanTapia1989, Dubus2008, Kosenkov2020, 2022Sci...375..874P, Kravtsov2023}. The low polarization levels during outburst complicate its association with a particular spectral component (inner accretion flow, jet, scattering at irradiated disk or in the wind), hence making any interpretation model-dependent.

A direct probe of the accretion geometry in the vicinity of the BH is provided by the X-ray polarization.
Its sensitivity to geometry enables constraints on the compactness of the X-ray source, thereby informing models of particle energization in the hard spectral state. Furthermore, polarization signatures shaped by spacetime curvature offer a potential diagnostic of BH spin \citep[e.g.,][]{Stark1977,PineaultRoeder1977b}. The first X-ray polarimetric observations of a BH X-ray binary were carried out by the OSO-8 satellite for Cyg X-1. A marginal detection of linear polarization at $2.6$\,keV, at about 2$\sigma$ significance level, indicated the ${\rm PD} \sim 3\%$ \citep{Weisskopf1977}. Subsequent balloon-borne experiments provided upper limits on hard X-ray polarization of $8.6\%$ \citep[PoGo+, $19-181$\,keV,][]{Chauvin2018} and $11.1\%$ \citep[XL-Calibur, $19-64$\,keV,][]{Awaki2025}. Additional efforts were directed into extracting the polarization information from the long-term observations of this source with INTEGRAL \citep{Laurent2011,Jourdain2012} and with AstroSAT \citep{Chattopadhyay2024}, both of which reported high polarization levels ($\gtrsim20\%$) at energies above $\sim100$\,keV in a direction different from the projected jet axis by $\sim 70^{\circ}$. The results from both INTEGRAL and AstroSAT should be considered with caution, as the instrument was not calibrated for polarimetric measurements.

The launch of the \textit{IXPE} \citep{Weisskopf2022} in December 2021 opened a new era of sensitive X-ray polarimetry. Operating in the $2$--$8$\,keV range, \textit{IXPE} represents a factor $\sim 100$ improvement in sensitivity to polarization over its predecessors, opening up the possibility of systematic measurements across multiple sources and accretion states \citep[a review of the results from the first two years of operations can be found in][]{Dovciak2024}. \textit{IXPE} has provided in-depth insights into the processes shaping the spectral energy distribution from accretion disks/flows in X-ray binaries, yet poses new challenges for our understanding of their internal structure and relevant processes. Below we review the results accumulated to date for different spectral states.

\subsection{Hard state and state transitions} \label{sec:polarhard}

In the hard state, the $2$--$8$\,keV flux is typically dominated by the Comptonized emission observed directly from the hot medium, whose geometry and compactness have been the subject of extensive debate (see Sect.~\ref{sec:hardXrays}). Considering photons in terms of electromagnetic waves, net polarization arises from individual scattering events because scattering suppresses the electric field component aligned with the direction of the outgoing photon (relative to that of the incoming radiation). This leads to the the observed partial polarization of scattered light, with an excess of electromagnetic waves oscillating perpendicular to the projected direction of the (unpolarized) incoming radiation. PD depends on the scattering angle (the angle between the direction of the incoming and outgoing photons) and approaches 100\% for the ideal case of Thomson scattering at $90^{\circ}$. 

Considerable polarization can be expected if the incoming photons stream from a particular direction, e.g., if they originate from an accretion disk below the hot medium. However, after the first scattering, the photon paths become considerably more isotropic. For subsequent scattering orders, when photons are scattered from various directions, net polarization can arise if the planes of the final scattering (i.e., those immediately preceding escape toward the observer) have a preferred orientation. This situation is expected to occur if the medium has an elongated shape, where the photons have a higher chance to stay in the system and get scattered if they fly along the major axis. The PD is generally higher for more axially symmetric and more elongated photosphere shapes of the medium. A number of factors, such as the aberration and light bending effects, can reduce the net observed PD from the hot medium \citep{Poutanen1994,Schnittman2010}.

\begin{figure}[h!]
\centering
\includegraphics[width=1.0\textwidth]{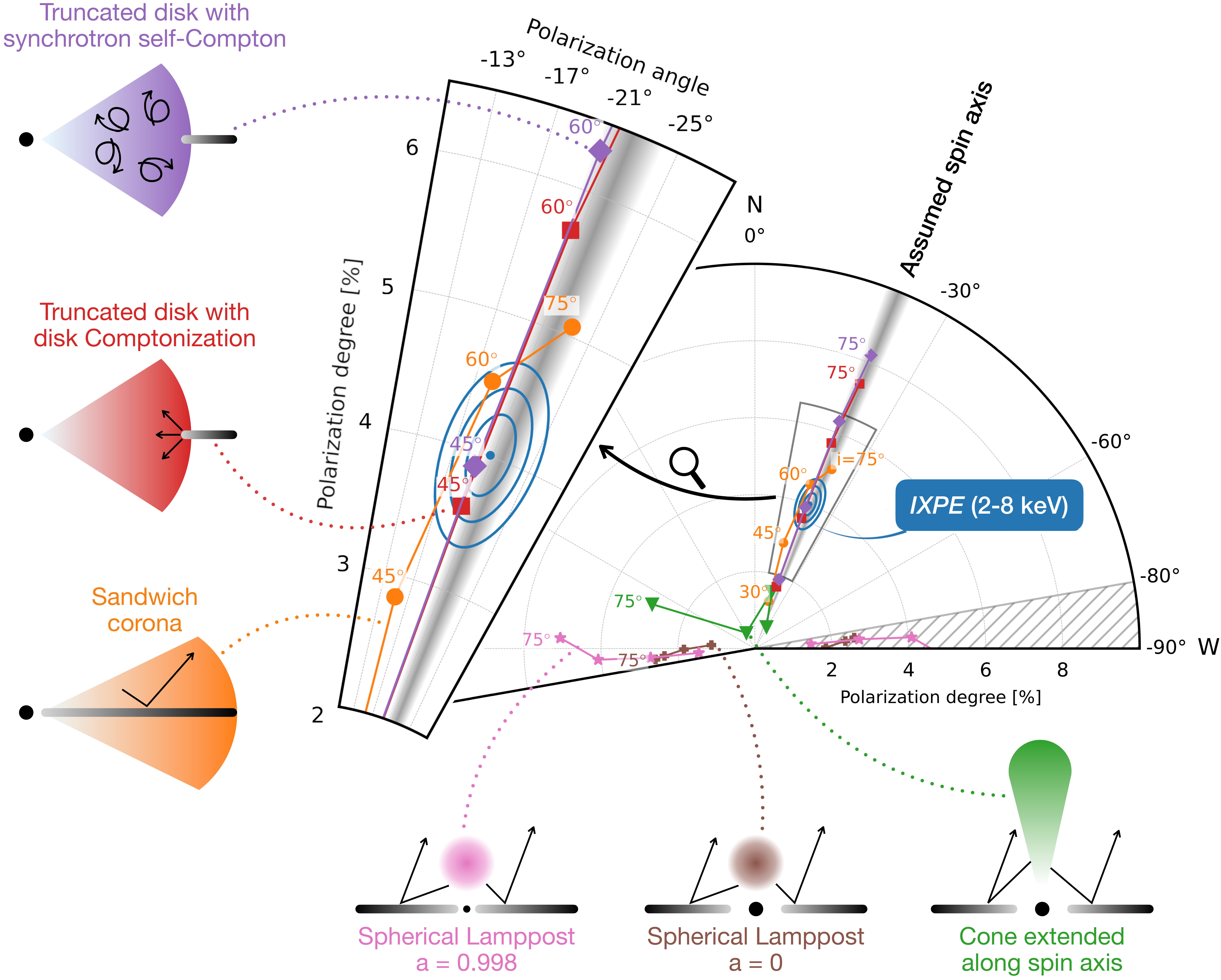}
\caption{Polar plot showing the observed PD and PA (1, 2 and 3$\sigma$ blue contours) of the first \textit{IXPE} observations of a stellar-mass BH (Cyg X-1) in the hard state. Observed PA aligns with the jet direction (grey area). Lines of different colors correspond to predictions of polarization from different geometries of the hot medium, numbers correspond to the inclinations. The geometries shown on the left are the ones consistent with the PA; the ones at the bottom are the inconsistent ones.
This figure was adapted from \citealt{Krawczynski2022}.}
\label{fig:CygX1pol}
\end{figure}

The first calculations of the expected polarization signatures were performed for the case of a slab geometry of various optical depths along the minor axis \citep{Sunyaev1985,Poutanen1996}. The PD generally increased with the increasing scattering order, reaching $\sim 15\%$ at high energies for the high-inclination sources and PA is aligned with the minor axis of the medium (along the disk axis) for high scattering orders. The spherical (lamppost) type of geometry can have non-zero polarization in the first scattering order due to the preferred direction of the incoming seed photons; PD decreases with energy due to spherical symmetry of the medium and PA is orthogonal to the first scattering plane (orthogonal to the disk axis). The compactness of the source and its location close to the BH are expected to lead to considerable light bending effects causing rotation of the PA \citep{Schnittman2010,Zhang2019,Niedzwiecki2026}. Finally, if the hot medium is elongated along the jet axis, the PA could be expected to be orthogonal to this axis \citep{Krawczynski2022a}. Colored lines in Fig.~\ref{fig:CygX1pol} give predictions for the PD and PA for different models and at various inclinations, denoted by numbers (note that the model PAs were rotated so that the disk axis aligns with the observed position angle of the jet in Cyg X-1).

Another important contribution to the net polarization can come from reflection, which is dominated by the Compton-scattered continuum and fluorescent iron line (see Sect.~\ref{sec:reflection}). Similar to the incident emission, the PD of the continuum reflected component depends on the viewing angle, however, also largely on the relative location of the  illuminator (hot medium) relative to the reflector (the accretion disk); with a point-like incident source illuminating a distant reflector generating a much higher polarization than a slab-like source illuminating the disk directly beneath it \citep{Matt1993, Poutanen1996refl, Dovciak2011, Podgorny2023}. The PA of reflected emission depends on the relative location of the incident source and reflector and can be both along the disk plane or axis. The fluorescent lines are produced unpolarized, but if these photons experience subsequent scattering in the disk atmosphere prior to escaping toward the observer, low polarization levels can be expected. Hence, the distinct signature of reflection contains the noticeable dip in PD around the Fe\,K$\alpha$ emission line.

The first hard-state BHXB to be observed by \textit{IXPE} was Cyg~X-1 \citep{Krawczynski2022}. The $2-8$\,keV polarization was observed to align with the resolved radio jet, with band-average ${\rm PD}=4.0\pm0.2\%$  (Fig.~\ref{fig:CygX1pol}) and an indication of an increasing trend with energy, which was confirmed in subsequent observations \citep{Kravtsov2025}. The considerable PD and the alignment of the PA with the jet axis indicate that the hot Comptonizing medium is extended perpendicular to the jet, presumably in the disk plane.
This result thus rules out models with geometries of either a compact spherical lamppost or a cone-like hot medium extended along the jet direction, and favors truncated disk and sandwich corona models (see Sect.~\ref{sec:hardXrays}). At the same time, the measured PD is much higher than the expected value for the low inclination angle of the Cyg~X-1 binary system \citep[$27.5^\circ$;][]{MillerJones2021}.  Even for a hot medium with a plane-parallel slab configuration, which maximizes the PD, the observed level can be achieved at a $\sim 45^\circ$--$60^\circ$ inclination. This result either indicates a higher inclination of the inner disk \citep[arising from a spin-orbit misalignment or from a superorbital precession,][]{Krawczynski2022,Kravtsov2025}, or the action of an additional mechanism that boosts the polarization \citep[bulk outflow or scattering in the disk wind,][]{Poutanen2023,Dexter2024,Tomaru2024,Nitindala2025}.

Reflection could also play a role in explaining the observed increase of PD with energy, despite contributing only at the level of $\sim 10\%$ of the $2-8$\,keV flux. Distant (lamppost+flared disk-type) reflection may give a boost to PD at the level up to a few percent, while local reflection (disk illuminated by the sandwich corona) instead leads to a reduction of the net polarization \citep{Poutanen1996refl,Podgorny2023}. However, it is unlikely that reflection plays a dominant role in explaining the observed polarization signatures, since no clear suppression of PD at the iron line energies has been observed for Cyg X-1-type sources.

\begin{figure}[h!]
\centering
\includegraphics[width=0.49\textwidth]{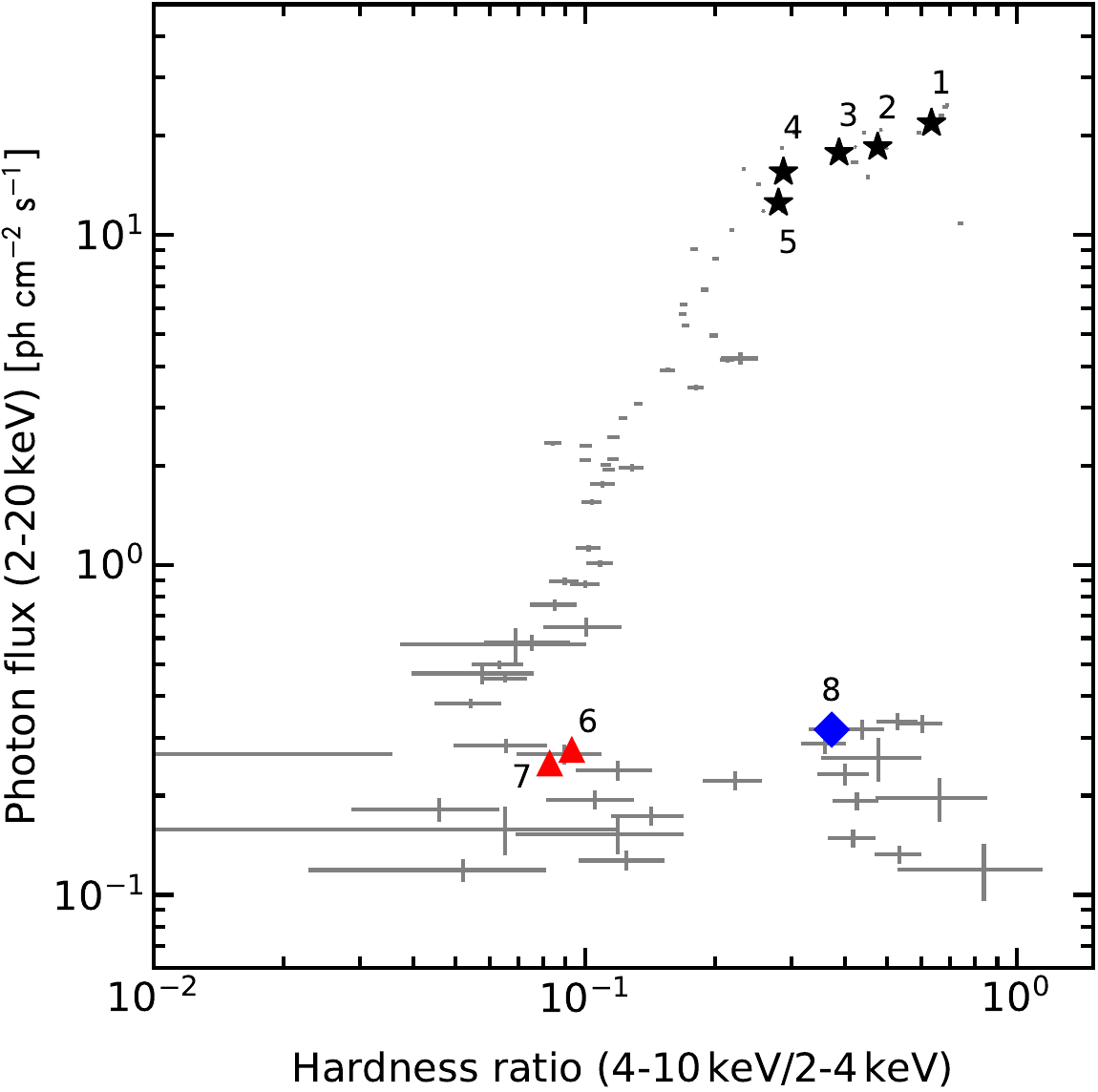}
\includegraphics[width=0.49\textwidth]{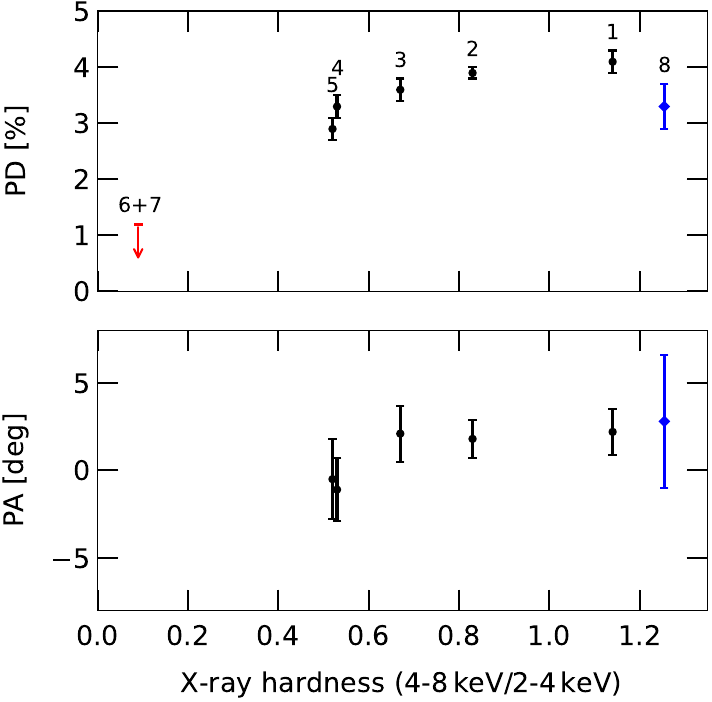}
\caption{Summary of the 2023--2024 outburst of Swift~J1727.8--1613. Left: The HID, showing that the first five observations were taken during the hard to soft state transition, the next two were taken during a dim soft state, and the final observation was during the soft to hard transition. Right: The PD (top) and PA (bottom) versus hardness. The PA is consistent with remaining constant, whereas the PD increases with hardness. Observation 8 follows the same relation as the others despite having a luminosity two orders of magnitude lower. Reproduced from \citet{Podgorny2024}.}
\label{fig:transpol}
\end{figure}

Subsequent observations of hard-state and intermediate-state sources have contributed to building a coherent picture of the geometry of the hot Comptonizing medium remaining extended in the plane orthogonal to the jet axis across various states. Whenever the detection or an indication of the jet direction is available (through the direct imaging of the extended compact jet or ejections; optical, sub-mm or radio polarization), the observed X-ray PA has remained aligned with the projected jet axis, with no detected energy dependence: Swift~J1727.8--1613 \citep{Veledina2023, Ingram2024, Podgorny2024}, GX~339--4 \citep{Mastroserio2025}, and IGR~J17091--3624 \citep{Ewing2025}. The remarkably stable corona geometry, indicated by the constant PA in the HIMS and SIMS \citep[Fig.~\ref{fig:transpol} and][]{Ingram2024,Mastroserio2025}, is in stark contrast to the dramatic increase in soft lag (see Fig.~\ref{fig:softlags}), which had previously been interpreted as evidence for a sharp increase in the light-crossing delay between direct and reflected components, implying a highly vertically extended geometry of the hot medium \citep[][see Sect.~\ref{sec:reflection}]{Wang2021, Wang2022, Mendez2022, Kylafis2024}. If the hot medium were to become vertically extended during the HIMS to SIMS transition, it would have caused a flip of PA by $90^\circ$, which is not observed.

\begin{figure}[h!]
\centering
\includegraphics[width=0.4\textwidth]{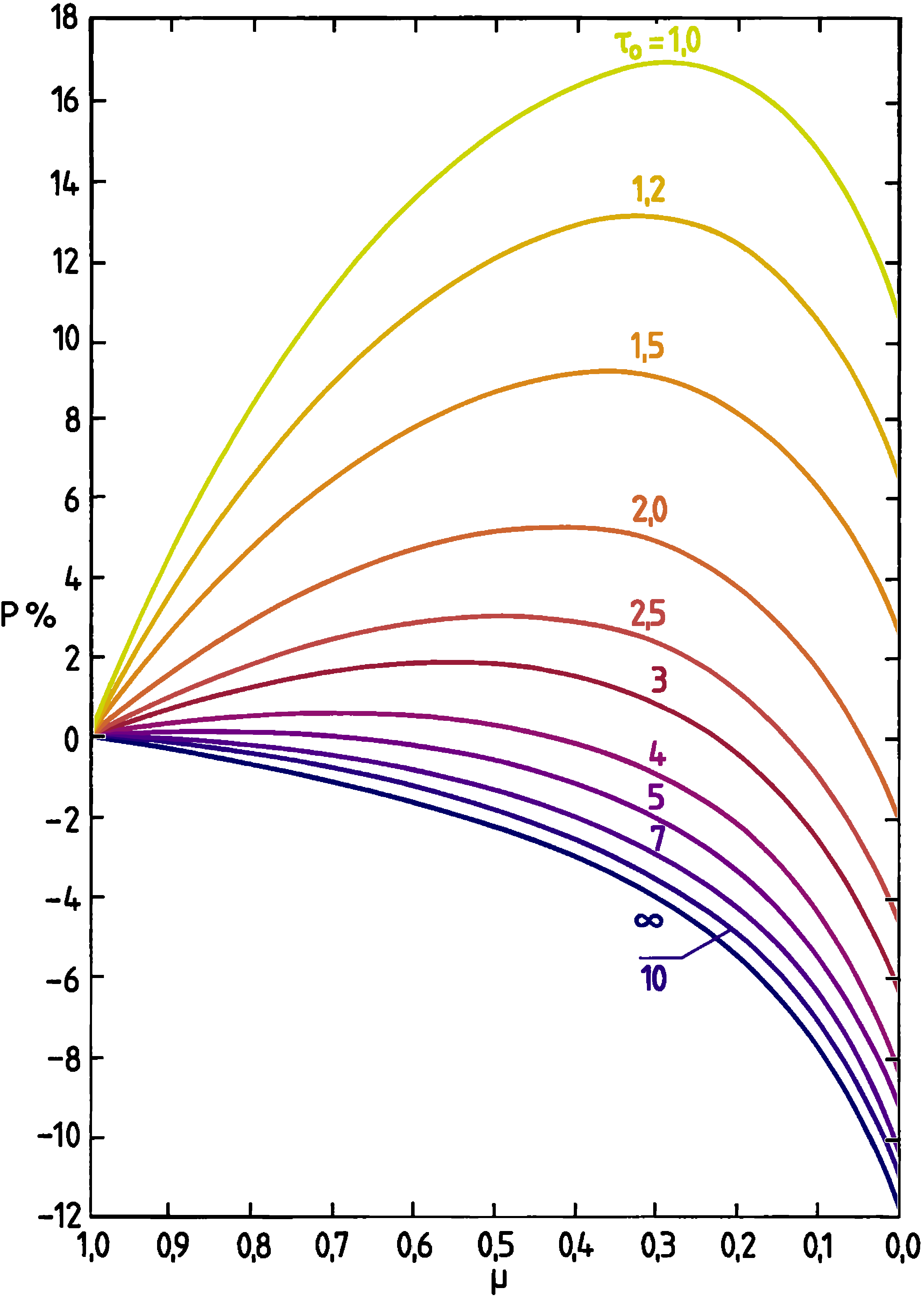}
\includegraphics[width=0.55\textwidth]{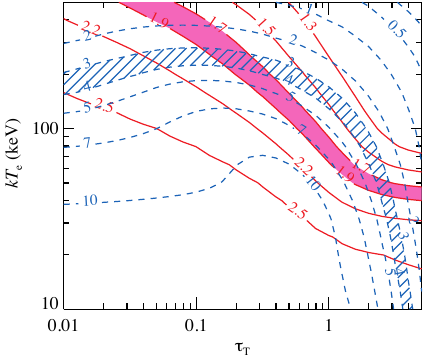}
\caption{Left: Predictions of PD for the case of a plane-parallel slab with Thomson scattering, as a function of cosine of the inclination ($\mu$) for different Thomson optical depth of the slab ($\tau_0$). Positive values of PD correspond to polarization along the slab normal. Right: Contours of constant spectral index ($\Gamma$, red) and PD (blue) for the case of Comptonization in a plane-parallel slab, as function of the electron temperature and vertical Thomson optical depth of the slab. Polarization is aligned with the slab normal. Adapted from \citet{Sunyaev1985} and \citet{Podgorny2024}.}
\label{fig:slab_pol}
\end{figure}

The detected X-ray polarization signatures were even more surprising for the observations of the reverse transition \citep[and Fig.~\ref{fig:transpol}]{Podgorny2024}, with both PA and PD following the same trend with spectral hardness as those obtained during HIMS observations of the outburst rise.  The two orders of magnitude lower luminosity in the decay stage translates to at least an order of magnitude lower optical depth. Therefore, the electron temperature must be higher for an observation in the decay stage than for an observation in the rising stage that exhibits the same Comptonizing continuum spectral slope (see the red solid contour lines in Fig \ref{fig:slab_pol}, right). In general, these differing conditions would be expected to result in different polarization signatures. Interestingly, it was possible to find two combinations of optical depth and electron temperature that yield consistent slopes and polarizations (Fig \ref{fig:slab_pol}, right), thus implying higher electron temperatures during the decay stage compared to the rise. However, quite \textit{why} this should be the case is still not clear.

An important prediction of the models with slab-type accretion geometry is the increase of PD with inclination. Detection of X-ray polarization in IGR~J17091--3624 provided the first opportunity to probe a high-inclination hard-state system: the source is thought to have an inclination in the range $\sim 60^\circ$-$80^\circ$ due to the presence of absorption dips in its light curve but the absence of eclipses. The detected PD of IGR~J17091--3624 was $\approx 9.1 \pm 1.6\%$ and aligned with the optical PA  \citep[although the jet has not been resolved so far,][]{Ewing2025}. It is just about possible to reproduce a PD of $9\%$ with a static Comptonizing slab model, but all parameters must be fine tuned to values that maximize the predicted PD, hence a boost of polarization may still be needed for this source as well \citep{Ewing2025}. 

X-ray polarization measurements also provide constraints on the line-of-sight magnetic field strength in the BH proximity. The Faraday effect in strong fields causes an energy-dependent rotation of the PA, which leads to depolarization at lower energies.  Thus, the observed upper limits on the energy dependence of the PA, together with the high (higher than predicted) observed PD, can be converted into upper limits on the magnetic field strength, yielding values $\lesssim 10^6-10^8$\,G, depending on the assumed field topology \citep{Barnier2024,Krawczynski2026}. However, these estimates rely on the simplified assumption of an external Faraday screen.
More accurate constraints on magnetic field threading the emission region require detailed radiative transfer calculations that self-consistently incorporate Faraday effects.

Together, the first X-ray polarimetric observations of hard- and intermediate-state systems are starting to build a picture that the PD is generally higher than that predicted by standard Comptonization models. On balance, it seems likely that a previously unconsidered physical mechanism is at play that boosts the PD, with the leading contenders being bulk outflow and scattering in a wind.

\subsection{Soft states}

In the soft state, the polarization is thought to be dominated by the thermal disk component, whose non-circular projection onto the plane of the sky leads to non-zero polarization signal. The emission of the disk that we see is coming from the upper layers of the disk atmosphere, the photosphere, hence predictions of polarization should rely on the mechanism of radiation production in this regime. Early works used the results of plane-parallel, pure electron scattering stellar atmospheres \citep{Chandrasekhar1960,Sobolev1963} to make first predictions of polarization from accretion disks \citep{Rees1975}. The inclination-dependent PD rises up to 11.7\% for edge-on systems and the PA remains perpendicular to the disk axis for all inclination angles (see the case of infinite optical depth in Fig.~\ref{fig:slab_pol}, left).  In fact, the potential detection of polarization was considered as an independent confirmation of accretion disk existence \citep{LightmanShapiro1975}.

An important addition to the theory of accretion disk polarization was consideration of the effects of strong gravity and fast matter motions. Aberration, light bending and frame dragging together alter the polarization orientation of different disk segments by a different amount, leading to distinctive signatures of depolarization and PA rotation with energy \citep{ConnorsStark1977,StarkConnors1977,PineaultRoeder1977a,PineaultRoeder1977b}. The rotation of PA with energy and, more broadly, the PD energy dependence, thus have been recognized as an important diagnostic of spacetime curvature, and in particular of BH spin \citep{Dovciak2008,Schnittman2009,LiNarayanMcClintock2009}. Additionally, extreme light bending close to the BH can cause some rays to return to the disk and scatter before eventually reaching the observer. This returning radiation can contribute to the bluest part of the disk spectrum and is strongly polarized in the direction of the symmetry (disk) axis \citep{Schnittman2009,Krawczynski2012}, causing an increasing PD with energy and, in the range where the polarization of this component dominates, a stable PA. The contribution of this component to the spectrum and the energy-dependent polarization crucially depends on the BH spin (practically, significant only for $a>0.9$) and the reflection albedo; we refer the reader to \citet{Niedzwiecki2026} for a more detailed discussion.

\textit{IXPE} has so far observed ten sources in the soft and soft-intermediate state. Polarization was not significantly detected for four of these: 
LMC~X-1 \citep{Podgorny2023a}, 
Swift~J1727.8--1613 \citep{Svoboda2024a},
Swift~J151857.0--572147 \citep{Ling2024}, and
MAXI~J1744--294 \citep{Marra2026}. 
It is likely that these sources are viewed from low inclination ($i \lesssim 60^\circ$), and thus the PD is too low for \textit{IXPE} to make a detection. There are four sources with firm detections in pure soft states 
LMC~X-3 \citep{Svoboda2024}, 
4U~1957+115 \citep{Marra2024}, 
and 4U~1630--47 \citep{Ratheesh2024}, and
GRS~1739--278 \citep{Zhao2026}. There are also two sources with peculiar detections. Cyg~X-1 shows a firm detection in its softest states \citep{Steiner2024,Kravtsov2025}, but this source never formally resides in a soft state and we thus classify it as a soft-intermediate (SIMS) detection. In turn, GX~339--4 is clearly polarized in the soft-intermediate state but has an upper-limit in the soft state \citep{Mastroserio2025}.

\begin{figure}[h!]
\centering
\includegraphics[width=1.0\textwidth]{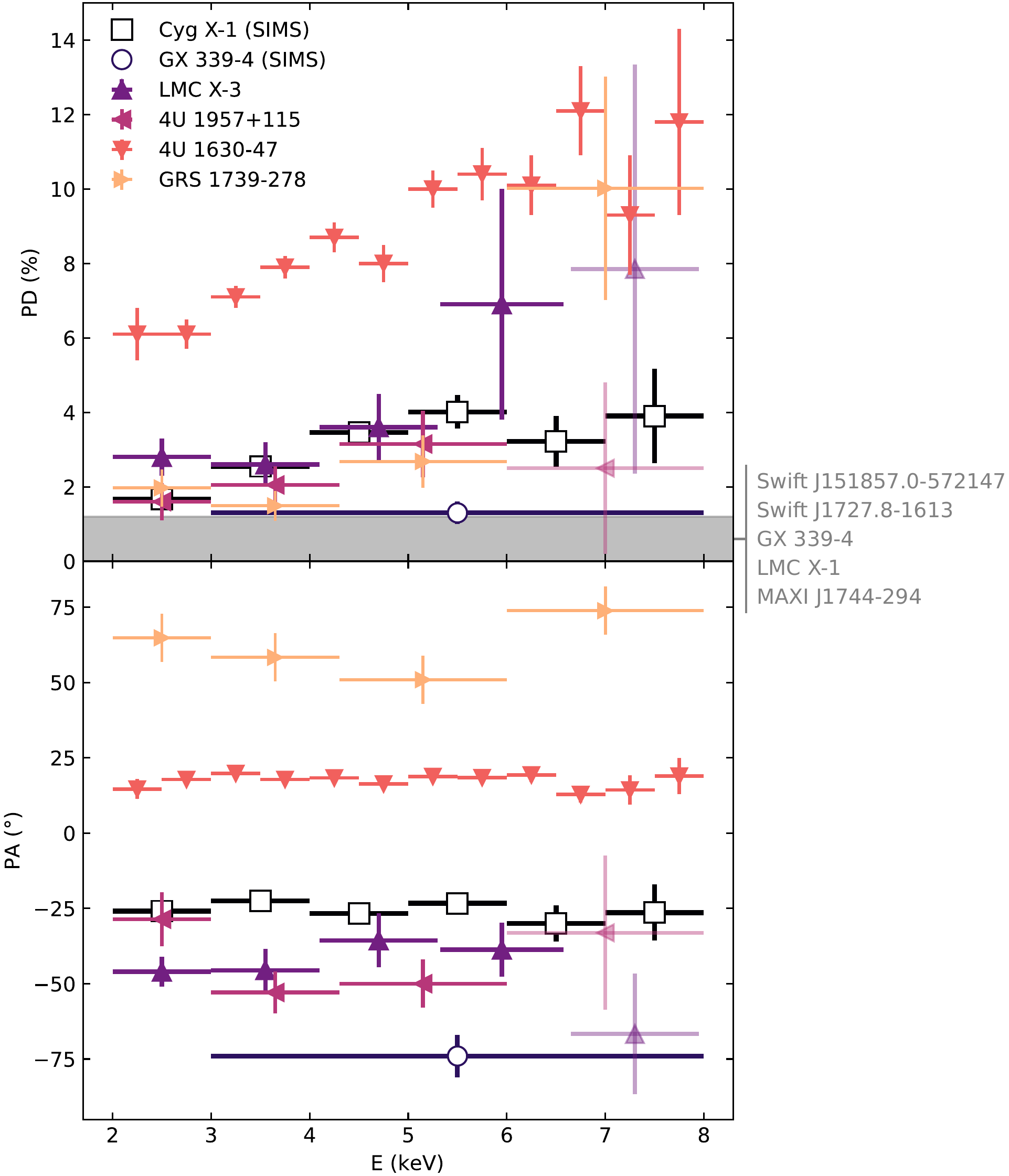}
\caption{PD and PA in the soft state of all sources observed with \textit{IXPE} so far. Sources with upper-limits in the soft-state are within the gray-shaded area (drawn at 1.2\% here), and sources observed in the soft-intermediate (SIMS) are shown with white markers. The two transparent points (LMC~X-3 and 4U~1957+115) are the ones below minimum detectable polarization at the $99\%$ confidence level (MDP99). The PD of 4U~1630--47 is far larger than that of the other sources observed in the soft state.}
\label{fig:SSpol}
\end{figure}

Fig.~\ref{fig:SSpol} illustrates the energy dependence of PD for sources in the soft- and soft-intermediate states. The majority of sources exhibit generally low PD, even in cases of high inclinations ($i\approx 70^\circ$ for LMC~X-3 and 4U~1957+115). At the same time, all sources show a tendency for PD to increase with energy, and in several cases this trend is statistically significant. This behavior stands in clear contrast to the early expectations of progressive depolarization with increasing energy, controlled by the BH spin. One possible explanation for this discrepancy is the oversimplified assumption of an energy-independent, pure electron-scattering disk. In more realistic disk atmospheres, absorption is expected to play a role; this can both enhance the PD and introduce an energy dependence, as well as align the polarization with the disk axis \citep{LoskutovSobolev1979,LoskutovSobolev1981,Taverna2021}. Alternatively, the observed increase of PD with energy may be attributed to the contribution of returning radiation, which would imply that sources exhibiting a statistically significant rise in PD host rapidly spinning BHs. Indeed, spectro-polarimetric modeling that includes returning radiation has yielded extreme spin estimates $a\gtrsim0.96$ for 4U~1957+115 \citep{Marra2024}, GRS~1739--278 \citep{Zhao2026} and Cyg~X-1 \citep[][but see \citealt{Niedzwiecki2026}]{Steiner2024}. Interestingly, these spin estimates are consistent with previous measurements based solely on spectral fitting \citep[see, however,][]{2026NewAR.10201746Z}, suggesting that they are primarily driven by the energy spectrum rather than by polarization constraints. A similar conclusion follows from the lower-spin system LMC~X-3 \citep[$a<0.7$,][]{Svoboda2024}, whose energy-dependent PD closely resembles that of 4U 1957+115, despite the latter being consistent with a near-maximally spinning BH.

The energy dependence of PA provides an independent probe of spacetime curvature. Regardless of mechanisms that may enhance locally produced polarization, the variation of PA with energy primarily traces the rotation of photon trajectories as they propagate toward the observer. Observations indicate that, whenever polarization is significantly detected, the PA remains consistent with being energy-independent (see Fig.~\ref{fig:SSpol}). This behavior can be interpreted either as evidence for a dominant returning-radiation component, implying an extreme BH spin, or, instead, can be used as a means to place an upper limit on the BH spin by requiring that PA rotation remains within the observational uncertainties.

The jet direction is known in two sources with significant detection of X-ray soft-intermediate state polarization: Cyg~X-1 and GX~339--4 \citep{Steiner2024,Mastroserio2025}. For the case of Swift~J1727.8--1613 with a formal upper limit on polarization, the preferred direction is likewise consistent with the jet axis \citep{Svoboda2024a}. This can be considered as a factor that favors the scenarios with returning radiation or a significant role of absorption in the disk atmospheres.

One system, 4U~1630--47, stands out as a spectacular outlier. The source was observed three times: once in the soft-state and twice in the very high state \citep{RodriguezCavero2023,Ratheesh2024}. In all cases, the PD was high, $>6\%$, and showed a highly significant increase with energy. The two main explanations that have been invoked to explain this unexpectedly high PD are essentially the same as those suggested for the hard state: a bulk outflow of the scattering material, this time of the electrons in the disk atmosphere \citep{Ratheesh2024}, or scattering in an accretion disk wind \citep{Tomaru2024,Nitindala2025}. Both scenarios can boost the intrinsically-polarized light of the source, yet to get to the observed values (up to $\sim 10\%$), the parameters need to be stretched. This may indicate that a fundamentally different geometry or mechanism is at play in this source.

\subsection{Obscured State}

\begin{figure}[h!]
\centering
\includegraphics[width=1.0\textwidth]{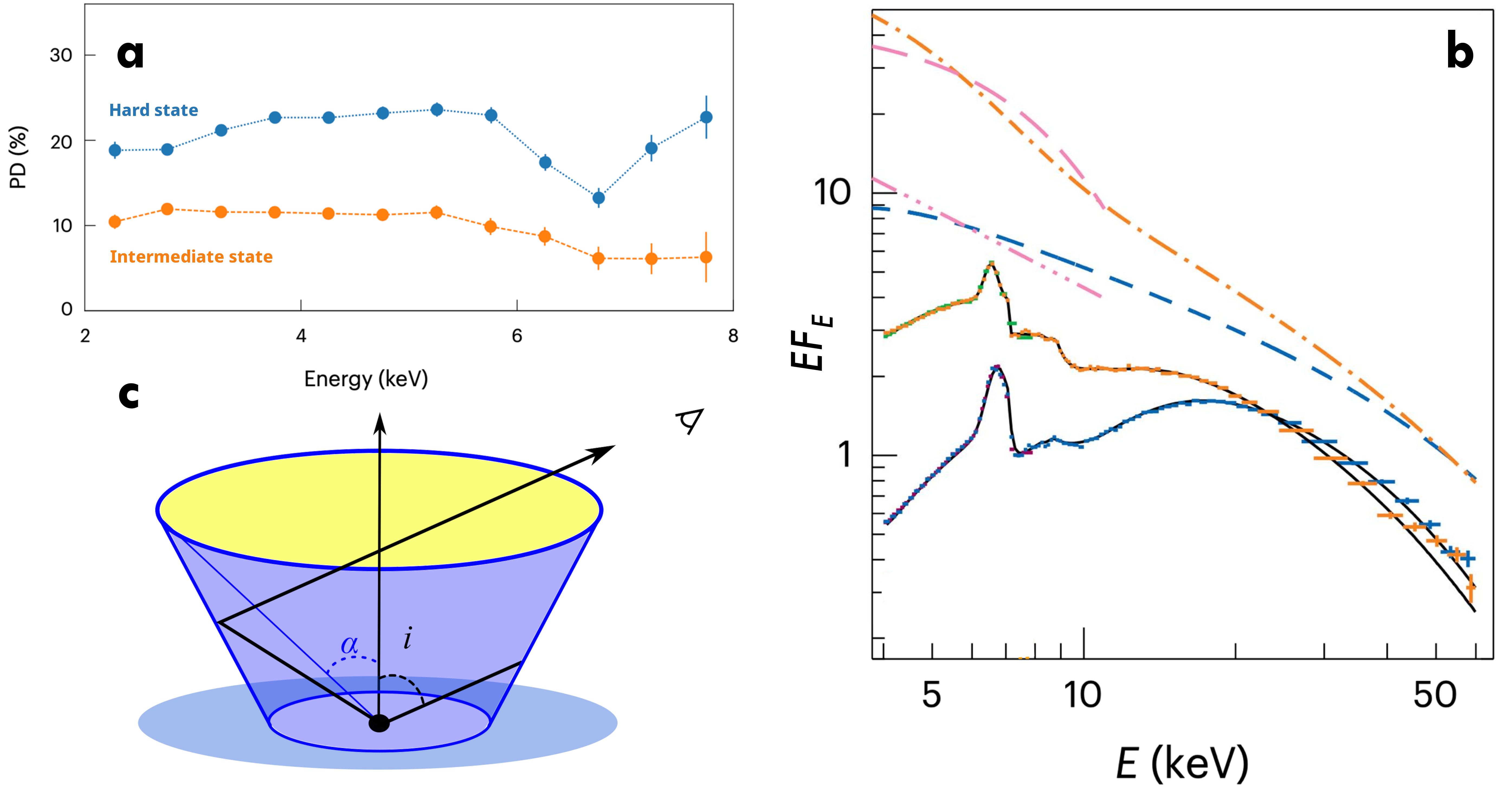}
\caption{\textit{IXPE} results of Cyg~X-3 observations and considered accretion geometry. Panel a: PD as a function of energy \textit{IXPE} during the hard-state (blue) and intermediate-state observation (orange). Panel b: broadband spectral modeling of \textit{IXPE} and \textit{NuSTAR} data and the inferred intrinsic spectrum of the source (blue dashed and orange dot-dashed). Intrinsic spectra are compared to known ULX spectra (pink). Panel c: Schematic of the inferred geometry, where $i > \alpha$ for Cyg~X-3. Adapted from \citet{Veledina2024}.}
\label{fig:CygX3}
\end{figure}

Of all the BHXBs observed by \textit{IXPE}, by far the most highly polarized is Cyg~X-3 \citep{Veledina2024, Veledina2024b, Mikuvincova2026}. The source was observed in three spectral states: hard, intermediate and soft, and in all cases the average PD exceeded $10\%$ with PA being perpendicular to the jet direction (see Fig.~\ref{fig:CygX3}). A clear drop in PD at the Fe K$\alpha$ line was observed during the hard state, becoming less significant or absent during the intermediate and soft states. Away from the iron line region, the hard-state PD reached $\sim 23\%$; a value that is by no means possible to achieve in Comptonization models, even in idealized settings. The high polarization is even more puzzling given the low inclination of the system, $i=29\fdg5\pm1\fdg2$ \citep{2022ApJ...926..123A}.

The high PD and the PA perpendicular to the jet direction, together with a distinct spectral shape with prominent iron line features, suggest that the continuum observed in the \textit{IXPE} band is dominated by reflected emission \citep[Fig.~\ref{fig:CygX3} and][]{Veledina2024}. The importance of reflection from an optically thick medium had previously been recognized in one of the possible scenarios for spectral formation in Cyg X-3 \citep{Hjalmarsdotter2008}, yet the apparent dominance of this component,  had not been realized before. To explain the observed polarization, the incident emission of the source must be hidden from direct view.  This, in turn, implies that the optically thick envelope, presumably the same structure responsible for the reflection features, extends to high elevations above the disk plane. In this picture, the central compact object in Cyg X-3 is embedded within an optically thick medium of roughly conical geometry, with radiation escaping only through a narrow axial cavity characterized by a grazing angle $\alpha$ (serving as a proxy for the funnel half-opening angle; Fig \ref{fig:CygX3}c). The observer viewing at $i>\alpha$ only sees light scattered from the wall of the cone, analogous to the geometry in Type-2 AGN.

Modeling of the PD in the single-scattering Thomson limit provides a constraint on the half-opening angle of the scattering cone of $\lesssim 15^\circ$, given the inclination of Cyg X-3, $i \approx 30^\circ$. Such a small opening angle implies that an observer with their line of sight within the cone would see a very bright source, with an apparent luminosity exceeding $5 \times 10^{39}~{\rm erg}~{\rm s}^{-1}$, rendering Cyg X-3 a Galactic ultraluminous X-ray source (ULX). This implies that at least a fraction of ULXs observed in other galaxies may be strongly geometrically beamed systems analogous to Cyg~X-3. It also implies that binaries in an obscured state like Cyg~X-3 can be common, with most of them viewed outside of the funnel and thus appearing as sub-Eddington sources, with only a small fraction ($f\sim 1-\cos\alpha \lesssim 3.4\%$) oriented sufficiently face-on to be seen as ULXs. Indeed, several other Galactic BHXBs show evidence of being intrinsically bright, yet heavily obscured: SS~433 \citep{Fabrika2004, Koljonen2020, Middleton2021}, V404~Cyg, V4641~Sgr and GRS~1915+105 \citep[][]{Koljonen2020}, making them promising targets for probing the obscuration scenario in future \textit{IXPE} observations.

\section{Conclusions and future prospects}

Six decades of multi-wavelength observations have established that black hole X-ray binaries cycle through distinct spectral states governed by the coupled dynamics of several components: the accretion disk, a hot inner flow (or corona), disk winds, and relativistic jets. Energy and momentum exchange between these regions leaves distinct, measurable signatures across the electromagnetic spectrum. Outbursts, state transitions, and the launching and quenching of winds and jets all arise from this coupling, even if the underlying microphysics remains contested. X-ray binaries provide access to accretion flows over the widest range of radii, with X-rays tracing the final stages of inflow onto the compact object and the highest-energy interactions that feed jets and winds. Black hole X-ray transients offer particularly sharp views of disk evolution over changing mass-transfer rates; in the last decade, the two nearby BHXBs MAXI\,J1820+070 and V404\,Cyg have enabled dense, multi-wavelength coverage on all timescales. In V404\,Cyg, pre-outburst disk evolution consistent with the Disk Instability Model could be followed only thanks to long-term robotic monitoring. Yet key questions remain unresolved. The physical origin of spectral state transitions is not understood from first principles. The corona -- its geometry, heating, and connection to winds and jets -- is still poorly constrained. While its name suggests similarities to the solar corona, this analogy is misleading as the conditions (temperature, densities, pressures) are vastly different. The conditions that regulate jet and wind production, and the reasons why outbursts differ so dramatically between sources, are likewise unsettled. These problems lie at the heart of accretion physics in the strong-gravity regime.

Rapid multi-wavelength variability is emerging as one of the most incisive probes of emission geometry, because temporal signatures can differ strongly even when spectra appear similar. Radio-to-optical time-series analysis of MAXI\,J1820+070 reveals characteristic break timescales that scale linearly with wavelength and much stronger variability at the shortest wavelengths, consistent with a jet in which the characteristic emission-region size grows in proportion to wavelength at roughly fixed opening angle and speed. In parallel, emission-line diagnostics are maturing into precision tools for mapping disk structure and outflows. Line profiles encode a velocity-space projection of the disk, and empirical correlations now allow the compact-object mass, mass ratio, and binary inclination to be inferred from profile shapes alone. Such indirect methods are crucial for expanding the census, since most systems reside in the heavily extincted Galactic Plane, where companion stars are too faint for classical dynamical studies even with the largest O-IR telescopes, leading to strong selection biases toward the brightest and nearest systems.

The nature and origin of the X-ray corona remain central open issues. Current evidence favors a hot, optically thin inner accretion flow close to the black hole, in line with ideas developed in the 1970s–1980s. X-ray polarimetry with \textit{IXPE} now adds powerful new constraints. In Cyg\,X-1’s hard state, the 2–8\,keV polarization (PD $=4\%$) aligns with the resolved radio jet, implying a corona extended perpendicular to the jet, likely in the disk plane. Across several sources, the measured polarization degree is systematically higher than predicted by standard Comptonization models; reaching the observed values with a slab geometry requires extreme, fine-tuned parameter choices. An additional mechanism must therefore boost the polarization, with bulk Comptonization and scattering in an outflowing wind among the leading candidates. In the soft state, 4U\,1630--47 stands out: the PD rises from $\approx 6\%$ to $\approx 10\%$ across the \textit{IXPE} bandpass, incompatible with standard disk-atmosphere models. The strikingly stable PD and PA across the state transition of Swift~J1727.8--1613 also contradict earlier interpretations of the large soft-lag increase as evidence for a vertically extended corona, which would produce a $90^\circ$ PA rotation that is not observed. These results challenge several established geometric pictures and highlight the diagnostic power of X-ray polarimetry.

The opening of X-ray polarization as a new observational window, together with increasingly capable multi-wavelength facilities in the radio (MeerKAT/SKA), IR (JWST), and optical (ELT, Rubin), will provide new opportunities to address these questions.

Looking ahead, several priorities stand out:
\begin{itemize}
\item Exploitation of the growing archive of X-ray transient outbursts, which now provides a rich, state-resolved database of spectra and their evolution, even though the theoretical interpretation of the underlying processes remains contentious,
\item Coordinated multi-wavelength monitoring to characterize variability on all timescales, from rapid QPOs that probe the inner disk to superorbital cycles tracing warps and tilts in the outer disk, which remain difficult to follow systematically,
\item High spectral- and time-resolution optical-infrared emission-line spectroscopy over full orbital and outburst cycles, enabling Doppler tomography of the disk and supporting indirect BHXB search strategies,
\item Sustained \textit{IXPE} operations and development of next-generation X-ray polarimeters to refine constraints on coronal and disk-atmosphere geometry, resolve the unexpectedly high polarizations, and discriminate between bulk Comptonization and wind scattering as the dominant PD-boosting mechanism,
\item Large-area X-ray timing missions (e.g., \textit{eXTP}, \textit{STROBE-X}) to detect high-frequency QPOs in a statistically meaningful sample, monitor low-frequency QPOs with all-sky instruments, and enable systematic detection of Fe K reverberation lags (not yet robustly detected) offering new routes to constraining black hole spins and testing strong-field gravity,
\item GRMHD and radiative transfer simulations capable of following full outburst cycles, including hysteresis, the diversity of outburst morphologies, and coupled spectral-timing-polarimetric behavior; closing the gap between such models and observations remains a central theoretical frontier.
\end{itemize}

%\clearpage
\vspace{1cm}

\backmatter

\begin{multicols}{2}

\bmhead{Supplementary information}

Not applicable

% If your article has accompanying supplementary file/s please state so here. 

% Authors reporting data from electrophoretic gels and blots should supply the full unprocessed scans for key as part of their Supplementary information. This may be requested by the editorial team/s if it is missing.

% Please refer to Journal-level guidance for any specific requirements.

\bmhead{Acknowledgements}

%Acknowledgements are not compulsory. Where included they should be brief. Grant or contribution numbers may be acknowledged.
%Please refer to Journal-level guidance for any specific requirements.

We thank the editor and the referees, including A. Zdziarski, for their careful reading and constructive comments that improved this manuscript.
All authors acknowledge the support of the International Space Science Institute (ISSI) in Bern through ISSI Workshop ``Accretion Disks: The First 50 Years''.
GM and AV acknowledge financial support the Academy of Finland grants 355672 and 372881. GM acknowledges financial support from the Polish National Science Center grant 2023/48/Q/ST9/00138.
PC thanks the Leverhulme Trust for the award of an Emeritus Fellowship.
%TJM thanks the Dublin International Airport, through which he travelled for this workshop, for making him appreciate the efficiency of Charles de Gaulle.
%
This work has made use of the HMXB (https://binary-revolution.github.io/HMXBwebcat/) and LMXB (https://binary-revolution.github.io/LMXBwebcat/) catalogues maintained by the Binary rEvolution team (https://github.com/Binary-rEvolution). This work was written during troubled times for many people around the world, and our thoughts go to the oppressed, whoever and wherever they may be.

\section*{Declarations}

Not applicable

\footnotesize
\setlength{\bibsep}{2pt}
\bibliography{sn-bibliography}
\end{multicols}

\end{document}